%% file: 0_NDSS2024_main.tex
\documentclass[conference]{IEEEtran}

\usepackage{diagbox}
\usepackage{multirow}
\usepackage{algorithm}
\usepackage{amsmath}
\usepackage{amssymb,amsfonts}
\usepackage{hyperref}
\usepackage{cleveref}
\usepackage[noend]{algpseudocode}
\usepackage{graphicx}
\usepackage{caption}
\usepackage{subcaption}
\usepackage{bbding}
\usepackage{makecell}
\newcommand*{\Scale}[2][4]{\scalebox{#1}{$#2$}}%
\usepackage{pifont}

\usepackage[square,sort,comma,numbers]{natbib}
\usepackage{booktabs} 
\usepackage{color,xcolor}
\usepackage{mathrsfs}  
\usepackage{mathtools}

\usepackage{stfloats}

\newtheorem{theorem}{Theorem}
\newtheorem{lemma}{Lemma}
\newtheorem{proof}{Proof}[section]
\usepackage[percent]{overpic}
\usepackage{tikz}
\usepackage{adjustbox}
\usepackage{threeparttable}

\begin{document}
\IEEEoverridecommandlockouts

\title{LADDER: Multi-objective Backdoor Attack via Evolutionary Algorithm}

\author{\IEEEauthorblockN{Dazhuang Liu\IEEEauthorrefmark{1}\thanks{\IEEEauthorrefmark{1}The first two authors contributed equally to this work.},
Yanqi Qiao\IEEEauthorrefmark{1},
Rui Wang, 
Kaitai Liang and
Georgios Smaragdakis}
\IEEEauthorblockA{\{d.liu-8, y.qiao, r.wang-8, kaitai.liang, g.smaragdakis\}@tudelft.nl}
\IEEEauthorblockA{% 
Delft University of Technology}
}

\maketitle

\input{1_abstract}

\input{2_introduction}

\input{3_RelatedWork}

\input{4_background}

\input{5_methodology}

\input{6_experiments}

\input{7_conclusion}

\bibliographystyle{IEEEtranS}
\bibliography{9_bib}

\input{8_appendix}

\end{document}

%% file: 1_abstract.tex
\begin{abstract}

Current black-box backdoor attacks in convolutional neural networks formulate attack objective(s) as \textit{single-objective} optimization problems in \textit{single domain}. 
Designing triggers in single domain harms semantics and trigger robustness as well as introduces visual and spectral anomaly. 
This work proposes a multi-objective black-box backdoor attack in dual domains via evolutionary algorithm (LADDER), the first instance of achieving multiple attack objectives simultaneously by optimizing triggers without requiring prior knowledge about victim model.
In particular, we formulate LADDER as a multi-objective optimization problem (MOP) and solve it via multi-objective evolutionary algorithm (MOEA). 
MOEA maintains a population of triggers with trade-offs among attack objectives and uses non-dominated sort to drive triggers toward optimal solutions. 
We further apply preference-based selection to MOEA to exclude impractical triggers. 
We state that LADDER investigates a new dual-domain perspective for trigger stealthiness by minimizing the anomaly between clean and poisoned samples in the spectral domain.
Lastly, the robustness against preprocessing operations is achieved by pushing triggers to low-frequency regions. 
Extensive experiments comprehensively showcase that LADDER achieves attack effectiveness of at least 99\%, attack robustness with 90.23\% (50.09\% higher than state-of-the-art attacks on average), superior natural stealthiness (1.12$\times$ to 196.74$\times$ improvement) and excellent spectral stealthiness (8.45$\times$ enhancement) as compared to current stealthy attacks by the average $l_2$-norm across 5 public datasets.

\end{abstract}

%% file: 2_introduction.tex
\section{Introduction}
\label{sec:intro}

Convolutional neural networks (CNNs) \cite{o2015introduction} have become an effective machine learning (ML) technique for image classification. 
They have proved to be vulnerable to backdoor attacks \cite{fiba,badnets,wanet}, allowing an attacker to mislead a victim model with incorrect yet desired predictions on poisoned images during inference while behaving normally on clean images.
These attacks pose severe risks to real-world applications, e.g., tumor diagnosis \cite{esteva2017dermatologist}, self-driving cars \cite{autodriving}. 

Some service providers of safety-critical applications may choose to collect data online to train a private model and prevent attackers from accessing their systems.
In this sense, backdoor attacks in the black-box setting are proposed \cite{Chen2017TargetedBA}. 
Under a black-box scenario, attackers do not have knowledge about the models and cannot manipulate training but they may poison training data by designing triggers. 

An ``ideal" trigger should satisfy \emph{stealthiness},
\emph{robustness}, attack \emph{effectiveness} and \emph{functionality} \cite{Wang2023RobustBA}. 
Stealthiness concerns the invisibility of trigger in the poisoned image to human visual perception; robustness is evidenced by its ability to withstand image preprocessing; 
effectiveness requires that backdoor to be successfully injected into the victim model; and functionality preservation
requires that inference accuracy on benign data remains unaffected.

Current designs for trigger stealthiness in the spectral domain are impractical. 
Conventional pixel-based backdoor attacks \cite{badnets,sig, wanet} inject triggers into spatial domain. 
Since spatial domain contains abundant semantic information, putting triggers into pixels can be easily detected by visual inspection.
Recent works \cite{ft,fiba,DUBA} thus design backdoor attacks by injecting triggers into spectral domain. 
Inspired by patch-based backdoor attacks, FTrojan \cite{ft} manipulates the mid- and high-frequency spectrum of images by inserting predefined perturbations to fixed frequency bands.
Manually crafting triggers in high-frequency components harms robustness, as most image preprocessing operations, e.g., low-pass filtering and JPEG compression, lead to greater information loss on these components.
Current spatial and frequency triggers \cite{badnets,sig,refool,ft,fiba,LFAP} introduce distinguishable artifacts in spectral and/or spatial domain (see \Cref{fig:freq_disparity_map} in \Cref{sec:exp}), which bear a high risk of existing attacks being detected.

\noindent{\bf A new perspective - starting with stealthiness.} Considering both spatial and spectral domains \cite{dualspace_concept1,dualspace_concept2} which we call dual domains hereafter, 
this work aims to achieve \emph{dual-domain stealthiness}: (1) spatial stealthiness, which guarantees the injection of trigger into the image does not harm cognitive semantics or introduce visual anomaly, and (2) spectral domain stealthiness, which avoids the disparities of frequency spectrum between clean and poisoned images.
In contrast, despite the stealthiness achieved by Wang et al. \cite{Wang2023RobustBA} at pixel level, we shed lights on stealthiness in the spectral domain as well as guaranteeing all the attack goals mentioned above.

\noindent{\bf Benefit the stealthiness and robustness in low-frequency domain.} Cox et al. \cite{watermark} claim that low-frequency components of natural images contain semantic information understandable to humans, whereas high-frequency ones stand for details and noise. 
Based on this, works \cite{watermark,LFAP} state two benefits of inserting triggers in low-frequency domain: 
(1) abundant information contained in low-frequency domain can provide a high perceptual capacity of accommodating trigger patterns without perceptual degradation, which improves trigger stealthiness; 
and (2) low-frequency components can bear better resilience in image compressions and are less prone to be removed by image filtering than mid- and high-frequency components, which guarantees a better attack robustness.

Achieving multiple attack objectives simultaneously in black-box backdoor attack is not trivial.  
Current backdoor attacks either adopt a fixed trigger pattern \cite{badnets, sig, ft,fiba}, or optimize triggers \cite{IBA, defeat, lira, wanet, color} by leveraging Lagrange multipliers to aggregate attack objectives into a single-objective problem (SOP) with gradient descent. 
{Conflicts between attack objectives (e.g., effectiveness and stealthiness) make tuning Lagrange coefficients challenging without prior knowledge. 
One lacking prior knowledge must repeatedly perform the single-objective optimization to identify a practical setting for Lagrange coefficients among objectives (see \Cref{fig:rela_disparity_alpha_asr}(a)).
Furthermore, applying Lagrange multipliers with SGD often fails to reliably produce practical triggers that optimally balance objectives 
(see \Cref{fig:rela_disparity_alpha_asr}(b)). 
}

\noindent{\bf Optimization to Multiple Objectives.} 
We aim to develop a backdoor attack to optimize triggers in the low-frequency region while ensuring attack effectiveness, functionality, dual-domain stealthiness, and robustness simultaneously without necessitating internal information of the victim model and in a Lagrange coefficient-free manner.
Developing such an optimal trigger that meets multiple objectives is non-trivial.
First, in the black-box setting where the target model is inaccessible to attackers, it is not possible to acquire gradient information and predict trigger performance on the victim model, which is therefore hard to find optimal trigger with gradient descent; 
also, improperly handcrafted fixed trigger (with a {predefined} magnitude of perturbations and locations) in the spectrum lead to improper signals in the spectrum and poor attack effectiveness. 
For example, large perturbation triggers like \cite{ft, fiba} disrupt invisibility and alter image semantics, while small perturbations could prevent the model from learning the trigger, reducing attack effectiveness. 

\begin{figure}[!bpt]
    \centering
    \scalebox{0.45}{\includegraphics{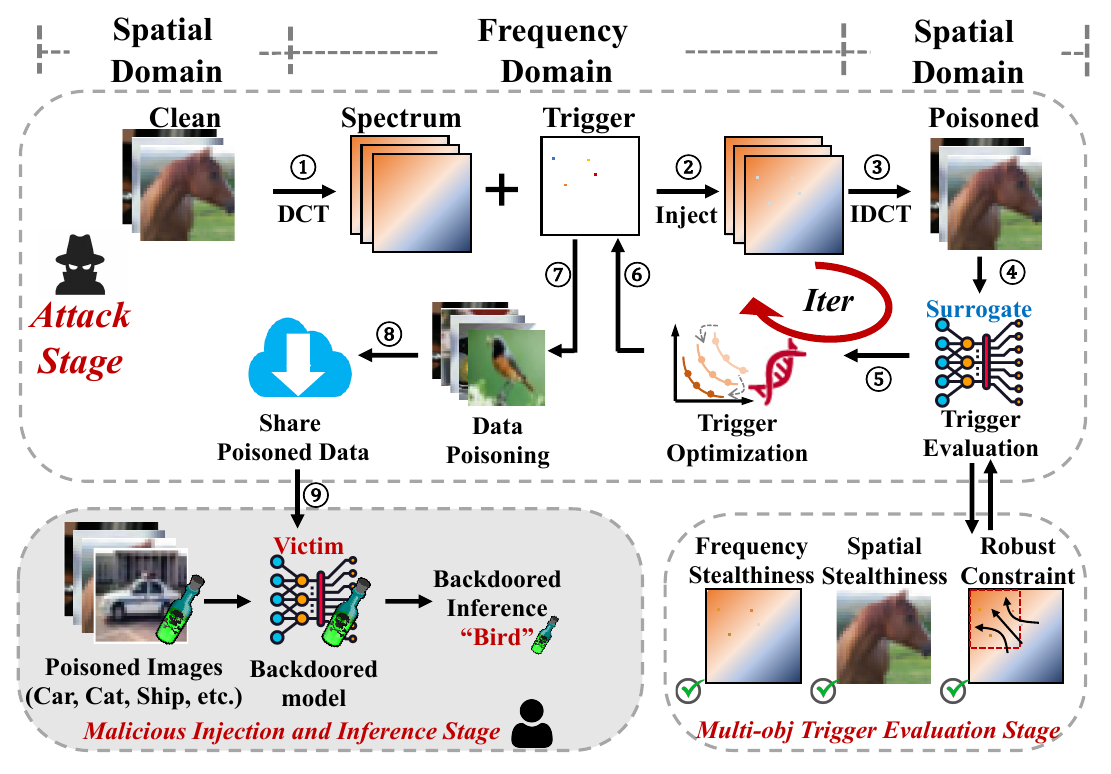}}
    \caption{The workflow of LADDER. 
    Step \ding{172}-\ding{174}: trigger injection;
    Step \ding{175}-\ding{177}: main loop for trigger optimization;
    Step \ding{178}-\ding{179}: poison dataset with trigger and release it to public;
    Step \ding{180}: the backdoor is injected when users download the poisoned data to train/tune their own model.
    The trigger optimization, evaluation, and injection are controlled by an attacker, whereas the malicious training and inference stage (marked in grey) are unseen to the attacker.
    }
    \label{fig:EMOBAF_workflow_chat}
\end{figure}

This work develops LADDER, a new black-box backdoor attack that leverages MOEA \cite{nsgaii}, a gradient-free optimization method, to effectively generate triggers in the spectral domain (see \Cref{fig:EMOBAF_workflow_chat} for its workflow). 
We maintain attack effectiveness, dual-domain stealthiness and robustness against image preprocessing operations simultaneously, obtaining practical triggers (see red dots in \Cref{fig:objective_conflicting}) without the need for tuning sensitive coefficients. 
{Specifically, we randomly initialize a population of triggers, each of which represents a unique trade-off across attack objectives. 
During optimization, we iteratively apply variations, such as crossover \cite{Deb1995SimulatedBC} and mutation \cite{Deb1996ACG}, to change the magnitude of perturbations and locations of triggers to produce a new set of candidate triggers. 
We then evaluate the performance of each trigger based on the values calculated by attack objectives (see \Cref{eq:moo_o1,eq:moo_o2,eq:moo_o3}). 
We also use non-dominated sort (NDSort) to drive the triggers toward optimal trade-offs.  
After that, we incorporate preference-based selection into MOEA to exclude impractical triggers (see red dots in \Cref{fig:ROI}).   
We note that the triggers to be excluded are considered as equal in quality to others during optimization, but they do not represent the practical solutions. 
Finally, we use our frequency trigger injection function to produce the adversary’s poisoned dataset with the most practical trigger.}

To evaluate triggers' performance, we construct a surrogate model (can be heterogeneous to the victim model) tuned on training data. 
Since the optimization direction guided by gradient descent from victim model is unknown, we improve triggers (concerning objective values) with variation (crossover and mutation) and selection pressure from NDSort which are inspired by the mating and survival of natural evolution. %(mitigation for \textbf{C2}).
We empirically confirm that the triggers' performance is independent to model structures, and in this way, a heterogeneous surrogate model is capable of approximating the victim model in practical accuracy and stealthiness. 

\noindent The \textbf{main contributions} are summarized as follows:
\\
\noindent$\bullet$ 
{We empirically demonstrate inherent conflicts among attack effectiveness, stealthiness and robustness, highlighting the difficulty in finding optimal Lagrange coefficients for balancing performance but also the unreliability in producing practical triggers (e.g., considering effectiveness and stealthiness) depending on coefficients.}
\\ 
\noindent$\bullet$ We formulate multiple attack goals (including effectiveness, dual-domain stealthiness and robustness) as a multi-objective problem (MOP) under the black-box setting. 
In MOP, we produce optimal triggers for all the objectives without using coefficients.
We leverage MOEA to optimize MOP, enhancing optimization efficacy as compared to SOP with gradient-based optimization. 
We also integrate the preference-based selection into MOEA to further filters out impractical triggers.
\\
\noindent$\bullet$ We conduct extensive experiments to show that LADDER achieves practical attack effectiveness $>$99\%, attack robustness with 90.23\% under image preprocessing operations, better natural stealthiness {(1.12$\times$ to 196.74$\times$ enhancement)}, and better spectral stealthiness (8.45$\times$ improvement), as measured by the average $l_2$-norm across five real-world datasets.

%% file: 3_RelatedWork.tex
\section{Related Work}
\label{sec:RelatedWork}

\subsection{Backdoor Attacks}
The first backdoor attack against CNNs is proposed by Gu et al. \cite{badnets}. 
It injects a patch-based pattern into a small fraction of clean data during  training process, triggering the victim model to misclassify those poisoned images to the attacker-desired label.
Since then, various attacks have been proposed to improve stealthiness through the design of triggers and training. 

\noindent\textbf{Spatial domain-based attacks}. 
To bypass human inspection, some works \cite{sig,refool,lira,issba,wanet,color} focus on stealthy backdoor attacks in spatial domain. 
For example, Barni et al. \cite{sig} use sinusoidal signals as triggers which results in only a slight varying backgrounds on the poisoned images.
Liu et al. \cite{refool} utilize natural reflection as triggers for backdoor injection in order to disguise triggers as natural light-reflection. 
Li et al. \cite{issba} leverage a CNN-based image steganography technique to hide an attacker-specified string into images as sample-specific triggers.
Besides visual stealthiness, several works \cite{wb,IBA,defeat,dfst} investigate the stealthiness in latent feature space.
Doan et al. \cite{wb} design a trigger generator to constrain the similarity of hidden features between clean and poisoned data via Wasserstein regularization. 
To improve the trigger stealthiness, Zhao et al. \cite{defeat} learn a generator adaptively to constrain the latent layers, which makes triggers more invisible in both input and latent feature space. 
Additionally, some studies focus on different aspects of attacks.
For example, Lv et al. \cite{lv2023data} propose an attack without leveraging original training/testing dataset.
Zeng et al. \cite{Narcissus} conduct clean-label backdoor attacks using knowledge of target class samples and out-of-distribution data.
While attacks in the spatial domain offer stealthiness, they often lack robustness against common image preprocessing operations, such as smoothing and compression.
Consequently, their effectiveness is significantly compromised by such operations.
Current spatial attacks customize triggers in a white-box setting, allowing attackers to access to the model’s structure and gradients, as well as the ability to manipulate the model arbitrarily.
These attacks often incorporate Lagrange multipliers, introducing additional coefficients and being sensitivity to the data, model, and optimization problem.
Besides, many spatial backdoor attacks exhibit severe mid- and high-frequency artifacts that can be easily detected in spectral domain. 

\noindent\textbf{Frequency domain-based attacks}. 
Due to the drawbacks of designing triggers in  spatial domain, studies \cite{cyo,ft,fiba,rethink,stealthy_freq} dive into backdoor attacks in frequency domain, naturally guaranteeing visual stealthiness by frequency properties. 
Wang et al. \cite{ft} handcraft two single frequency bands with fixed (predefined) perturbations as triggers.
Feng et al. \cite{fiba} poison a clean image by linearly combining the spectral amplitude of a trigger image with the clean one. 
Unfortunately, both of them, although maintaining stealthiness in spatial domain, introduce distinguishable frequency artifacts (see \Cref{fig:freq_disparity_map}) that can be detected via frequency inspection. 
Furthermore, they focus on natural (spatial) stealthiness yet do not consider robustness against image preprocessing operations. 
Moreover, due to lacking gradients, existing frequency backdoor attacks in black-box setting adopt fixed trigger pattern and consequently fail to achieve stealthiness in spectrum. 
In contrast, we leverage the evolutionary algorithm, a gradient-free optimization to design triggers in the spectral domain, which, for the first time,  achieves advanced imperceptibility in dual domains but also improves the attack robustness against image preprocessing-based defenses.  
\emph{We briefly compare the SOTA backdoor attacks in \Cref{tab:comparison_considered_factors} based on various attack attributes. 
For experimental comparisons, please refer to \Cref{sec:exp}.} 

\noindent\textbf{Other backdoor attacks.}
There are other types of attacks tailored to different scenarios. 
For instance, Lan et al. \cite{lan2024flowmur} introduce a stealthy and practical backdoor attack on speech recognition tasks.
Abad et al. \cite{abad2024sneaky} propose a stealthy attack against spiking neural networks.
Zhang et al. \cite{zhang2024badmerging} present the first backdoor attack for model merging scenario. 
We note that these attacks aim for different tasks, models and do not consider spectral domain stealthiness. 
We do not include them as baselines in the experiments.

\begin{table}[htb]
\caption{Critical attack attributes among LADDER and other attacks in spatial (\emph{S}) and frequency (\emph{F}) domains. The attack task is formulated as a single-objective problem (SOP) or a multi-objective problem (MOP).}
\centering
%\resizebox{\columnwidth}{!}{%
\scalebox{0.92}{
\begin{tabular}{@{}ccccccc@{}}
\toprule
 Attributes$\rightarrow$ & \multirowcell{2}{Attack\\ Domain} & \multirowcell{2}{Attack\\ Scenario} & \multicolumn{2}{c}{Stealthiness} & \multicolumn{1}{c}{\multirow{2}{*}{\begin{tabular}[c]{@{}c@{}}Attack\\ Robustness\end{tabular}}} & \multicolumn{1}{c}{\multirow{2}{*}{\begin{tabular}[c]{@{}c@{}}Optimization\\ Task Type\end{tabular}}} \\ \cmidrule(lr){4-5}
Attacks\ $\downarrow$ & & & \multicolumn{1}{c}{S}  &\multicolumn{1}{c}{F} & \multicolumn{1}{c}{} & \multicolumn{1}{c}{} \\ \midrule
Input-aware \cite{inputAware} & \emph{S} & White-box &\multicolumn{1}{c}{\XSolidBrush} & \multicolumn{1}{c}{\XSolidBrush}  & \multicolumn{1}{c}{\XSolidBrush} & SOP \\
ISSBA \cite{issba} & \emph{S} & White-box &\multicolumn{1}{c}{\XSolidBrush} & \multicolumn{1}{c}{\XSolidBrush}  & \multicolumn{1}{c}{\XSolidBrush} & SOP \\
LIRA \cite{lira} & \emph{S} & White-box &\multicolumn{1}{c}{\XSolidBrush} & \multicolumn{1}{c}{\XSolidBrush}  & \multicolumn{1}{c}{\XSolidBrush} & SOP \\
DFST \cite{dfst} & \emph{S} & White-box &\multicolumn{1}{c}{\XSolidBrush} & \multicolumn{1}{c}{\XSolidBrush}  & \multicolumn{1}{c}{\XSolidBrush} & SOP \\
WB \cite{wb} & \emph{S} & White-box &\multicolumn{1}{c}{\XSolidBrush} & \multicolumn{1}{c}{\XSolidBrush}  & \multicolumn{1}{c}{\XSolidBrush} & SOP \\
IBA \cite{IBA} & \emph{S} & White-box &\multicolumn{1}{c}{\XSolidBrush} & \multicolumn{1}{c}{\XSolidBrush}  & \multicolumn{1}{c}{\XSolidBrush} & SOP \\
BadNets \cite{badnets} & \emph{S} & Black-box &\multicolumn{1}{c}{\XSolidBrush} & \multicolumn{1}{c}{\XSolidBrush}  & \multicolumn{1}{c}{\XSolidBrush} & SOP \\
SIG \cite{sig} & \emph{S} & Black-box &\multicolumn{1}{c}{\Checkmark} & \multicolumn{1}{c}{\XSolidBrush} & \multicolumn{1}{c}{\XSolidBrush} & SOP \\
ReFool \cite{refool} & \emph{S} & Black-box & \multicolumn{1}{c}{\Checkmark} & \multicolumn{1}{c}{\XSolidBrush} & \multicolumn{1}{c}{\XSolidBrush} & SOP \\
WaNet \cite{wanet} & \emph{S} & Black-box & \multicolumn{1}{c}{\Checkmark} & \multicolumn{1}{c}{\XSolidBrush} & \multicolumn{1}{c}{\XSolidBrush} & SOP \\
Narcissus \cite{Narcissus} & \emph{S} & Black-box & \multicolumn{1}{c}{\XSolidBrush} &  \multicolumn{1}{c}{\XSolidBrush}& \multicolumn{1}{c}{\Checkmark} & SOP\\
FTrojan \cite{ft} & \emph{F} & Black-box & \multicolumn{1}{c}{\Checkmark} & \multicolumn{1}{c}{\XSolidBrush} & \multicolumn{1}{c}{\XSolidBrush} & SOP \\
FIBA \cite{fiba} & \emph{F} & Black-box & \multicolumn{1}{c}{\Checkmark} &  \multicolumn{1}{c}{\XSolidBrush}& \multicolumn{1}{c}{\XSolidBrush} & SOP\\
DUBA \cite{DUBA} & \emph{S+F} & Black-box & \multicolumn{1}{c}{\Checkmark} &  \multicolumn{1}{c}{\Checkmark}& \multicolumn{1}{c}{\XSolidBrush} & SOP\\
LADDER (Ours) & \emph{S+F} & Black-box & \multicolumn{1}{c}{\Checkmark}  & \multicolumn{1}{c}{\Checkmark} & \multicolumn{1}{c}{\Checkmark} & MOP\\ \bottomrule
\end{tabular}}
%}
\label{tab:comparison_considered_factors}
\end{table}

\subsection{Backdoor Defense}
Backdoor defense can be roughly divided into detection \cite{strip,activation_clustering,spectral,ulp,rethink} and defensive \cite{fine_pruning,nc,deepinspect,generative_distribution_modeling,nad,Rethinking} mechanisms. 
Typical detection methods include STRIP \cite{strip}, which deliberately perturbs clean inputs to identify potential backdoored CNN models during inference. 
Spectral Signature \cite{spectral} detects outliers using latent feature representations, while Zeng et al. \cite{rethink} propose a method that discriminates between clean and poisoned data in the frequency domain using supervised learning.   
Image preprocessing-based methods \cite{Rethinking,ft,deepsweep} have recently been explored to remove backdoors using techniques such as transformations and compression. 

Defensive methods aim to detect potential backdoor attacks but also to actively mitigate their effectiveness. 
For instance, 
fine-pruning \cite{fine_pruning} reduces the impact of backdoors by trimming dormant neurons in the last convolution layer, based on the minimum activation values of clean inputs. 
Neural Cleanse \cite{nc} leverages reverse engineering to reconstruct potential triggers for each target label and eventually renders the backdoor ineffective by retraining patches strategy.
Neural Attention Distillation \cite{nad} utilizes a ``teacher" model to guide the fine-tuning of the backdoored ``student" network to erase backdoor triggers. 
In this work, we showcase that the proposed attack can evade the defenses including frequency inspection, image preprocessing operations, and mainstream backdoor defenses.

Recently, several state-of-the-art backdoor defenses have been proposed. 
For example, Gao et al. \cite{ASD} introduce a training-time defense that separates training data into clean and poisoned subsets. 
Zhu et al. \cite{NEURIPS2023_03df5246} purify poisoned models by incorporating a learnable neural polarizer as an intermediate layer. 
Shi et al. \cite{NEURIPS2023_b36554b9} mitigate backdoor attacks through zero-shot image purification.

%% file: 4_background.tex
\section{Background}
\label{sec:background}
\noindent{\textbf{Preliminary Notations on CNN.}}
CNN is a cutting-edge ML architecture that achieves striking performance, especially for tasks with high-dimensional input space, such as image classification.
Given a CNN-based image classification model $f_\theta$: $\mathcal{I}^{S}\in[0,1]^{S}\rightarrow \mathbb{R}^{K}$ that takes an image \emph{x}$\in\mathcal{I}^{S}$ as input, and outputs an inference label \emph{y} $\in$ $\mathbb{R}^{K}$, where $\mathcal{I}^{S}$ represents the input space with dimension $S=H\times W\times C$ (\underline{H}eight, \underline{W}idth and \underline{C}hannels).
The $\mathbb{R}^{K}$ is the classification space which is divided into \emph{K} categories, the label \emph{y} $\in\mathbb{R}^{K}$ indicates the category where image \emph{x} belongs to, i.e., \emph{y} $\in\left\{0, 1, \cdots, \emph{K}-1\right\}$. 

\noindent{\textbf{Backdoor Attacks and Data Poisoning.}}
In a standard backdoor attack, the attacker crafts a subset of the clean training set (which contains \emph{N} samples) $D_{c}=\{(x_{i},y_{i})\mid x_{i}\in \mathcal{I}^{S}, y_{i}\in \mathbb{R}^{K} \}_{i=1}^{N}$ with a poison ratio $r\in(0,1]$ to produce a poisoned dataset: $D_{bd}=\{(x_{j}^{\prime},y_{j}^{\prime})\mid x_{j}^{\prime}\in \mathcal{I}^{S}, y_{j}^{\prime}\in \mathbb{R}^{K} \}_{j=1}^{\lceil N\times r\rceil}$ in which each poisoned image $(x_{j}^{\prime},y_{j}^{\prime})\in D_{bd}$ is obtained by applying a trigger function $\mathcal{T}$ and target label function $\eta$ on the image and label of counterpart clean sample $(x_{j},y_{j})\in D_{c}$:
\begin{equation}
\begin{aligned}
    x_{j}^{\prime} &=\mathcal{T}(x_{j},m,t)\triangleq x_{j}\cdot(1-m)+t\cdot m,\\
    y_{j}^{\prime} &=\eta(y_{j})\triangleq y_{tgt},
\end{aligned}
\label{eq:trigger_function_spatial}
\end{equation}
where $m\in[0,1]$ is a scaling parameter and $y_{tgt}$ is the attacker-desired target label.

Backdoor attack aims to inject a trojan into a CNN model $f_{\theta}$ by tuning model parameters $\theta$ on $D_{c}$ and $D_{bd}$ so that the poisoned model misclassifies any poisoned images in $D_{bd}$ into target (attacker-desired) class while behaving normally on clean data in $D_{c}$ without sacrificing benign accuracy. 
Details about the formulation of $D_{bd}$ with frequency triggers generated by LADDER are provided in \Cref{subsec:problem_formulation}.
Given a loss function $\mathcal{L}$, backdoor attack is commonly defined as an optimization task 
$\underset{\theta}{min}\sum\nolimits_{(x,y)\in D_{c}\cup D_{bd}}{\mathcal{L}(f_{\theta}(x,y))}.
\label{eq:loss_weight_sum}$

\noindent{\textbf{Discrete Cosine Transform\footnote{We choose commonly used type-II DCT and its inversion in this work.}(DCT)}} is a widely used transformation that represents a finite sequence of image pixels as a sum of cosine functions oscillating at various frequencies. 
In the spectrum, most of the semantic information of images tends to be concentrated in a few low-frequency components on the top-left region, where the $(0,0)$ element (top-left) is the zero-frequency component.
DCT and its inverse (IDCT) are channel-wise independent and can be applied to each channel of color images independently. Therefore, we simply introduce the DCT/IDCT operation on a single-channel image. 
The relationship between a single-channel image \emph{x} $\in[0,1]^{H\times W}$ (height $H$, width $W$) in spatial domain and its correspondent frequency spectrum \emph{X}$^{H\times W}$ can be described by type-II DCT and its inverse (IDCT) \cite{DCT}, denoted as $\mathcal{D}(\cdot)$ and $\mathcal{D}^{-1}(\cdot)$ respectively as follows:
\begin{align}
& \Scale[0.88]{\mathcal{D}(u,v) = N_u N_v\sum^{H-1}_{i=0}\sum^{W-1}_{j=0}x(i,j) cos\frac{(2i+1)u\pi}{2H}cos\frac{(2j+1)v\pi}{2W} \label{eq:DCT}},\\ 
& \Scale[0.84]{\mathcal{D}^{-1}(i,j) = \sum^{H-1}_{u=0}\sum^{W-1}_{v=0}N_u N_v  X(u,v) cos\frac{(2u+1)i\pi}{2H}cos\frac{(2v+1)j\pi}{2W}}, 
\label{eq:IDCT}
\end{align}
where $u,i\in\{0,1,\cdots,H-1\}$, and $v,j\in\{0,1,\cdots,W-1\}$. 
A pair $(u,v)$ refers to a specific frequency \emph{band} of spectrum of an image. 
$\mathcal{D}(u,v)$ defines the \emph{magnitude} of frequency component in a frequency band $(u,v)$. 
The value $x(i,j)\in[0,1]$ indicates the pixel value of location $(i,j)$ in an image $x$ in spatial domain. 
$N_u$ and $N_v$ are normalization terms, $N_u\triangleq \sqrt{1/H}$ if $u=0$ and otherwise $N_u\triangleq\sqrt{2/H}$. 
Similarly, $N_v\triangleq\sqrt{1/W}$ if $v=0$ and otherwise $N_v\triangleq \sqrt{2/W}$. 
We introduce $N_u$ and $N_v$ in order to ensure the DCT and its inverse are both isometric under $l_2$-norm so that $\|x\|_2\equiv\|DCT(x)\|_2$ is guaranteed for a given image $x$.

\noindent\textbf{Multi-objective Optimization (MOP).}
An MOP refers to an optimization task involving two or more conflicting objectives that cannot be optimal simultaneously to a single optimal solution. 
MOP is best addressed by generating a set of solutions, each reflecting different trade-offs among the objectives. 
Under MOP, multi-objective optimization (MOO) is the process of optimizing these multiple conflicting objectives concurrently to obtain an optimal set of solutions.

\noindent{\textbf{Multi-objective Evolutionary Algorithm (MOEA).}} 
MOEA \cite{nsgaii} is one of the commonly used MOO methods to solve MOP. 
It is a gradient-free optimization approach inspired by biological evolution. 
It explores the search space with a population of candidate solutions, drives the population toward promising areas with variation operators such as crossover \cite{Deb1995SimulatedBC} and mutation \cite{Deb1996ACG}, and eventually leads to high-quality solutions. 
Specifically, MOEA maintains a set of non-dominated solutions known as the Pareto (approximation) front, which is determined by the domination relationship between the objectives.  
Since the objectives in MOP cannot achieve optimal at the same time, each solution in the Pareto front represents a unique trade-off between the objectives.
MOEA is highly effective in solving MOP, as its variation operators can explore large solution spaces more thoroughly, without the need for gradient information and Lagrange coefficients tuning.

\section{Threat model}
\label{sec:threat_model}

\noindent\textbf{Attacker Capability.}  Similar to  \cite{ft, color, Narcissus}, we assume the attacker acts as a malicious data provider who can only embed a trigger into samples from the training set for public use. 
But it has no control over the training process and lacks any knowledge of the victim model. 

\noindent\textbf{Attacker Goals.} The attacker tricks the victim into training a backdoored CNN model for an image classification task, so that (1) the compromised CNN model outputs a target label desired by the attacker with high probability for any input containing the embedded trigger, while maintaining high inference accuracy on benign data; (2) \emph{dual-domain trigger stealthiness} can be guaranteed, preventing any noticeable anomaly in both the spatial and spectral domains of the input images; (3) the attack achieves \emph{robustness},  ensuring that the backdoor remains effective even after image preprocessing is applied to the poisoned data.

\noindent\textbf{Performance Metrics.} We introduce metrics to quantitatively measure our attack performance in three aspects: effectiveness, stealthiness, and robustness. 
\\
(1) For attack effectiveness and functionality preservation: we empirically evaluate the effectiveness with \emph{attack success rate} (ASR), which computes the ratio of poisoned samples misclassified by the poisoned CNN model as the attacker desires.
We further use the \emph{accuracy} (ACC) to evaluate the ratio of benign samples correctly classified as indicated by its ground-truth label by the victim model.  
ACC (ASR) $\in [0,100]$ is a scalar value reflecting the proportion of samples (\%) being successfully classified (attacked) among a given set of samples. 
The attacker wishes to achieve high ASR and ACC when a user trains its private model with the provided poisoned dataset. 
\\
(2) For stealthiness: we use PSNR, SSIM and LPIPS \cite{lpips} that can reflect human vision on images to evaluate spatial invisibility between clean and poisoned data.
LPIPS utilizes deep features of CNNs to identify perceptual similarity, while SSIM and PSNR are calculated based on the statistical pixel-wise similarity. 
Besides, since $l_2$-norm is often used \cite{Li2019InvisibleBA, LFAP} to evaluate the trigger stealthiness, we also include it in experimental comparison. 
For frequency inspection, we draw the residual map between the spectrum of clean and poisoned images. 
Ideally, a stealthy backdoor trigger should almost introduce nothing to the residual map, leading to almost no anomaly in the frequency and pixel domains.
\\ 
(3) For robustness: 
we define the robustness on any maliciously backdoored image $x_{bd}$ and its target label $y_{tgd}$ against a backdoored model $f_{\theta_{bd}}$ as follows:
\begin{equation}
    \begin{aligned}
        f_{\theta_{bd}}(Trans(x_{bd})) = y_{tgt},
    \end{aligned}
    \label{eq_robustness_def}
\end{equation}
where $Trans(\cdot)$ refers to any preprocessing operations and $f_{\theta_{bd}}$ has been well poisoned so that for any poisoned images, $f_{\theta_{bd}}(x_{bd}) = y_{tgd}$. 
To quantitatively measure the robustness, we record the ASR before and after the image preprocessing.
We also investigate the attack robustness against various preprocessing techniques \cite{jahne2005digital}, including JPEG compression, Gaussian filter, Wiener filter, and image brightness, which are commonly used in real-world applications.

%% file: 5_methodology.tex
\section{Objectives Conflict}
\label{subsec:objconflict}
One may apply a stealthy attack, e.g., FTrojan \cite{ft}, in a low-frequency region to achieve practical attack objectives (robustness, stealthiness and effectiveness), without considering trigger optimization. 
In contrast, this work aims to search a trigger that balances multiple objectives. 
In such a scenario, the conflict among objectives refers to the fact that attack objectives cannot achieve optimal simultaneously.

\begin{figure}[t]
    \centering
    \scalebox{0.44}{\includegraphics{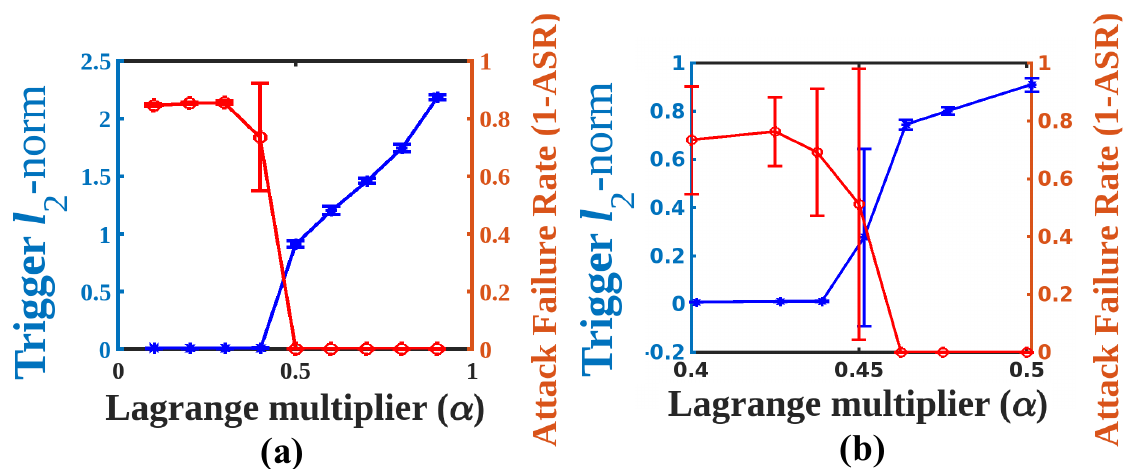}}
    \caption{The impact of Lagrange coefficient $\alpha$ in backdoor attack formulated with Lagrange multipliers and solved by SGD concerning trigger perceptibility and attack failure rate.}
    \label{fig:rela_disparity_alpha_asr}
\end{figure}

In a backdoor attack, effectiveness and trigger stealthiness are mutually conflicting objectives.
We confirm the conflict by formulating a simple optimization problem with the Lagrange multipliers under the control of two coefficients $\alpha,\beta$:
\begin{equation}
\label{obj}
\centering
\Scale[0.9]{\underset{\theta,t}{min}\ \alpha \sum\limits_{(x,y)\in D_c\cup D_{bd}} \mathcal{L}(f_{\theta}(x),y)  + \beta \Vert t\Vert_2},
\end{equation}
where $\alpha$, $\beta\in$[0,1], $D_{bd}$ is a set of poisoned images produced by the spatial domain-based trigger function in \Cref{eq:trigger_function_spatial} with trigger $t$, and $\alpha$+$\beta$=1, $t$ is a trigger initialized with random noise.
With stochastic gradient descent (SGD) \cite{sgd}, we update $t$ first while remaining $\theta$ unchanged, and then update model parameters $\theta$ with the optimal $t^*$. 
The results on CIFAR-10 with PreAct-ResNet18 are in \Cref{fig:rela_disparity_alpha_asr}.

In \Cref{fig:rela_disparity_alpha_asr}(a), we show the stealthiness measured by $l_2$-norm (a lower value indicates better stealthiness) marked in blue, and attack failure rate (AFR=1.0-ASR, a lower AFR indicates better attack effectiveness) marked in red with bars indicating the standard deviation of 10 repetitions under parameter $\alpha$ uniformly sampled between 0 and 1 with an interval of 0.1. 
As $\alpha$ increases, greater emphasis is placed on the attack effectiveness, while the trigger stealthiness is not considered critical.
Therefore, the attack failure rate (AFR) drops with the increase of $l_2$-norm. In other words, a stealthy trigger (i.e., with low $l_2$-norm) always achieves unsatisfied ASR (i.e., high AFR). 
While the curves of AFR and trigger stealthiness exhibit nearly monotonic changes along the increase of $\alpha$, we note a drastic variation within 0.4~$\leq\alpha\leq$~0.5.
These results highlight the conflict between effectiveness and stealthiness, indicating the significance of locating the best alpha.
We further sample $\alpha$ in this range, and present the result in \Cref{fig:rela_disparity_alpha_asr}(b). 
Similarly, the trigger norm and AFR exhibit an almost monotonic but opposite trend, providing strong evidence of the inherent conflict among objectives.
However, the standard deviation of trigger stealthiness is remarkably enlarged in this range, while the trigger norm and AFR change rapidly within the range of $\alpha$ between 0.425 and 0.475. 
Outside this range, the objectives exhibit minimal response to changes in $\alpha$.
\Cref{fig:rela_disparity_alpha_asr}(b) shows significant variances under alpha=0.45, indicating Lagrange multipliers+SGD cannot stably produce stealthy/effective triggers.

\begin{figure}
    
     \centering
     \scalebox{0.25}{\includegraphics[width=\textwidth]{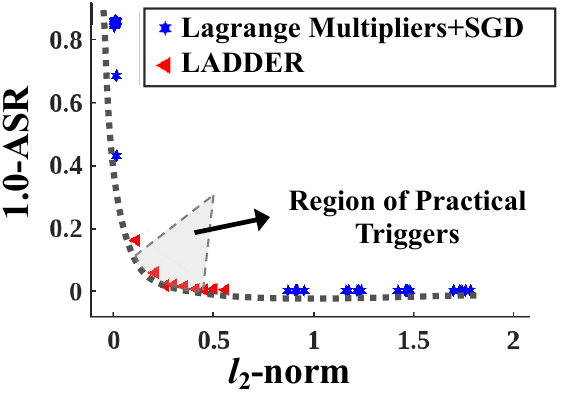}}
     \label{fig:paretoFront}
     % \end{subfigure}
    %\hfill
    \caption{Explanation of objective conflicting in backdoor attack, where red and blue dots represent 
    the triggers obtained by LADDER and SGD in victim model. The grey region indicates the objective value of triggers that we prefer to achieve.
    In this case we reflect our preference by ASR$\leftarrow$0.9 and $l_{2}\leftarrow$0.4.
    }
    \label{fig:objective_conflicting}
\end{figure}

Due to the conflict of objectives and instability of the gradient-based optimization process, formulating multiple attack objectives in a single-objective manner with the Lagrange multipliers and solving it with SGD leads to unsatisfied attack performance. 
In \Cref{fig:objective_conflicting}, we showcase the triggers produced by Lagrange multipliers+SGD and LADDER to illustrate that LADDER can find more practical triggers than the conventional method.
The dashed line demonstrates the expectation of trigger distribution which illustrates natural conflict between the two objectives, and the grey region includes the desired triggers. 
For example, the attacker aims to achieve a practical ASR ($>99\%$) while maintaining an $l_2$-norm below $0.4$ in CIFAR-10.
Triggers obtained by the Lagrange multipliers+SGD method (marked in blue) are notably distant from the grey region, as they tend to lack either stealthiness or effectiveness. 
In contrast, most of the triggers generated by LADDER remain within the grey region, ensuring both stealthiness and attack effectiveness.

\section{Evolutionary Multi-objective Backdoor Attack}
\label{sec:methodology}

\subsection{Problem Formulation}
\label{subsec:problem_formulation}
We formulate our backdoor attack as an MOP.
The main task of solving the MOP is to search an optimal trigger, which is patched to images to create a poisoned dataset.

\begin{figure}[htbp]
    \centering
    \scalebox{0.30}{
    \includegraphics{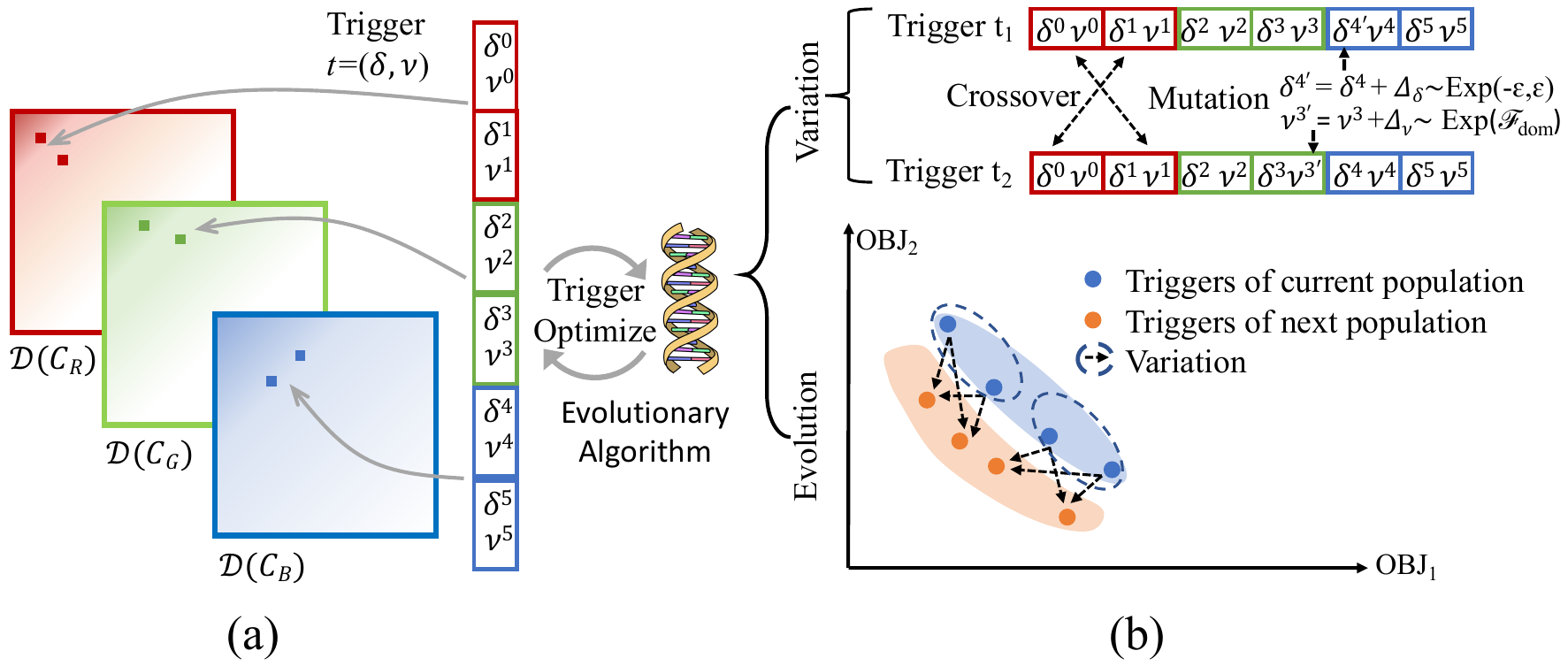}
    }
    \caption{The workflow of (a): Patching a trigger $t$=$(\delta,\nu)$ into the spectrum of each channel of an RGB image. $\mathcal{D}$ denotes the DCT function in \Cref{eq:DCT}. $C_R$, $C_G$ and $C_B$ denotes the R, G and B channel. (b): Optimizing trigger via MOEA. 
    {Exp($\cdot$) denotes sampling from an exponential distribution.}
    }
    \label{fig:trigger_encoding}
\end{figure}

\noindent\textbf{Frequency Trigger Injection Function.} 
Formally, a frequency trigger $t=(\delta,\nu)$ where $\delta=\{\delta^0, \delta^1, \cdots, \delta^{n-1}\}$ is a series of magnitude of perturbations, $\nu$=$\left\{\nu^0, \nu^1,\cdots, \nu^{n-1}\right\}$ describes the frequency bands to insert the correspondent perturbations on, and $n$ is the number of manipulated frequency bands.
We describe the trigger patching operation $\odot$ in \Cref{fig:trigger_encoding}(a).

In order to inject our trigger $t$ into an image $x$ in the spectral domain, we obtain the spectrum of \emph{x} via DCT ($\mathcal{D(\cdot)}$) and put the trigger optimized by LADDER in it. 
Finally, the poisoned spectrum is inverted to the spatial domain using IDCT ($\mathcal{D}^{-1}(\cdot)$), while we reset the label to an adversary-desired target.
Our trigger injection function $\mathcal{T}$ and target label function $\eta$ on a given sample $(x,y)$ are formally defined as: 
\begin{equation}
\label{eq:frequency_poisoning}
\begin{aligned}
    x' &= \mathcal{T}(x,t)\triangleq
    \mathcal{D}^{-1}(\mathcal{D}(x)\odot t),\\
    y' &= \eta(y)\triangleq y_{tgt}.
\end{aligned}
\end{equation}

\noindent{\textbf{Dual-domain Stealthiness.}} 
We pioneer the consideration of stealthiness in both spatial and spectral domains highly desired in backdoor attacks, since the former ensures the poisoned image evades human inspection while the latter mitigates the anomaly of frequency disparities between benign and poisoned images.
Given the widespread use of $l_p$-norm to evaluate the perturbation strength of the designed trigger \cite{lira,hiddentriggerbd,defeat}, we adopt this measurement to calculate the spatial stealthiness between clean image $x$ and poisoned image $x^{\prime}$ (obtained with trigger t and injection function $\mathcal{T}$): 
\begin{equation}
\emph{Stealthiness}_{spatial} \coloneqq \Vert \mathcal{T}(x,t)-x \Vert_p,
\label{eq:l2norm_spatial}
\end{equation}
while the frequency stealthiness is reflected by the $l_p$-norm of trigger perturbations as:
\begin{equation}
\emph{Stealthiness}_{freq} \coloneqq \Vert \delta\Vert_p.
\label{eq:l2norm_spectral}
\end{equation}
This work selects $p=2$, i.e., the $l_2$-norm as a measurement of trigger stealthiness for two reasons: 
(1) since the $l_2$-norm of disparity between the clean and poisoned images in dual domains is consistent, measuring the $l_2$-norm of trigger perturbation in the spectral domain can reflect dual-domain stealthiness;
(2) we empirically demonstrate that common visibility metrics, such as PSNR, SSIM, LPIPS, cannot properly evaluate frequency stealthiness (see natural stealthiness in \Cref{sec_attack_performance} for details).
We evaluate $l_2$-norm of triggers in the spectral domain due to the benefit of injecting triggers in this domain (see low-frequency robustness below).

\noindent{\textbf{Robustness in the Low-frequency Spectrum.}}
%Furthermore, 
Low-frequency components show great resilience to image preprocessing operations such as lossy compression and low-pass filtering since these operations are all designed to destroy the mid- and high-frequency components first. 
Therefore, we constrain our manipulated frequency bands $\nu$ in the low-frequency domain $\mathscr{F}_{dom}$, i.e. $\nu_{k}\in\mathscr{F}_{dom}, \forall k\in\{0,1,\dots,|\nu|-1\}$.
Within $\mathscr{F}_{dom}$, we minimize the distance between the location of each frequency band of a trigger and the zero-frequency band as:
\begin{equation}
\Scale[0.9]{
Robustness:=||\sum\nolimits_{i=0}^{n-1} (loc(\nu_{i})-loc(\emph{min}(\mathscr{F}_{dom})))||_2},
\label{eq:robustness_obj}
\end{equation}
where $\mathscr{F}_{dom}$ is the low-frequency domain, $\emph{min}(\mathscr{F}_{dom})$ is the zero-frequency band, and the function $loc(\cdot)$ is to find the vertical and horizontal index values for a given frequency band.

We provide a thorough analysis of the trade-off in terms of low-frequency regions and stealthiness.
We also investigate the impact on attack effectiveness and robustness of our trigger design. 
Please see \Cref{sec:objconflict_rob_stealth} for the details.

\noindent{\textbf{Multi-objective Backdoor Attacks Formulation.}}
Current backdoor attacks, even when addressing multiple attack objectives, are typically formulated by linear combination with the Lagrange multipliers.
As a result, excessive number of Lagrange coefficients are involved, complicating the parameter-tuning process.
In contrast, we formulate the objectives simultaneously as an MOP and optimize a set of triggers (each trigger represents a unique trade-off among the objectives) without aggregating the objectives into an SOP. 
Considering the above objectives and constraints while maintaining the functionality (benign accuracy) of backdoored model, we formulate a multi-objective black-box backdoor attack as: 
\begin{subequations}
\begin{align}
    \Scale[0.95]{(\delta^*,\nu^*)} & \Scale[0.95]{=  \underset{\delta,\nu}{\operatorname{argmin}}  \ O(\delta,\nu) =   (O_{1}, O_{2}, O_{3})}, \label{eq:moo_overall} \\
    \textrm{where}\ O_{1}(\delta,\nu) & = \Scale[1.0]{\sum\nolimits_{(x,y)\in D_{c}\cup D_{bd}}^{}\mathcal{L}(f^s_{\theta}(x),y)},
    \label{eq:moo_o1}
    \\
     O_{2}(\delta,\nu) & = \left\| \delta \right\|_{p=2}, \label{eq:moo_o2}
     \\
     O_{3}(\delta, \nu) & = \Scale[0.88]{||\sum\nolimits_{i=0}^{n-1} (loc(\nu_{i})-loc(\emph{min}(\mathscr{F}_{dom})))||_2}, 
     \label{eq:moo_o3}
     \\
     \textrm{s.t.}\quad |\delta_{k}|& \leq \Scale[1.0]{\epsilon, \  \forall k\in\{0,1,\cdots,|\delta|-1\}}, \label{eq:moo_st1}\\
     \nu_{k} &\in \Scale[1.0]{\mathscr{F}_{dom}, \ \forall k\in\{0,1,\cdots,|\nu|-1\}}, \label{eq:moo_st2}\\
     \textrm{Pref:}\quad O^{*} &\rightarrow O_{pref}, \label{eq:moo_pref}
\end{align}
\label{eq:MOP}
\end{subequations}
The task of our attack is formulated in \Cref{eq:moo_overall}, which contains three objectives, $O_1$ of \Cref{eq:moo_o1} that ensures a practical ACC and ASR, where $f^s_{\theta}$ is the surrogate model to evaluate trigger performance since the adversary cannot access a victim model, and the set of poisoned images $D_{bd}$ is obtained with frequency trigger function in \Cref{eq:frequency_poisoning}; $O_2$ of \Cref{eq:moo_o2} that ensures the dual-domain stealthiness and $O_3$ of \Cref{eq:moo_o3} which seeks triggers robust against image preprocessing within $\mathscr{F}_{dom}$.  
We introduce two constraints, Constraint (\ref{eq:moo_st1}) ensuring the magnitude of perturbation for each manipulated frequency band is within a reasonable range; and Constraint (\ref{eq:moo_st2}) restricting trigger to design in the low-frequency region $\mathscr{F}_{dom}$.
Finally, a preference-based selection  (see \Cref{algo:NDSort}) is considered in \Cref{eq:moo_pref} to reflect the preferred range of objective values.  

\subsection{Evolutionary Multi-objective Trigger optimization}
\label{subsec:emot_rigger_optimization}
\begin{figure}[]
    \centering
    \includegraphics[width=0.29\textwidth]{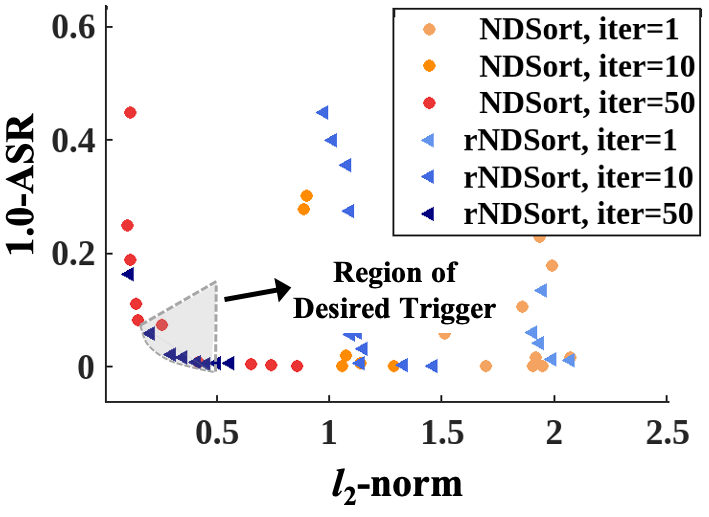}
    \caption{
    Comparison of triggers on MOEA with/without preference-based selection in CIFAR-10 on VGG11. Compared to NDSort, rNDSort pulls LADDER triggers closer to the attacker-desired region.
    }
    \label{fig:ROI}
\end{figure}

Solving an MOP (with conflicting objectives) by using SGD+Lagrange multipliers often leads to suboptimal attack performance (see \Cref{subsec:objconflict}). 
We introduce an MOEA to solve the problem. 
Our MOEA-based approach leverages crossover and mutation operators, avoiding the need to tune sensitive coefficients required by SGD+Lagrange multipliers. 
However, applying MOEA directly could produce impractical triggers (see those points outside the grey region, in \Cref{fig:ROI}).
To address this, we integrate MOEA with preference-based selection to prioritize practical triggers (those in the grey region).

Specifically, we leverage an MOEA to search the optimal trigger that can maximize performance of all the objectives in \Cref{eq:moo_overall}.
The workflow of trigger optimization is described in \Cref{fig:trigger_encoding}(b). 
MOEA estimates the performance of candidate triggers across objectives simultaneously in each iteration, without incurring the problems (in \Cref{fig:rela_disparity_alpha_asr,fig:objective_conflicting}).
It initializes random triggers (Step 1 in \Cref{alg_topLevel}), iteratively optimizes them with variation (\Cref{fig:trigger_encoding} (b)), evaluates triggers' objective values (\Cref{alg_Eval}) and selects non-dominated triggers by preference-based selection (\Cref{algo:NDSort}).
We introduce the details of LADDER optimization in \Cref{alg_topLevel}.

\begin{algorithm}[htbp]
    \caption{LADDER Optimization via MOEA}
    \begin{algorithmic}[1]
        \Require{A subset of training data $\mathcal{D}_{c}$, Poison Ratio $r$, Total optimization generations \emph{Gen}, Maximum frequency perturbation \emph{$\epsilon$},  
        Number of retrain epoch $E_{re}$,
        Surrogate model $f^s_{\theta}$, Population size \emph{P}}
        \Ensure{Poisoned Dataset $D_{bd}$ injected by $t^*=(\delta^*,\nu^*)$}
        \Statex \underline{Step\,1: Initialization}
        \State \scalebox{0.9}{$T_{popu}$: \{$(\delta_0,\nu_0),(\delta_1,\nu_1),\cdots,(\delta_{P-1},\nu_{P-1})\} \gets$ RandomInit()}
        \Statex \underline{Step\, 2: Evaluation}
        \State $\{O\}_{popu}$ = Eval($T_{popu}, f^{s}_{\theta}, \mathcal{D}_{c},r$)
        \Statex \underline{Step\,3: Trigger Optimization}
        \For{$gen$ in [0,1,$\cdots$,$Gen$-1]}
        \State $T$$_{\rm{\emph{offsp}}}$: $\{(\delta_{0}^{'},\nu_{0}^{'}),\cdots,(\delta_{P-1}^{'},\nu_{P-1}^{'})\} \gets$Variation($T_{popu}$)
        \State $\{O\}$$_{\rm{\emph{offsp}}}$ = Eval($T$$_{\rm{\emph{offsp}}}$$, f^{s}_{\theta}, \mathcal{D}_c,r$)
        \State $T_{popu}$ $\leftarrow$ rNDSort($T_{popu}\cup$ $T$$_{\rm{\emph{offsp}}},\{O\}_{popu}\cup \{O\}$$_{\rm{\emph{offsp}}}$)
        \EndFor
        \Statex \underline{Step\,4: Trigger Selection \& Data Preparation}
        \State $(\delta^*,\nu^*) \gets$ SelectTrigger($T_{popu}$)
        \State $D_{bd}$ = Poison($D_c$, $r$, $(\delta^*,\nu^*)$)
        \State \Return $D_{bd}$
    \end{algorithmic}
    \label{alg_topLevel}
\end{algorithm}

\noindent\textbf{\underline{Step 1: Initialization.}} MOEA initializes a population \scalebox{0.93}{$popu=$ $\{t_0,t_1, \cdots, t_{P-1}\}=\{(\delta_0,\nu_0),(\delta_1,\nu_1),\cdots,(\delta_{P-1},\nu_{P-1})\}$} of triggers, where $P$ is the population size. 
Besides, the triggers are generated under the constraints in \Cref{eq:moo_st1,eq:moo_st2}. 
Then, we evaluate the initialized triggers on the objectives.

\noindent\textbf{\underline{Step 2: Trigger Evaluation.}}
The idea of trigger evaluation is to calculate the objective values $O_1$, $O_2$ and $O_3$ in \Cref{eq:moo_o1}, \ref{eq:moo_o2} and \ref{eq:moo_o3} for each candidate trigger in lines 2 and 5 of \Cref{alg_topLevel}.
The trigger evaluation is described in \Cref{alg_Eval}. 
To evaluate the attack effectiveness for each trigger, we poison a subset of data with it and train the backdoor task (\Cref{eq:moo_o1}) on a surrogate model. 
A surrogate model refers to a CNN model %(usually heterogeneous to the victim model) 
to approximate the victim model. 
We use this approach because the attacker has no knowledge about the victim model in the black-box setting. 
Also, evaluating triggers by training a model from scratch is computationally expensive.
We hereby employ a pre-trained surrogate model on clean data, fine-tuning it through a limited number of retraining epochs, achieving evaluation efficiency.
One may argue that the heterogeneous model structures between the surrogate and victim model may cause a bias of trigger performance in the evaluation process.
To address this concern, we experimentally assess the trigger performance between various combinations of surrogate and victim model structures and demonstrate the high consistency among them (see \Cref{subsec:transferability}).

\begin{algorithm}[t]
    \caption{Eval: Evaluate Triggers in LADDER}
    \begin{algorithmic}[1]
        \Require{A set of triggers \emph{T}, Surrogate model $f_\theta^s$, A subset of training data $\mathcal{D}_c$, Poison Ratio $r$, Population size \emph{P}}
        \Ensure{The objective values $\{O\}$ of each trigger in \emph{T}}
        \For{$(\delta_i,\nu_i)$ $\textbf{in}$ \emph{T}}
            \State $\mathcal{D}_{bd}\gets$Poison($D_c$, $r,(\delta_i,\nu_i)$)
            \State $f_{\theta^{'}}^s \gets$Train($f_{\theta}^{s}, \mathcal{D}_{bd}$)
            \State $O_1^i = \Scale[1.0]{\sum\nolimits_{(x,y)\in D_{bd}}^{}\mathcal{L}(f_{\theta^{'}}^s(x),y)}$
            \State $O_2^i=\left\| \delta \right\|_{2}$
            \State $O_3^i=||loc(\nu_{i})-loc(\emph{min}(\mathscr{F}_{dom}))||_2$
            \State Rollback $f_{\theta^{'}}^{s}\leftarrow$ $f_{\theta}^{s}$
        \EndFor
        \State \scalebox{0.8}{$\{O\}=\{(O_1^0,O_2^0,O_3^0),(O_1^1,O_2^1,O_3^1),\cdots,(O_1^{P-1},O_2^{P-1},O_3^{P-1})\}$}
        \State \Return \emph{$\{O\}$}
    \end{algorithmic}
    \label{alg_Eval}
\end{algorithm}
\noindent\textbf{\underline{Step 3: Trigger Optimization.}}
After initializing and evaluating the triggers, MOEA iteratively optimizes the triggers by applying variation to the triggers in the population to generate offsprings, evaluating their quality, and finally selecting well-performing triggers among all of them. 
Through this process, triggers gradually converge toward an optimal balance of stealthiness, attack effectiveness, and robustness. 

\noindent\emph{Trigger variation and evaluation.}
The variation process is used to generate offspring triggers from population, which involves two procedures, simulated binary crossover (SBX) \cite{Deb1995SimulatedBC} and polynomial mutation (PM) \cite{Deb1996ACG}.
{The former 
randomly generates offspring triggers by exchanging a specific component (such as a perturbation or band) between two triggers from the population; the latter is to randomly alter the magnitude of frequency perturbations or shift the location of bands based on the perturbation sampled from a specific exponential distribution (see details in \Cref{fig:trigger_encoding}(b)).}
The variation is repeatedly applied for each trigger in each iteration until the produced offsprings satisfy the restrictions in \Cref{eq:moo_st1,eq:moo_st2}. 
After that, we evaluate newly generated triggers in the same way as described in line 2 of  \Cref{alg_topLevel}.

\noindent\emph{Next population formulation with rNDSort.} 
In each iteration, after generating offsprings from parents and evaluating their performance on $O_1$, $O_2$ and $O_3$, we combine the population with offsprings and leverage the proposed rNDSort (see \Cref{algo:NDSort}) to pick up superior triggers survival into the next iteration while eliminating inferior triggers.

rNDSort includes two components, NDSort and \emph{preference-based selection}, which drives triggers to converge toward the attacker-desired region and maintain a stable number of triggers in the population per iteration.
We first introduce the dominance relationship for non-dominated sort.
Given two triggers $t_1$ and $t_2$ along with their objective values $O^{1}$=\{$O_{1}^{1},O_{2}^{1},O_{3}^{1}$\} and $O^{2}$=\{$O_{1}^{2},O_{2}^{2},O_{3}^{2}$\} (recall smaller objective value leads to a better trigger), we say $t_1$ \emph{dominates} $t_2$, denotes as $t_1 \prec t_2$ iff. 
$\forall k\in[1,3], O_{k}^{1}\leq O_{k}^{2}$ and $\exists k\in[1,3]$ s.t. $O_{k}^{1}<O_{k}^{2}$.
In this case, $t_1$ is a \emph{non-dominated} trigger among \{$t_1$, $t_2$\}.
With the help of the dominance relationship, the non-dominated sort repeatedly moves non-dominated triggers from a trigger set $T$ to a new set $T_{r}$, until adding non-dominated triggers in $T_r$ results in $|T_r|>P$.
Finally, the remaining triggers in $T$ with the largest k-nearest sparsity \cite{nsgaii} concerning objective values are selected to fill in $T_r$ until $|T_r|=P$. 
This step ensures that triggers are searched along the entire objective space. 
However, impractical triggers (see those points which are out of the grey region in \Cref{fig:ROI}) may be still acquired, as triggers searched by NDSort are non-dominated to attacker-desired triggers. 
This means they are considered of equal quality from the MOO perspective, even though they may not be practical. 
To alleviate locating impractical triggers caused by NDSort, we fill in $T_r$ to the size $P$ with preference-based selection.
Specifically, we calculate the Euclidean distance of remaining triggers in $T$ to the attacker-desired region in terms of objective values and select triggers with the smallest distance until $|T_r|=P$.

\begin{algorithm}[t]
\caption{\scalebox{0.93}{rNDSort: Preference-based NDSort}}
\begin{algorithmic}[1]
\Require{Population size $P$, a trigger set \emph{T}=$\{t_0,t_1,\cdots,t_{2P-1}\}$, the objective value set \{\emph{O}\} = $\{(O^0_1,O^0_2,O^0_3)$,$(O^1_1,O^1_2,O^1_3)$,$\cdots$,$(O^{2P-1}_1,O^{2P-1}_2,O^{2P-1}_3)\}$, attacker preferred region of objective values \emph{O}$_{pref}$}
\Ensure{The trigger set $T_{r}$ ranked by the distance of their objective values to preference}
\State $T_{r}\gets \emptyset$, $list \gets$[],\ \emph{order}$\gets$0
\While{$|T_{r}|\leq P$ and $|T_{r}|$+NonDom($T,\ \{O\}$) $\leq P$}
    \State $T_{r}$.append(NonDom($T,\ \{O\}$))
    \State $T'$=NonDom($T,\{O\}$)
    \State $\{O\}$=$\{O\}\backslash \{O\}^{T'}$, \ $T$=$T\backslash T'$
\EndWhile
\For{\emph{i} \textbf{in} $\{0,1,\cdots,|T|\}$}
\State \emph{d} = Euc(\emph{O}[\emph{i}],\ \emph{O$_{pref}$})
\State \emph{list}.append($<$\emph{d},\ \emph{T}[\emph{i}]$>$) \Comment{$<\cdot,\cdot>$ is a pair}
\EndFor
\State $list\leftarrow Sort_{ascend}$($list$) by $d$
\While{$|T_{r}|< P$}
    \State $T_{r}$.append($list$[$order$++].SecondElem)
\EndWhile
\State \Return $T_{r}$
\end{algorithmic}
\label{algo:NDSort}
\end{algorithm}

To validate the efficacy of rNDSort, we compare the triggers obtained by NDSort and rNDSort of $1^{st}$, $10^{th}$ and $50^{th}$ iterations and visualize their objective values in \Cref{fig:ROI}.
We can observe that after 50 iterations, the triggers obtained by rNDSort are mostly located within the attacker-desired region (marked in grey). In contrast, triggers obtained by NDSort span a wider range, including impractical ones.

\noindent\textbf{\underline{Step 4: Trigger Selection and Data Preparation.}} 
After the trigger optimization, we obtain a set of practical triggers (see those points in or close to the grey region in \Cref{fig:ROI}) and choose the best trade-off under our attack scenario among them.
Specifically, we choose the most practical trigger from the population based on whose objective values are closest to the best values for each objective. 
Finally, we release a poisoned dataset injected by the trigger.

%% file: 6_experiments.tex
%%%%%%%%%%%%%%%%%%%%%%%%%%%%%%%%%%%%%%%%%%%%%%%%%%%%%%%%%%%%%%%%%%%%%%%%%%%%%%%%
\section{Experiments}
\label{sec:exp}
\subsection{Experimental Setup}
\label{subsec:exp_env_setting}
\noindent\textbf{Experimental Environment and Settings.}
Our LADDER is implemented\footnote{The code is available at github.com/dzhliu/LADDER} on Python, PyTorch \cite{paszke2019pytorch} and Ubuntu. 
All the experiments are conducted on a workstation with Ryzen 9 7950X, 2$\times$32GB DDR5 RAM, and NVIDIA GeForce RTX 4090.
For the default training, we learn the classifiers by SGD optimizer with the initial learning rate of 0.01 and a decay of 0.1 per 50 epochs. 
We set the batch size to 64 and the total number of epochs to 200 for all the datasets to train surrogate and victim models.
When evaluating triggers on the surrogate models, the number of retraining epochs is set to 20.
For the default attack setting, we search triggers in low-frequency regions. 
Following Sharma et al. \cite{sharma2019OnTheEffe}, we use around 18.3\% of the whole frequency spectrum on the top-left region to search the low-frequency trigger. 
{Meanwhile}, we manipulate 3 frequency bands per channel for our attack in all the datasets. 
For a fair comparison, the poison ratio and target label are set to 5\% and 7, unless otherwise specified.  
For the default MOEA setting, 
we set the population size to 10, the optimization iterations to 20.
We set the $O_{pref}$ in \Cref{eq:moo_pref} as: 0.9 for $O_{1}$ and 0.4 for $O_{2}$; $O_3$ is 8 for images of size 32$\times$ 32 and 12 for images of size 64$\times$ 64. 
Note LADDER is label-independent, i.e., the attack effectiveness and trigger stealthiness remain high when attacking other labels (see \Cref{fig:multiTargets}). 
\\
\noindent\textbf{Datasets and Models.} 
We evaluate LADDER on five benchmark tasks including digit recognition on SVHN \cite{SVHN}, object classification on CIFAR-10 \cite{cifar10}, real objects on Tiny-ImageNet \cite{tiny}, traffic sign recognition on GTSRB \cite{gtsrb} and face attribute recognition on CelebA \cite{celeba}.
For CelebA, we follow \cite{dynamic,wanet} to select the top three most balanced attributes including Heavy Makeup, Mouth Slightly Open, and Smiling. 
Then, we concatenate them to create an eight-label classification task.
We evaluate LADDER on both small- and large-scale datasets to confirm its scalability across various image and dataset sizes.
The five datasets chosen for this paper span a remarkably broad scope of typical real-world scenarios, underscoring the practicality of LADDER.
Following \cite{bypassing,spectral,activation_clustering,wanet,marksman}, we consider various network architectures for the image classifier. 
Specifically, we employ a classic CNN model \cite{wanet,marksman} for SVHN, PreAct-ResNet18 \cite{resnet} for CIFAR-10 and GTSRB, as well as ResNet18 \cite{resnet} for Tiny-ImageNet and CelebA.
In contrast to victim models, we choose surrogate models from a series of VGGs \cite{vgg}, whose structures are heterogeneous against ResNet models. 
For example, we utilize VGG11 for CIFAR-10 and GTSRB, VGG16 for SVHN and CelebA, as well as VGG19 for Tiny-ImageNet.
The details of computer vision tasks, datasets, and models are outlined in \Cref{tab:task_dataset_model} of \Cref{sec:task_data_model_details}. 
It is important to emphasize our intentional use of heterogeneous structures between victim and surrogate models. 
This approach effectively demonstrates that the model mismatch between the victim and surrogate models does not hinder the efficacy and practicality of our attack. 

\subsection{Attack Performance}
\label{sec_attack_performance}
We compare LADDER with popular spatial attacks, such as BadNets \cite{badnets}, ReFool \cite{refool}, SIG \cite{sig}, WaNet \cite{wanet}, Narcissus \cite{Narcissus}, HCB \cite{watchout}, and BELT \cite{BELT} as well as frequency attacks such as FIBA \cite{fiba}, FTrojan \cite{ft} and DUBA \cite{DUBA} as baseline methods to showcase the attack performance in: attack effectiveness, natural (spatial) and spectral (frequency) stealthiness. 
Note that several white-box attacks \cite{defeat,inputAware,IBA,wb}, although achieving practical effectiveness, require access and manipulation of victim models. 
They are not included in the experiments.

\noindent\textbf{Attack Effectiveness.}
We evaluate the effectiveness of 10 attacks against 5 datasets via ACC and ASR.  
\begin{table*}[t]
\centering
\caption{Attack performance measured by ACC (\%) and ASR (\%) for 10 attacks against 5 datasets. The number in the brackets indicates the differences between clean ACC and the correspondent ACC on the backdoored model.
Our method achieves comparable or superior performance on ACCs/ASRs compared to other attacks, with the exception of the SVHN dataset, where our ACC/ASR are only 0.48\% and 0.21\% lower than the best results, respectively.
}
\scalebox{0.9}{
\begin{threeparttable}[]

\label{attack_performance}

\begin{tabular}{@{}ccccccccccc@{}}
\toprule
      \multirow{2}{*}{Attack} & \multicolumn{2}{c}{SVHN} & \multicolumn{2}{c}{GTSRB} & \multicolumn{2}{c}{CIFAR-10} & \multicolumn{2}{c}{Tiny-ImageNet} & \multicolumn{2}{c}{CelebA} \\ \cmidrule(l){2-3} \cmidrule(l){4-5} \cmidrule(l){6-7} \cmidrule(l){8-9} \cmidrule(l){10-11}
        
  & ACC           & ASR        & ACC           & ASR        & ACC            & ASR          & ACC    & ASR   & ACC    & ASR           \\ \midrule
Clean   & 92.81   & -    & 98.55    & -   & 93.14          & -            & 54.60        & - & 79.20    & -  \\ \midrule
\textsc{BadNets \cite{badnets}}  & \textbf{92.67} (0.14)    & 99.14      & 97.91 (0.64)     & 96.67          & 92.05 (1.09)      & 98.24   &  51.90 (2.70)    & 97.82  &  76.54 (2.66)    & 99.35  \\
\textsc{SIG \cite{sig}}     & 92.45 (0.36)        & 99.87          & 97.90 (0.65)         & 99.87          & 92.14 (1.00)      & 99.98      & 51.98 (2.62)    & 99.49   &  77.90 
 (1.30)    & 99.85           \\
\textsc{Refool \cite{refool}}  & 92.24 (0.57)        & 99.31        & 97.94 (0.61)         & 98.51          & 91.09 (2.05)    & 97.03       & 48.37 (6.23)      & 97.32 &  77.53 (1.67)    & 98.09             \\
\textsc{WaNet \cite{wanet}}   & 92.33 (0.48)         & 99.17         & 98.19 (0.36)        & 99.83          & 92.31 (0.83)     & 99.94     & 52.85 (1.75)     & 99.16 &  77.99 (1.21)    & 99.33       \\
\textsc{FTrojan \cite{ft}} & 92.63 (0.18)       & 99.98        & 96.63 (1.92)         & 99.25          & 92.53 (0.61)  & 99.82   & 53.41 (1.19)       & 99.38  &  76.63 (2.87)    & 99.20            \\
\textsc{FIBA \cite{fiba}}    &     91.10 (1.71)     &     96.91       & 96.73 (1.82)         & 98.88          & 91.13 (2.01)   & 97.60   &   51.11 (3.49)   & 92.14 & 75.90 (3.30)   & 99.16  \\
\textsc{DUBA \cite{DUBA}}    &     91.23 (1.58)  &   99.79 & 96.90 (1.65)      & 98.32   & 91.97 (1.17)   & \textbf{99.99}   &   52.74 (1.86)   & \textbf{99.99}  & 77.30 (1.90) & \textbf{99.99} \\
\textsc{Narcissus-D \cite{Narcissus}\tnote{\scalebox{1.2}{$\star$}}}    &     91.94 (0.87)   &   99.97 & 97.47 (1.08)  & \textbf{99.99}  & 92.17 (0.97)  & \textbf{99.99}  &   54.17 (0.43) & \textbf{99.99}  & 77.85 (1.35) & \textbf{99.99} \\
\textsc{BELT}\cite{BELT}    &   91.64 (1.17)  & \textbf{99.99}  & 95.58 (2.97)   & \textbf{99.99}  & 91.58 (1.56)  & 99.93 &  52.51 (2.09)  & 99.12 & 77.59 (1.61) & 99.88 \\
\textsc{HCB}\cite{watchout}    &  91.43 (1.38)    & 98.29  & 95.15 (3.40)    & 97.49  & 91.17 (1.97)  & 98.42 &  52.59 (2.01)  & 98.31 & 77.16 (2.04) & 98.25 \\
\textsc{Ours}    & 92.19 (0.62)   & 99.77   & \textbf{98.37} (0.18)   & 99.93     & \textbf{92.82} (0.32)    & \textbf{99.99}  & \textbf{54.20} (0.40)             & 99.54 &  \textbf{79.57} (0.37)    & 99.90       \\ \bottomrule
    \end{tabular}
     \begin{tablenotes}
     \item[\scalebox{1.2}{$\star$}] Narcissus is a clean-label backdoor attack, which does not align with the dirty-label attack framework of this paper. Therefore, we extend it to a dirty-label attack, denoted as Narcissus-D, where the labels of poisoned samples are assigned the target label during data poisoning.
   \end{tablenotes}
\end{threeparttable}}
\end{table*}
\begin{table*}[]
\centering
\caption{Natural stealthiness (PSNR $\uparrow$, SSIM $\uparrow$, LPIPS $\downarrow$) as well as $l_2$-norm $\downarrow$ of trigger pattern.
Across 4 metrics and 5 datasets, LADDER consistently demonstrates superior stealthiness compared to 10 attacks, with the only minor exception where LADDER has a 0.081 difference on $l_2$-norm on SVHN and a 0.0007 gap on LPIPS on Tiny-ImageNet.}
\label{natural_stealthiness}
\scalebox{0.75}{
\begin{tabular}{@{}ccccccccccccccccccccc@{}}
\toprule
\multirow{2}{*}{Attacks}  & \multicolumn{4}{c}{SVHN} & \multicolumn{4}{c}{GTSRB} & \multicolumn{4}{c}{CIFAR-10} & \multicolumn{4}{c}{Tiny-ImageNet} & \multicolumn{4}{c}{CelebA}  \\ \cmidrule(l){2-5} \cmidrule(l){6-9} \cmidrule(l){10-13} \cmidrule(l){14-17}  \cmidrule(l){18-21}
 & $l_2$ & PSNR   & SSIM   & LPIPS & $l_2$ & PSNR   & SSIM   & LPIPS   & $l_2$ & PSNR    & SSIM  & LPIPS  & $l_2$  & PSNR    & SSIM   & LPIPS  & $l_2$ & PSNR  & SSIM   & LPIPS  \\ \midrule
Clean       &    0.0000    &   Inf     &    1.0000     &   0.0000      &    0.0000    &   Inf     &    1.0000     &   0.0000      &  0.0000       &  Inf        & 1.0000           & 0.0000     & 0.0000     &Inf    &  1.0000    & 0.0000 & 0.0000 &  Inf     &    1.0000     &   0.0000   \\ \midrule
\textsc{BadNets \cite{badnets}}   & 2.9363 & 27.49  & 0.9763   & 0.0187    &    3.8479        &    27.18    &     0.9754   &     0.0059  &    2.7358  &    36.67     &    0.9763     &     0.0012  &    2.9737   &      36.35     &    0.9913       &     0.0006 &    3.2871 &  32.50  &  0.9951   &  0.0005    \\
\textsc{SIG \cite{sig}}   & 3.0525 & 25.18   & 0.7490   & 0.0706     &    3.0113    &    25.32    &    0.7313    &   0.0766  &    3.0259    &    25.26     &     0.8533    &    0.0289   &    6.0205   &     25.36      &      0.8504     &     0.0631  &    5.9627 &  25.38  &  0.7949   &  0.0359   \\
\textsc{Refool \cite{refool}}    & 4.8254 & 21.61   & 0.8511   & 0.0456   &    5.0275             &    20.57    &     0.7418   &    0.3097  &    5.9169   &    18.37     &    0.6542     &      0.0697  &    6.4901  &    20.42       &      0.8564     &      0.4574 &    7.0494 &  23.72  &  0.8359   &  0.2134    \\
\textsc{WaNet \cite{wanet}}    & \textbf{0.1969} & 37.72   & 0.9905   & 0.0016     &    0.4280            &    30.11    &    0.9669    &   0.0584   &    1.9397   &    19.30     &    0.8854     &    0.0090   &    1.4926   &     29.59      &     0.9359      &      0.0360 &    0.7880 &  30.42  &  0.9175   &  0.0530    \\
\textsc{FTrojan \cite{ft}}   & 0.4866 & 41.13   & 0.9896   & 0.0002    &    0.4874            &    41.11    &    0.9885    &    0.0007  &    0.4850   &     41.16    &   0.9946      &     0.0006  &    0.8553   &     42.28      &     0.9931      &     \textbf{0.0003} &    0.8568 &  42.25  &  0.9904   &  0.0003    \\
\textsc{FIBA \cite{fiba}}    & 1.9250 & 29.67   & 0.9782   & 0.0044   &    1.8693               &    29.74    &   0.9589     &    0.0083 &     1.8437   &    29.69     &    0.9858     &     0.0024  &    3.7459   &      29.39     &    0.9755       &     0.0080 &    4.0548  &  29.25  &  0.9592   &  0.0057    \\
\textsc{DUBA} \cite{DUBA}   &     0.9574     &    35.71    & 0.9721      & 0.0028   & 1.5812   & 31.82   &   0.9376   & 0.0034  & 1.9642 & 29.35 & 0.9415 & 0.0027 & 5.2490 & 26.83 & 0.8815 & 0.0256 & 3.3136 & 30.51 & 0.9191 & 0.0210 \\
\textsc{Narcissus-D} \cite{Narcissus}    &    6.6200    & 18.45  & 0.5952     & 0.1704  & 5.5698  & 19.94 &  0.5795  & 0.0925 & 6.5335 & 18.56 & 0.7137 & 0.0324 & 3.3335 & 30.44 & 0.9328 & 0.0170 & 4.5943 & 27.65 & 0.9278 & 0.0637 \\
\textsc{BELT}\cite{BELT}    &    3.8757    & 23.17  & 0.9337    & 0.0266  & 4.5140  & 21.83 &  0.9243  & 0.0249 & 4.1914 & 22.57 & 0.9413 & 0.0061 & 4.3731 & 28.22 & 0.9836 & 0.0078 & 4.6113 & 27.82 & 0.9802 & 0.0107 \\
\textsc{HCB}\cite{watchout}    &    3.8060    & 24.00  & 0.8911    & 0.0469  & 4.9650  & 21.23 &  0.8069  & 0.0361 & 3.7938 & 23.49 & 0.9294 & 0.0070 & 5.7686 & 26.07 & 0.9264 & 0.0131 & 6.0281 & 25.56 & 0.8968 & 0.0163 \\
\textsc{Ours}    & {0.2781} & \textbf{45.99}   & \textbf{0.9973}   & \textbf{0.0003}   &    \textbf{0.3406}               &    \textbf{44.23}    &   \textbf{0.9943}     &    \textbf{0.0002}  &    \textbf{0.3183}   &   \textbf{44.81}      &    \textbf{0.9976}     &     \textbf{0.0001} &    \textbf{0.6132}    &     \textbf{45.14}      &     \textbf{0.9976}      &     0.0010 &    \textbf{0.4132} &  \textbf{48.57}  &  \textbf{0.9974}   &  \textbf{0.0002}    \\ \bottomrule
\end{tabular}}
\end{table*}
Based on the results given in \Cref{attack_performance}, LADDER achieves ASRs exceeding 99\% on all poisoned CNN models. 
Meanwhile, its drop of ACCs after the backdoor attack is limited to only 0.23\% on average, while the compared attacks yield larger ACC drops. 
This confirms that LADDER delivers practical attack performance under various attack tasks. 
Recall that we consider attack effectiveness as one of the objectives when formulating the multi-objective attack problem, ensuring that the triggers searched by LADDER are oriented towards maximizing effectiveness. 
We also note that heterogeneous network structure settings between surrogate and victim models do not affect the attack effectiveness of LADDER.

\noindent\textbf{Natural (Spatial) Stealthiness.}
Natural stealthiness is vital for backdoor attacks, guaranteeing that poisoned images remain imperceptible to human inspection. 
We quantitatively compare the differences between poisoned and clean images against four popular visual stealthiness measurements, including $l_2$-norm, PSNR, SSIM, and LPIPS.
All metric values are averaged over 1,000 randomly selected samples from the test dataset.  
In \Cref{natural_stealthiness}, we list the ideal metric values on clean images under 5 datasets, then show metric values on the poisoned samples under various attacks.
SSIM cannot precisely capture minor differences of trigger perturbation (e.g., in CIFAR-10, a 5.79$\times$ difference for $l_2$-norm between $0.3183$ and $1.8437$ results in only a $1.1\%$ difference for SSIM); trigger perturbation is inconsistent with PSNR results (e.g., in CelebA, an increase of $l_2$-norm from 0.8568 to 0.7880 leads to a decline of PSNR from 42.25 to 30.42).
We can observe that LADDER achieves superior spatial stealthiness in 18 out of 20 cases, underscoring its significantly enhanced natural stealthiness compared to others. 
Note frequency attacks such as FIBA and FTrojan still show better stealthiness than those spatial attacks.
LADDER achieves such a practical stealthiness  because 
(1) the perturbations produced by LADDER are minimal due to our trigger stealthiness objective; 
(2) since the LADDER trigger is inserted in the spectral domain, the intensity of the trigger pattern spreads across the entire spatial domain and (3) we pose the perturbation in the low-frequency domain where large magnitude of frequency information exists, which provides capacity to hide small perturbations.
LADDER triggers induce less perturbation on each pixel in the spatial domain.

\begin{figure*}[]
    \centering
    \scalebox{0.38}{\includegraphics{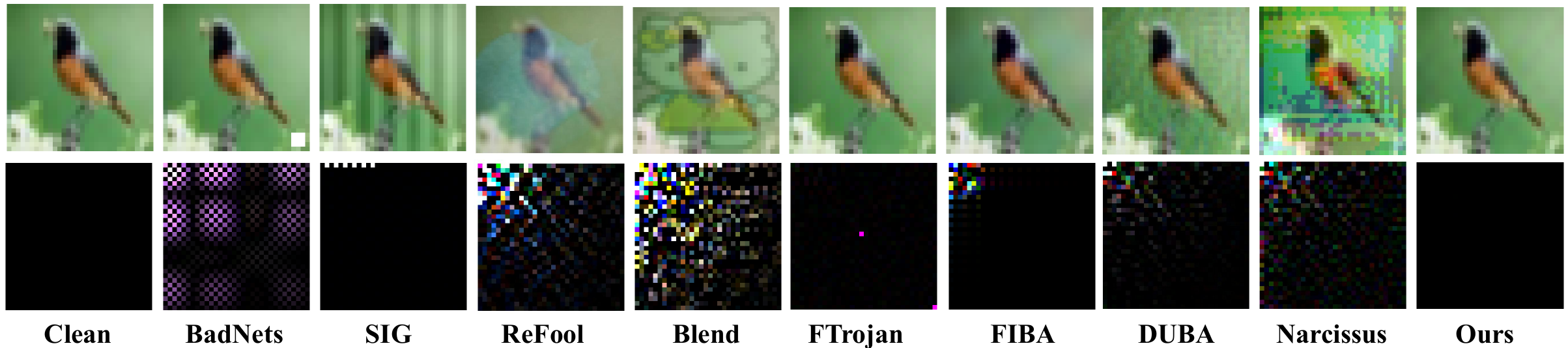}
    }
\caption{Comparing poisoned and clean images in the frequency domain reveals disparities caused by backdoor attacks on CIFAR-10 dataset. The top row displays clean and poisoned images, while the bottom row illustrates the spectrum disparity of each poisoned image compared to the clean spectrum. The disparity of clean image is black since disparity does not exist. Also, the sound dual-domain stealthiness remains the same when we attack different labels, see  \Cref{fig:multiTargets}.}
    \label{fig:freq_disparity_map}
\end{figure*}

To visually confirm the superior trigger stealthiness achieved by LADDER in \Cref{natural_stealthiness}, we plot the clean and poisoned images in the first row in \Cref{fig:freq_disparity_map}.
We see that the image poisoned by LADDER is undetectable, so that its trigger achieves equal and superior stealthiness to other frequency and spatial backdoor attacks.
More poisoned samples produced by LADDER across 5 datasets (see \Cref{fig:visual_inspect_images} in Appendix) can further confirm its practical and natural stealthiness.

\noindent\textbf{Spectral Stealthiness.}
This work represents the first instance to consider stealthiness in dual domains.
To confirm the trigger stealthiness in the spectral domain, we visualize the residual map of the frequency disparities between clean and poisoned images in the second row in \Cref{fig:freq_disparity_map}.
The frequency disparity is derived by subtracting the spectrum of the poisoned sample from its clean counterpart.
Bright pixels emerge in the residual map of the compared attacks, indicating a notable frequency disparity between clean and poisoned samples.
Different from that, our residual map is almost black, where disparity exists yet is not visible.
Through the results, we can draw a solid conclusion that LADDER, compared to those eight black-box attacks, achieves a remarkably better spectral stealthiness. 
Note along with the natural stealthiness, LADDER obtains a solid \emph{dual-domain stealthiness}.
\begin{table*}[h]
\centering
\caption{Attack robustness (\%) of various triggers against preprocessing-based defenses. To illustrate the robustness of our low-frequency trigger, we introduce various variants of LADDER for comparison, named LADDER-Mid, LADDER-High and LADDER-Full, which search the triggers across different regions in spectrum with the same attack settings. 
Our low-frequency trigger design achieves an average ASR of 90.23\%, which is 50.09\% higher than the ASR averaged by seven popular attacks and three variations of LADDER targeting different spectral regions. This demonstrates our superior robustness against preprocessing.}
\label{preprocess_defense}
\scalebox{0.7}{
\begin{tabular}{ccccccccccccccccccccccc} 
\toprule
Attacks $\rightarrow$       & \multicolumn{2}{c}{\textsc{BadNets} \cite{badnets}} & \multicolumn{2}{c}{\textsc{FTrojan} \cite{ft}}& \multicolumn{2}{c}{\textsc{FIBA} \cite{fiba}} & \multicolumn{2}{c}{\textsc{DUBA} \cite{DUBA}} & \multicolumn{2}{c}{\textsc{Narcissus-D} \cite{Narcissus}} & \multicolumn{2}{c}{\textsc{BELT} \cite{BELT}} & \multicolumn{2}{c}{\textsc{HCB} \cite{watchout}} & \multicolumn{2}{c}{\textsc{LADDER-Mid}} & \multicolumn{2}{c}{\textsc{LADDER-High}} & \multicolumn{2}{c}{\textsc{LADDER-Full}} & \multicolumn{2}{c}{\textsc{LADDER-Low}}        \\ 
\cmidrule(l){2-3} \cmidrule(l){4-5} \cmidrule(l){6-7}\cmidrule(l){8-9}\cmidrule(l){10-11}\cmidrule(l){12-13} \cmidrule(l){14-15} \cmidrule(l){16-17} \cmidrule(l){18-19} \cmidrule(l){20-21} \cmidrule(l){22-23}
Methods $\downarrow$ &ACC &ASR   & ACC & ASR     & ACC   & ASR                 & ACC   & ASR                 & ACC   & ASR                 & ACC  & ASR                   & ACC   & ASR   & ACC   & ASR    & ACC   & ASR   & ACC & ASR   & ACC  & ASR \\ 
\midrule
Original      & 92.02 & 98.78          & 92.53 & 99.82 & 91.13 & 97.60  & 91.97 &     99.99  & 92.17 &  99.99  & 91.58 & 99.93 & 91.17 & 98.42 & 91.51 & 99.49               & 92.33 & 99.99                  & 92.54 & 99.94 & 92.82 & 99.95         \\
Gaussian Filter ($w=(3,3)$) & 66.17 & 15.11               & 67.80 & 6.47 & 61.99 & 94.48 & 65.30 &    6.31 & 65.19 &   4.42 & 60.83 & 6.36 & 64.10 & 17.54        & 67.45 & 11.79               & 67.04 & 5.92                  & 64.29 & 6.32  & 66.41 & 95.17         \\
Gaussian Filter ($w=(5,5)$) & 39.81 & 6.88                & 45.03 & 3.25 & 46.00  & 93.71 & 44.37 &      3.44  & 45.21 &   0.61 & 33.96 & 1.89 & 36.95 & 10.09     & 42.76 & 3.18                & 42.90 & 2.20                  & 40.12 & 2.52  & 61.21   & 94.33         \\
Wiener Filter ($w=(3,3)$)   & 69.53 & 88.11       & 69.11 & 10.54 & 58.72 & 95.17 & 65.10 &     53.42 & 64.27 &    4.85  & 63.35 & 93.12 & 62.43 & 92.04   & 65.81 & 9.82                & 67.87 & 6.23                  & 63.95 & 8.56  & 67.11 & 94.83         \\
Wiener Filter ($w=(5,5)$)   & 52.18 & 96.43               & 49.20 & 5.28 & 37.67 & 94.79 & 45.22 &     92.40  & 45.01 &   5.87  & 41.33 & 48.35 & 42.44 & 96.89    & 44.92 & 3.86                & 50.18 & 2.24                  & 43.78 & 4.49  & 47.15 & 92.65         \\
Brightness ($1.1$)    & 81.14 & 97.27    & 82.86 & 74.83 & 71.39 & 44.19 & 69.75 &    95.15  & 75.18 &   84.64  & 91.09 & 99.88 & 90.87 & 95.53  & 71.64   & 9.08                 & 77.12 & 10.74                  & 76.57 & 8.81 & 80.36   & 91.94           \\
Brightness ($1.5$)         & 82.08 & 91.76        & 79.24 & 75.52 & 70.43 &  38.67 & 67.07 &   99.46  & 70.28 &    83.71  & 86.75 & 74.89 & 85.90 & 83.94      & 73.54   & 9.83                 & 71.44 & 13.37                  & 78.64  & 8.77  & 77.15   & 83.32           \\
JPEG (quality = $90\%$)     & 88.98 & 97.85       & 89.22 & 9.36 & 67.06 & 82.18 & 88.34 &      11.18  & 89.15 &    89.33 & 88.27 & 9.12 & 88.39 & 97.72    & 89.56 & 9.72                & 89.75 & 9.15                  & 90.35 & 9.57  & 91.72   & 89.86           \\
JPEG (quality = $50\%$)     & 78.84 & 92.59       & 79.66 & 8.58 & 70.43 & 38.67 & 73.83 &     8.80 & 75.42 &     70.08  & 76.22 & 8.37 & 77.79 & 95.08   & 80.39 & 9.10                & 79.21 & 8.40                  & 80.20 & 6.45  & 76.09   & 79.79           \\ 
\midrule
Average ASR    &       & 73.25       &       & 32.63  & & 72.73 &  &       46.27  &  &   42.94  &  & 42.75 & & 73.60 &       & 18.43                 &      & 17.58                  &       & 17.27  &       & \textbf{90.23}  \\
\bottomrule
\end{tabular}}
\end{table*}

\Cref{fig:freq_disparity_map} clearly emphasizes the necessity of designing triggers in the \textbf{dual domains} in order to avoid anomaly in both domains.
Taking two frequency-domain backdoor attacks FTrojan and FIBA for example, their triggers achieve almost perfect visual stealthiness in the spatial domain; 
but they cannot eliminate the anomaly in the spectrum domain. 

\begin{figure}
    \centering
    \scalebox{0.35}{
    \includegraphics{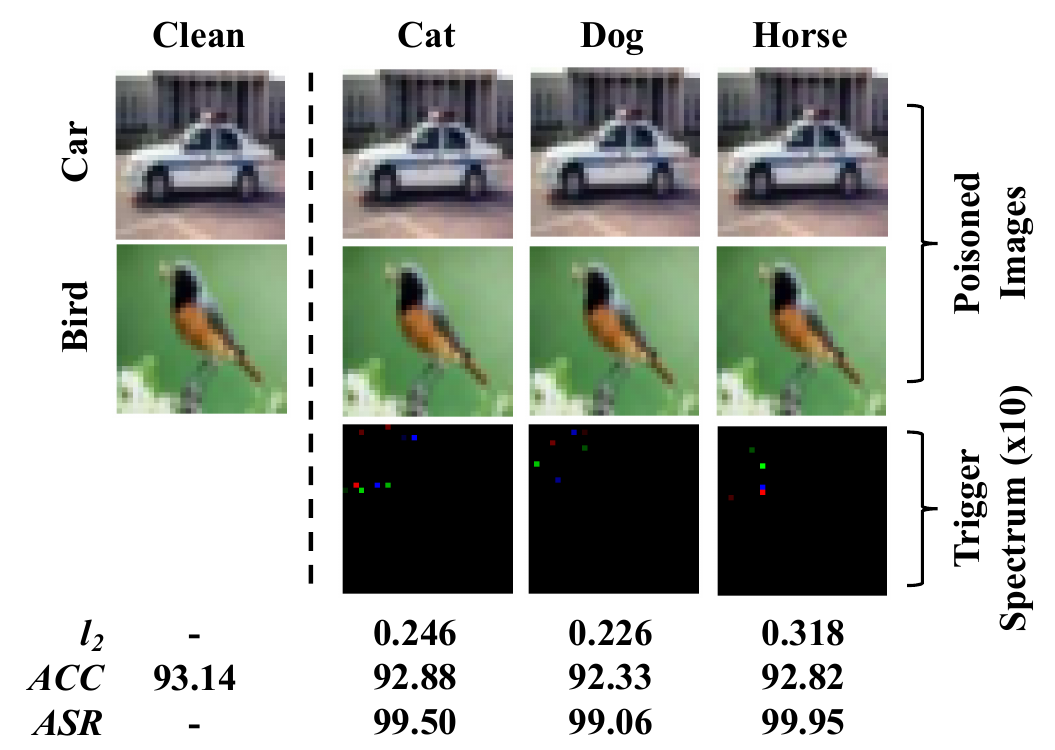}}
    \caption{
    Visualization of triggers along with their $l_2$, ACC (\%), and ASR (\%) on CIFAR-10 for arbitrary target labels.
    We select clean samples from two classes, "car" and "bird," and amplify the trigger strength by 10$\times$ to make triggers perceptible to humans.
    As shown in the bottom, LADDER can deliver high attack effectiveness with any chosen target label without degradation of trigger stealthiness in dual domains.}
    \label{fig:multiTargets}
\end{figure}

In default, we set target label to the 7th class for all the datasets (which is ``horse" in CIFAR-10).
To confirm the ability to generate dual-domain stealthy triggers on arbitrary classes, we showcase the poisoned images and triggers under target labels of 3rd (``cat") and 5th (``dog") in \Cref{fig:multiTargets}.

\subsection{Attack Performance Against Defenses}
\label{Robustness Against Defenses}
We evaluate the attack effectiveness against mainstream detection defenses, such as the network inspection \cite{gradcam} and STRIP \cite{strip}; and defensive-based defenses, such as Neural Cleanse \cite{nc} and Fine-pruning \cite{fine_pruning}. 
We further test the effectiveness of LADDER against the SOTA defenses including ASD \cite{ASD}, CBD \cite{CBD}, DBD \cite{DBD}, FIP \cite{fip}, MM-BD \cite{MM-BM} and MOTH \cite{moth} in \Cref{sec:asd_defense}.
We also confirm the dual-domain stealthiness of LADDER via frequency artifacts inspection \cite{rethink} on poisoned images.
Besides, we evaluate our attack under preprocessing-based operations as in works \cite{color,ft} to yield a solid confirmation of robustness.

\begin{figure}
    \centering
    \scalebox{0.31}{
    \includegraphics{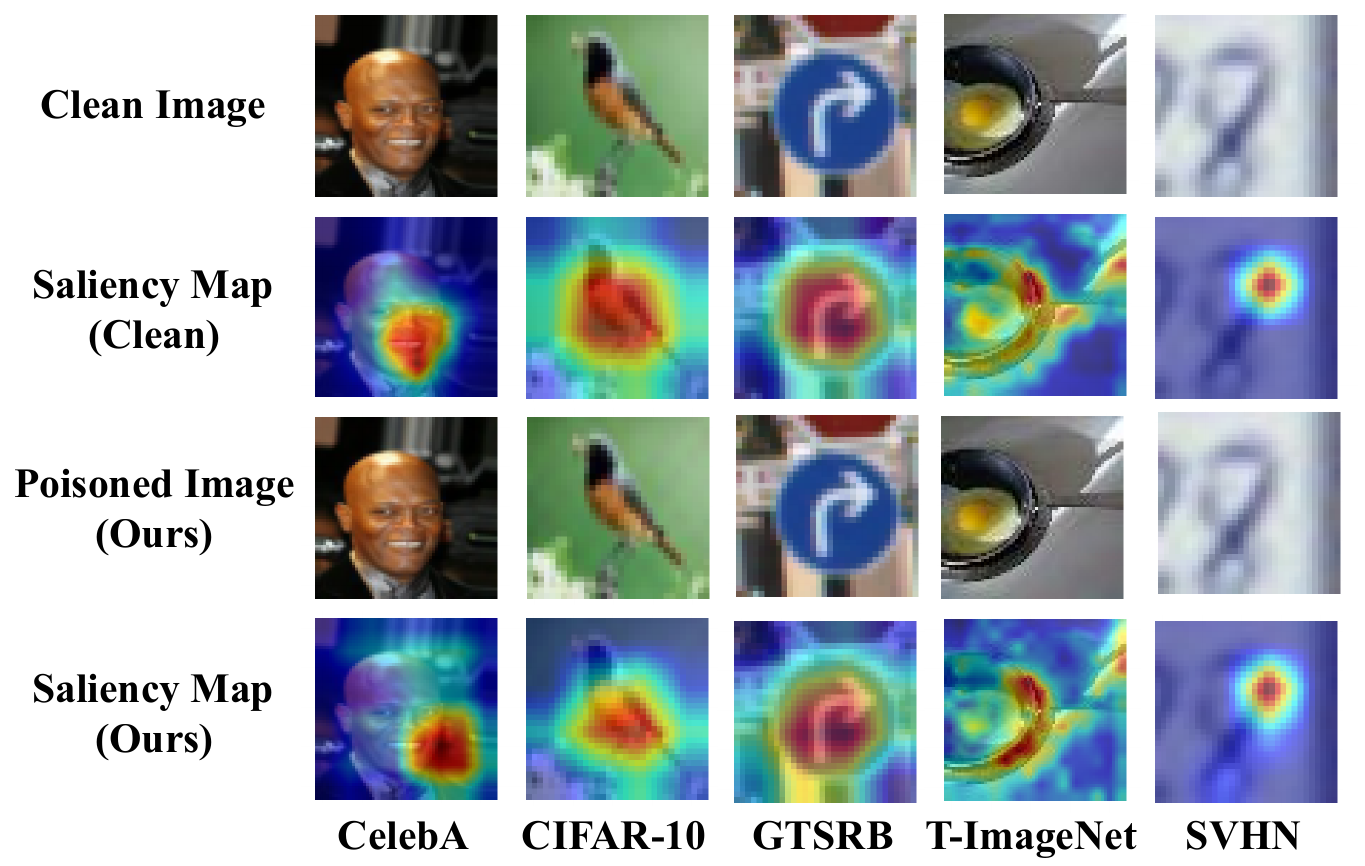}
    }
    \caption{Visualization of network attention (saliency map) via Grad-CAM from clean and poisoned images denoted in \Cref{tab:task_dataset_model}. 
    The region masked by red indicates a strong contribution toward model prediction.
    }
    \label{fig:gradCam}
\end{figure}

\noindent{\textbf{Against Network Inspection.}}
Grad-CAM \cite{gradcam} visualizes the critical regions of an input image that can mostly activate the prediction, which helps understand the features a CNN model learned.
It has been reported \cite{inputAware,fiba,wanet} that a backdoored CNN model tends to show an attention shift on poisoned images compare to clean ones. 
We showcase the network attention map on benign and victim models against CelebA, CIFAR-10, GTSRB, Tiny-ImageNet, and SVHN datasets, in \Cref{fig:gradCam}.
According to the results, we see that the attention of the model on benign and poisoned images almost remains the same, indicating LADDER does not cause severe attention anomaly. 
We insert our triggers in the low-frequency region where abundant semantic information exists. 
Thus, the trigger pattern is obfuscated within the original semantics, ensuring that LADDER does not introduce any anomalous regions. 

\begin{figure}
\begin{subfigure}[b]{0.15\textwidth}
         \centering
        \includegraphics[width=\textwidth]{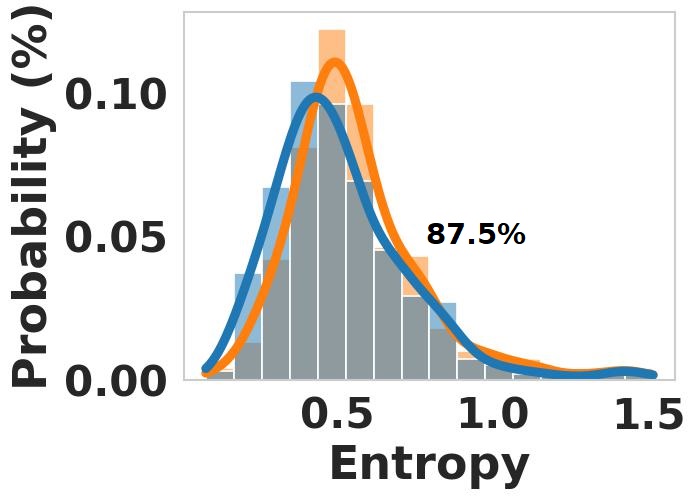}
         \caption{SVHN}
         \label{fig:strip_SVHN}
     \end{subfigure} 
    \begin{subfigure}[b]{0.15\textwidth}
         \centering
   \includegraphics[width=\textwidth]{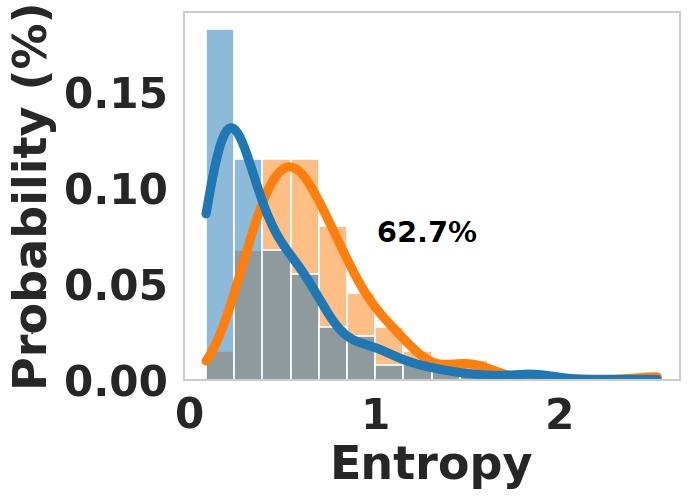}
         \caption{GTSRB}
         \label{fig:strip_gtsrb}
   
     \end{subfigure}
    \begin{subfigure}[b]{0.15\textwidth}
         \centering
    \includegraphics[width=\textwidth]{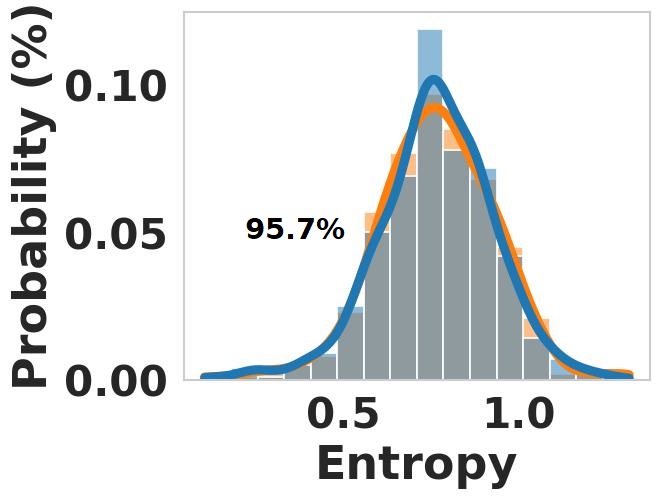}
         \caption{CIFAR-10}
         \label{fig:strip_cifar10}
    \end{subfigure}
     \centering
     \begin{subfigure}[b]{0.15\textwidth}
         \centering
    \includegraphics[width=\textwidth]{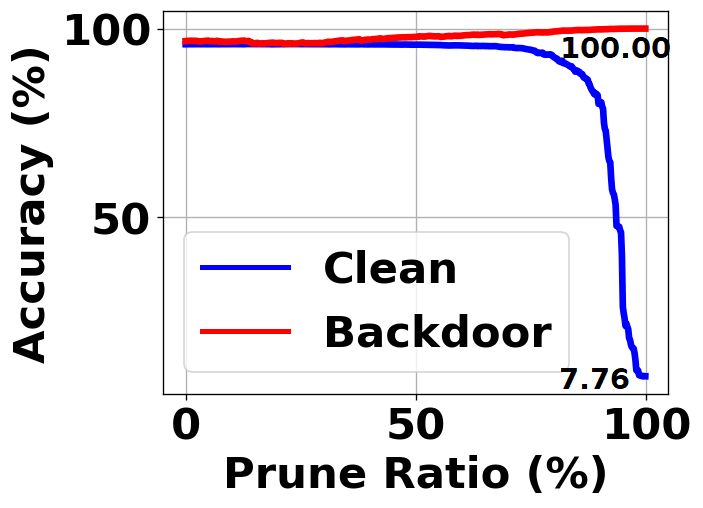}
         \caption{SVHN}
         \label{fig:prune_SVHN}
     \end{subfigure} 
     \begin{subfigure}[b]{0.15\textwidth}
         \centering
    \includegraphics[width=\textwidth]{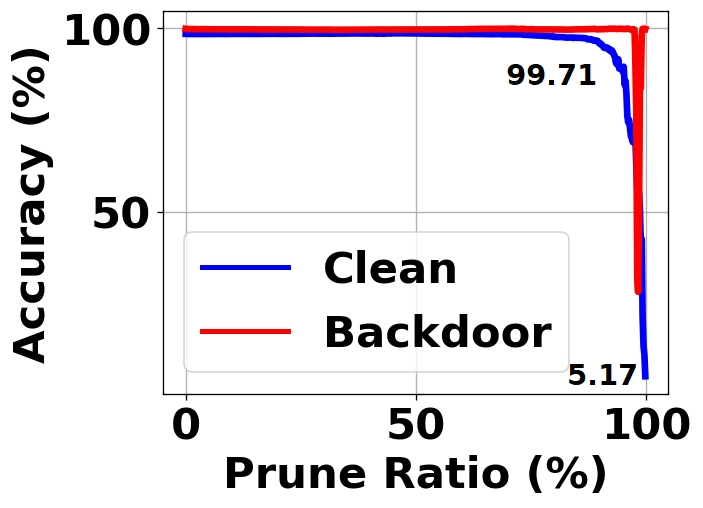}
         \caption{GTSRB}
         \label{fig:prune_GTSRB}
     \end{subfigure}
    \begin{subfigure}[b]{0.15\textwidth}
             \centering
    \includegraphics[width=\textwidth]{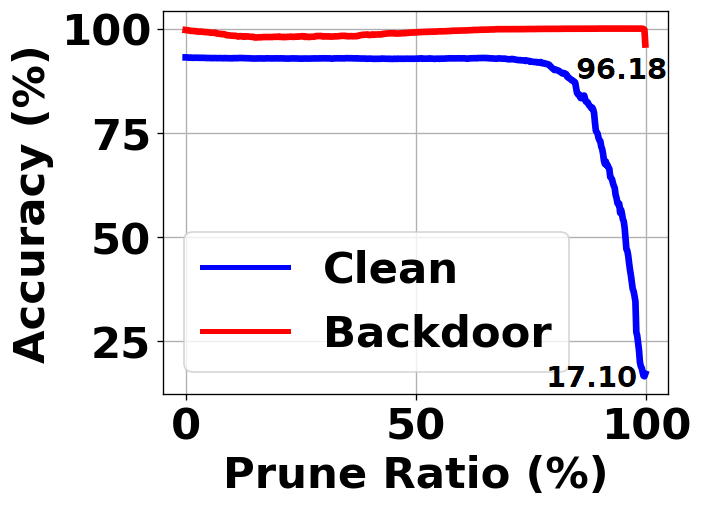}
             \caption{CIFAR-10}
             \label{fig:prune_cifar10}
    \end{subfigure}
    \caption{(a)-(c): The entropy distribution obtained with model poisoned by LADDER against STRIP. The distributions marked in blue and orange are obtained with benign and poisoned (by LADDER) testing data. Each curve is fitted to its respective distribution, with the annotated numbers representing the proportion of the overlapped areas relative to the backdoor distributions; (d)-(f): The ASR and ACC of LADDER against Fine-pruning after the correspondent percentage of pruned neurons. The final ACCs and ASRs are annotated after the defense.}
    \label{fig:strip+fp}
\end{figure}

\noindent{\textbf{Against STRIP.}} STRIP is a well-established backdoor defense strategy based on the assumption that poisoned data in a backdoored model consistently produces the target label and cannot be easily altered. 
Under this assumption, STRIP poisons samples by assessing the entropy of classification, achieved by overlaying randomly selected clean images onto the test samples. 
It expects the resulting entropy distribution to resemble that of the entropy distribution obtained with only clean images, thereby identifying and mitigating poisoned samples. 
We test the images poisoned by LADDER against STRIP, and visualize the entropy distribution among samples. 
The results are in \Cref{fig:strip+fp} (a)-(c), in which blue and orange bars indicate the (normalized) probability of the correspondent entropy on clean and poisoned images respectively, while the curves are the respective fitted distributions. 
The overlapped areas of distributions reflect the difficulty of poisoned samples being detected.
We observe that LADDER achieves almost perfect entropy probability distributions as clean samples  on SVHN, GTSRB and CIFAR-10 since their distributions are almost overlapped.
This is so because superimposing random images in the spatial domain destroy low-frequency components (containing LADDER trigger pattern) of our poisoned images.
Therefore, the predictions of superimposed images also undergo significant changes, which is similar to the clean cases. 
In conclusion, STRIP cannot effectively identify the difference between clean and poisoned samples by LADDER.

\begin{figure}
    \centering
    \scalebox{0.13}{\includegraphics{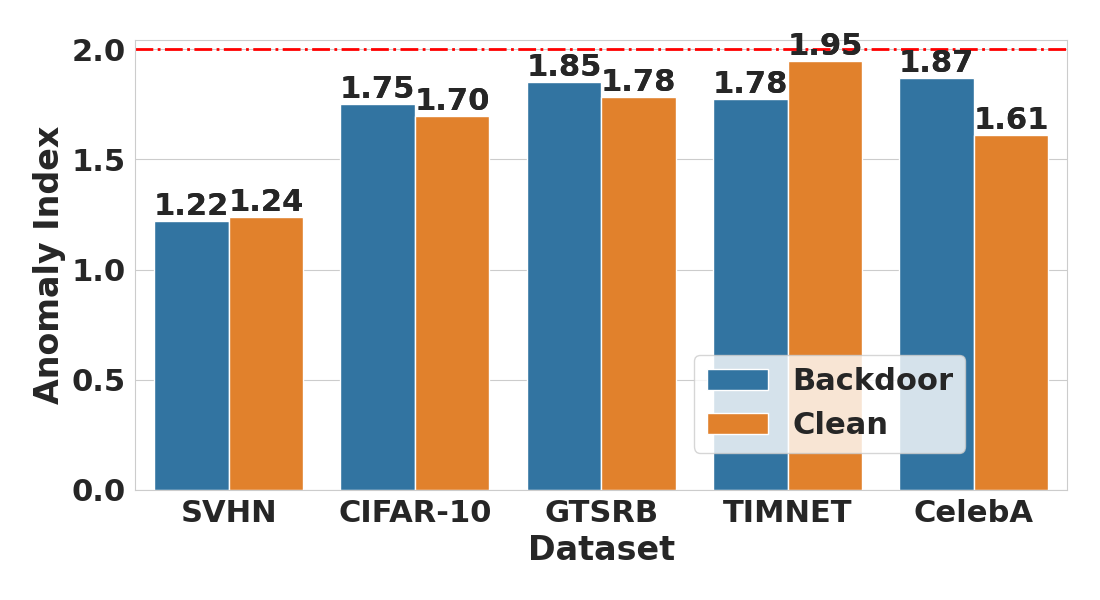}}
    \caption{The results of LADDER under Neural Cleanse on different datasets. The dotted line marks the threshold, below which a model is regarded as clean.}
    \label{fig:NC}
\end{figure}

\noindent{\textbf{Against Neural Cleanse (NC).}}
The insight behind NC is that any samples with a backdoor trigger result in a misclassification to the target label in the victim model.
NC reverses the possible triggers to detect backdoors on an unverified model by checking if the reversed trigger can possibly cause misclassification on the test dataset. 
It determines if a model has been compromised with an anomaly index. 
The index exceeding 2 indicates a high-risk level of model poisoning.
We test LADDER against NC on five datasets. 
The results are in \Cref{fig:NC}, in which the x-axis indicates different datasets and the y-axis records the anomaly index produced by NC. 
The blue and orange color bars represent the anomaly index with clean and poisoned data, respectively, on the poisoned model with LADDER.
We see that the results are within the threshold of 2.0, showcasing LADDER successfully evades NC. 
Recall that NC focuses on small and fixed backdoor patches. 
Triggers produced by LADDER in the low-frequency region of spectrum spread across the entire spatial domain, rendering the perturbation visually imperceptible due to the small number of magnitudes in the triggers. 
As a result, NC cannot discern the trigger pattern, leading to the failure of detecting poisoned samples.

\noindent{\textbf{Against Fine-pruning.}}
Fine-pruning, which is an effective and widely adopted backdoor defense, iteratively removes neurons to eliminate potential backdoor triggers inserted during training, thus fortifying the model against backdoor attacks without severely compromising its primary performance on clean data. 
The results are given in \Cref{fig:strip+fp} (d)-(f), in which neurons are iteratively pruned from 0 to 100\%, as indicated in x-axis; red and blue curves indicate the ASR and ACC for backdoor and benign data respectively.
We see that with the increase of pruning ratio, the benign accuracy drops more quickly than that of backdoor. 
Till the end of the pruning process, the ACC falls to almost zero whereas ASR is still valid, making backdoor mitigation by fine-pruning impossible. 
Thus, fine-pruning is ineffective to mitigate the attack effectiveness of LADDER.

\noindent{\textbf{Against Image Preprocessing.}}
It has been demonstrated in \cite{color,LFAP} that image preprocessing is effective in filtering trigger patterns and can mitigate backdoor attacks. 
To demonstrate the robustness of our attack against preprocessing-based defenses, we apply typical preprocessing methods \cite{jahne2005digital} such as the brightness adjustment, Gaussian filter, Wiener filter 
and JPEG compression on the poisoned images before inference, and demonstrate the robustness of LADDER quantitatively in \Cref{preprocess_defense} and visually in \Cref{fig:smoothing_disparity} of \Cref{sec:Disparity of spectrum smoothing}.

In \Cref{preprocess_defense}, we record the ACC and ASR achieved by BadNets, FTrojan, FIBA, DUBA, Narcissus as well as LADDER on low, mid, high and full frequency regions after image preprocessing. 
We clearly see that the average ASR achieved by LADDER in low-frequency region is 90.23\%, significantly outperforming others. 
This is because most preprocessing operations focus on either spatial domain (e.g., brightness adjustment) or high-frequency artifacts (such as filter and compression), which does not destroy LADDER's trigger pattern in low-frequency region. 
We can expect to achieve a similar performance (by LADDER) on other similar preprocessing-based defenses.

\begin{figure}[]
    \centering
    \scalebox{0.18}{\includegraphics{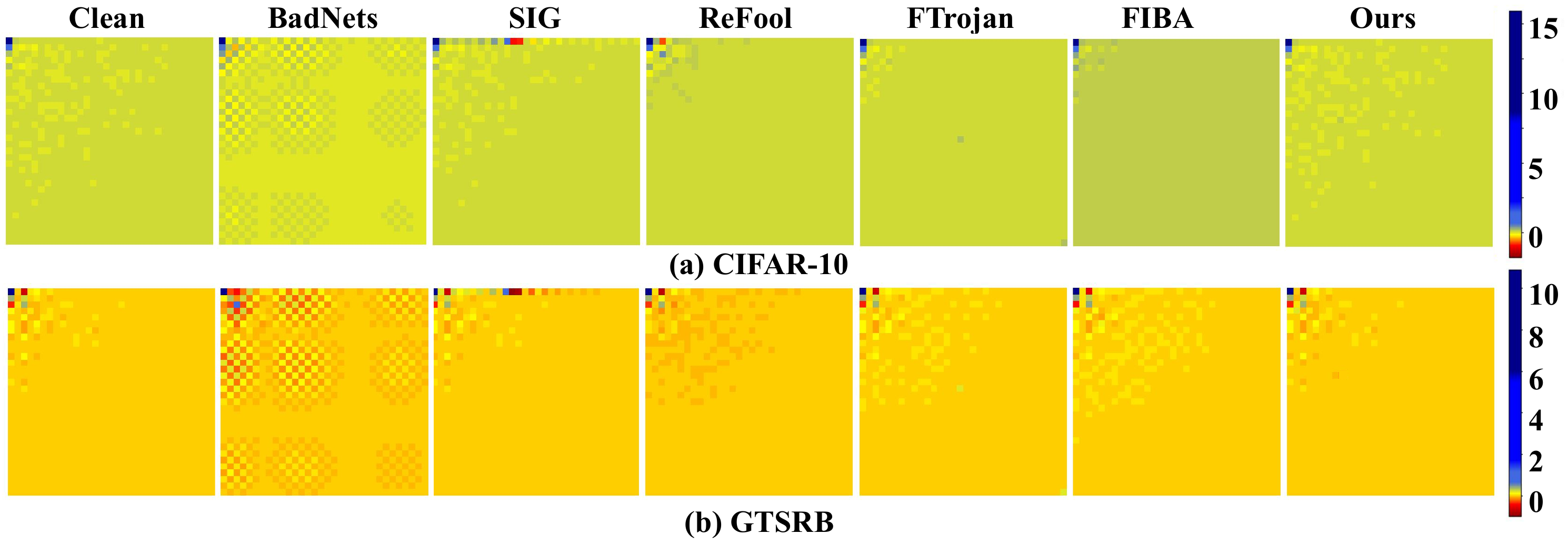}}
    \caption{Visualization of the averaged spectrum obtained by DCT on 5000 randomly selected clean samples and their poisoned counterparts under various spatial and frequency attacks on (a) CIFAR-10 \cite{cifar10} and (b) GTSRB \cite{gtsrb}.}
    \label{fig:freqInspec}
\end{figure}

\noindent\textbf{Against Frequency Artifacts Inspection.}
We inspect the spectrums of clean and poisoned samples against clean images, poisoned images from LADDER and other attacks, and report the spectrum averaged on 1000 randomly selected test images in \Cref{fig:freqInspec}. 
The average spectrum between clean images and the images poisoned by LADDER does not produce an anomaly, while a significant anomaly exists in the spectrum of poisoned images obtained by other attacks. 
This is because LADDER minimally perturbs low-frequency components within a few frequency bands, yet others introduce anomaly frequency artifacts without spectral stealthiness. 

\noindent{\textbf{Against Other SOTA Defenses.}} 
We analyse the attack effectiveness of LADDER and other attacks against three recent SOTA white-box defenses (see \Cref{sec:asd_defense}).

\noindent\textbf{Adaptive Defense under Black-box Setting.} To mitigate backdoor attacks that take spectral stealthiness into account, we tentatively propose frequency domain anomaly detection that distinguishes poisoned images by the parameter $p$ of the distribution $1/r^{p}$ concerning  magnitudes $r$ in the averaged spectrum. 
Please see \Cref{sec:possible_data_defense} for methodology and results. 

\section{Ablation Study}
\label{ablation_study}

\subsection{On the Transferability of Surrogate Model}
\label{subsec:transferability}
We assume the role of a malicious data provider who has access to a dataset but lacks access to the target model. 
To evaluate the quality of the trigger during its optimization, we introduce a surrogate model for injecting the trigger and assessing attack effectiveness. 
We test LADDER's attack transferability on CIFAR-10 dataset across a range of typical surrogate and victim (target) model architectures including VGG16 \cite{vgg}, ViT \cite{ViT}, ResNet18 \cite{resnet} and Google-Net \cite{googlenet}. 
We search for triggers based on different surrogate model architectures, and choose the optimal trigger on surrogate model among those triggers. 
Finally, we inject the trigger obtained from each surrogate model into the corresponding target models and record the ACC and ASR after 50 epochs. 

In \Cref{tab:transferrability}, we verify that LADDER is transferable between heterogeneous model architectures in practical attack scenarios. 
The mismatch between surrogate and victim models does not degrade the ACC, while a high ASR is maintained across all mismatched cases. 
Additionally, using the same surrogate and victim models does not always guarantee the best performance, as seen with ResNet18 and GoogleNet. 
We conclude that LADDER's effectiveness is not sensitive to the specific combination of surrogate and target models. 
Therefore, optimal ASR can be achieved without requiring a specific model structure pairing between surrogate and target models.

\emph{Transferability discussions}. 
Recall that this work (in the context of CNNs and computer vision) addresses backdoor attacks with three objectives: attack effectiveness, stealthiness, and robustness. 
Among them, stealthiness (\Cref{eq:l2norm_spatial}) is model-independent and can be directly calculated based on trigger perturbation.
Similarly, robustness (\Cref{eq_robustness_def}) depends on how well the trigger perturbation retains its effectiveness after image preprocessing.
This perturbation depends on the specific preprocessing and the design of the trigger itself—independent of model architectures—and its effectiveness is largely determined by the norm of the perturbation. 
Attack effectiveness, on the other hand, is guaranteed by training a model with the trigger injected into the data. 
This objective is influenced primarily by: the number of feature vectors, the poison ratio, and the norm of the trigger perturbation \cite{litheoretical}. 
For example, using the CIFAR-10 dataset, with ResNet18 as the surrogate model and GoogLeNet as the target model, both models leverage the same dataset, ensuring an identical number of feature vectors. 
The optimal trigger generated via optimization is used directly in the actual attack phase, which maintains the same trigger perturbation. 
Finally, both models use the same poison ratio to create the poisoned dataset. 
Controlling these factors enables us to yield consistent attack effectiveness across models, thus providing transferability.

\subsection{Ablation Study of Trigger Design}
We conduct an ablation study (see \Cref{sec:ablation}) to confirm the importance of all the components in our trigger design.

\begin{table}[]
\centering
\caption{Transferability of LADDER across different surrogate and target model architectures (ASR/ACC)(\%).
The ASRs close to 100 indicate a tiny discrepancy in backdoor performance between the surrogate and victim model.
}
\scalebox{0.95}{
\begin{tabular}{ccccc}
\toprule
 
 {\diagbox{Sur}{Tar}}
 & VGG16 \cite{vgg} & ResNet18 \cite{resnet} & Google-Net \cite{googlenet} & ViT \cite{ViT} \\ \midrule
VGG16      &  99.86 / 91.87    &    99.97 / 93.54  &   99.82 / 93.41  &  99.48 / 83.34     \\
ResNet18   &   99.07 / 91.51    &   99.42 / 92.74 &    99.62 / 93.40  & 99.93 / 82.69 \\
Google-Net &   99.51 / 91.88   &   99.58 / 92.91     &     99.17 / 93.78 & 99.61 / 82.02  \\ 
ViT & 99.52 / 91.10 & 99.88 / 92.75 & 99.40 / 93.77 & 99.66 / 82.74\\
\bottomrule
\end{tabular}}
\label{tab:transferrability}
\end{table}

\subsection{Scalability Analysis} 

We evaluate time and resource usage (CPU/GPU utilization, RAM, GPU memory) of trigger optimization across various models, datasets and the number of objectives in \Cref{sec:complexity_analysis}.

%% file: 7_conclusion.tex
\section{Ethical Consideration}
This work exposes the vulnerability of deep learning models to practical, stealthy and robust backdoors and can inspire follow-up studies that enhance the security of deep learning. 
In this sense, this work has a positive impact on the future research of AI safety.
In the following, we discuss the intellectual property, intended usage, potential misuse, risk control and human subject.
\\
\textbf{Intellectual property.}  
All comparative attacks and defenses, models, datasets and implementation libraries are open-source. 
We believe that the datasets are well-desensitized. 
We strictly comply with all applicable licenses for academic use.
\\
\textbf{Intended Usage.} 
We expose the vulnerability of current deep learning models to practical stealthy and robust backdoor triggers.
We encourage researchers to use our findings to assess the security of their models and hope that this work will inspire development of robustness against backdoor attacks.
\\
\textbf{Potential Misuse.}
This work could be exploited to produce stealthy and robust poisoned datasets for real-world applications, which potentially leads to more covertly malicious models.  
To maintain safety of deep learning models, we propose an adaptive defense in~\Cref{sec:possible_data_defense}. 
\\
\textbf{Risk Control.} 
To further mitigate potential risks, we will release the code used in this work.
By doing so, we believe that transparency will reduce the risks related to our work, encourage responsible use and foster further advancement of secure techniques for deep learning models.
\\
\textbf{Human Subject.} 
We do not involve any human subjects in this work. 
Instead, we rely solely on mathematical models and metrics to simulate human visual inspection, thereby eliminating the need for human participation.

\section{Limitations}
The effectiveness of LADDER against white-box defenses is naturally reduced. 
Recall that this work designs triggers under a black-box attack scenario.  
Unlike white-box attacks which can directly manipulate model parameters, black-box variants could naturally not perform well against some specific white-box backdoor defenses. 
This work focuses on computer vision using CNNs and has not yet been extended to other tasks or network architectures. 
It is interesting to make such an extension, e.g., for natural language processing.

\section{Conclusion} 
\label{sec:conclusion}

This work introduces LADDER, a multi-objective backdoor attack that effectively searches for backdoor triggers via an evolutionary algorithm. It achieves effectiveness, dual-domain stealthiness, and robustness, instilling confidence in its capabilities.  
First, we observe the conflict between trigger stealthiness and attack performance and find the sensitivity of solving multiple attack goals with the Lagrange multipliers and SGD.
Then, we improve the trigger robustness by designing triggers in the low-frequency domain while extending the trigger stealthiness to 
the dual domains. 
We also design a new multi-objective backdoor attack problem to capture the objectives simultaneously.
Finally, we leverage the evolutionary algorithm to solve the proposed problem in a black-box setting without tuning Lagrange coefficients. 
Experimental results confirm the practical performance of LADDER.

%% file: 8_appendix.tex
\appendix

\subsection{Summary of Computer Vision Tasks}
\label{sec:task_data_model_details}
We adopt 5 real-world computer vision tasks and 8 commonly used CNN architectures. We briefly list the details of the datasets and the victim/surrogate model structures in \Cref{tab:task_dataset_model}.

\begin{table*}
\centering
\caption{Summary of tasks and models.}
\scalebox{0.9}{
\begin{tabular}{@{}ccccccc@{}}
\toprule
Task    & Dataset  & \# of Training/Test Images & \# of Labels & Image Size                                        & Victim Model   & Surrogate Model              \\ \midrule
House Number Recognition & SVHN    & 73,257/26,032                & 10           & 32$\times$32$\times$3   & 3 Conv $+$ 2 Dense  & VGG16 \\
Object Classification         & CIFAR-10 & 50,000/10,000                & 10           & 32$\times$32$\times$3   & PreAct-ResNet18 & VGG11\\
Traffic Sign Recognition      & GTSRB    & 39,209/12,630                & 43           & 32$\times$32$\times$3   & PreAct-ResNet18  & VGG11                         \\
Object Classification         & Tiny-ImageNet & 100,000/10,000                        & 200           & 64$\times$64$\times$3 & ResNet18 & VGG19                       \\ 
Face Attribute Recognition         & CelebA & 162,770/19,962               & 8           & 64$\times$64$\times$3 & ResNet18   & VGG16                    \\ \bottomrule
\end{tabular}}
\label{tab:task_dataset_model}
\end{table*}

\subsection{Disparity of spectrum smoothing}
\label{sec:Disparity of spectrum smoothing}
\begin{figure*}
    \centering
        \scalebox{0.32}{\includegraphics{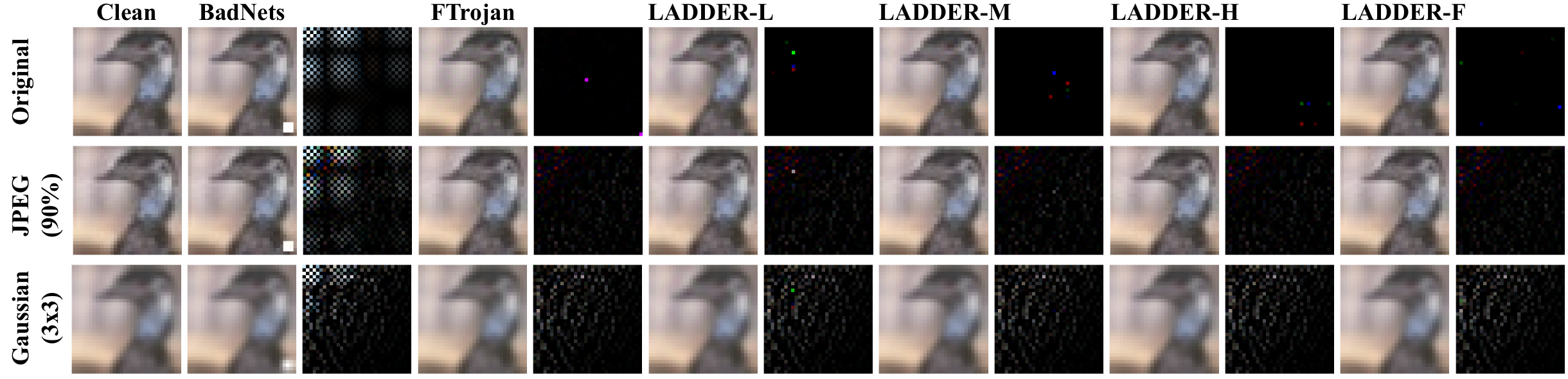}}
    \caption{Comparison of poisoned images with their corresponding frequency disparities (amplified by $5\times$) to clean images of existing attacks under different image pre-processing based defenses.
    Each frequency disparities spectrum is calculated based on the original clean image's spectrum.
    These image transformations can effectively remove the trigger pattern through spectral domain, while the disparities spectrums of LADDER-Low attack still contain original frequency backdoor patterns.}
    \label{fig:smoothing_disparity}
\end{figure*}

We present poisoned images alongside their frequency disparities related to clean images under various image preprocessing methods, as illustrated in \Cref{fig:smoothing_disparity}.
We observe that the frequency disparities of BadNets remain consistent with the original patterns after JPEG compression, whereas the Gaussian filter effectively disrupts the BadNets patterns.
These results corroborate the findings demonstrated in \Cref{preprocess_defense} that BadNets proves resilient to JPEG compression but fails to survive after Gaussian filtering. 
For FTrojan and LADDER-Full, frequency patterns are destroyed after the preprocessing operations. 
Even after undergoing the image preprocessing operations, the frequency disparities of LADDER-Low remain conspicuous, suggesting the robustness of our low-frequency attack against preprocessing-based defenses.
We note that low-frequency components exhibit greater resilience to image preprocessing than mid- and high-frequency components.

\begin{table}[]
\centering
\caption{Time and resource usage across various datasets on ResNet18.}
\label{tab:resource_dataset}
\scalebox{0.95}{
\begin{tabular}{cccccc}
\toprule
{\diagbox{Resource}{Dataset}} & SVHN   & CIFAR-10 & GTSRB   & T-ImageNet & CelebA \\ 
\midrule

CPU util. (\%) & 50.1                    & 50.2                      & 36.2                     & 24.9                        & 28.2                    \\
GPU util. (\%) & 59.3                    & 44.7                      & 54.8                     & 77.9                        & 48.1                    \\
RAM (GB)             & 6.17                    & 6.05                      & 7.55                     & 12.19                       & 6.02                    \\
GPU Mem (GB)    & 4.07                    & 4.02                      & 4.06                     & 7.65                        & 7.67                    \\
Time (s)          & 329 & 421  &  576 & 1970    & 2080                    \\ \bottomrule
\end{tabular}}
\end{table}

\begin{table}[]
\centering
\caption{Time and resource usage across different model architectures on CIFAR-10.}
\scalebox{0.95}{
\begin{tabular}{cccccc}
\toprule
\diagbox{Model}{Resource} & GoogLeNet & ResNet18 & ViT & VGG11 & VGG16 \\ \midrule
CPU util. (\%) & 38.1                       & 50.2                      & 50.6                 & 51.0                   & 50.3                   \\
GPU util. (\%) & 82.1                       & 44.7                      & 63.5                 & 20.7                   & 33.7                   \\
RAM (GB)             & 6.07                       & 6.05                      & 4.48                 & 6.04                   & 6.05                   \\
GPU Mem (GB)    & 13.43                      & 4.02                      & 5.01                 & 3.48                   & 3.76                   \\
Time (s)          & 1197                       & 421                       & 537                  & 411                    & 431                    \\ \bottomrule
\end{tabular}}
\label{tab:resource_models}
\end{table}

\begin{table}[]
\centering
\caption{Time and resource usage across various objectives on CIFAR-10 and ResNet18.}
\scalebox{0.95}{
\begin{tabular}{cccccc} \toprule
{\diagbox{Resource}{Objectives}} & Eff+Ste+Rob & Eff+Ste & Eff+Rob & Ste+Rob& Eff \\ \midrule
CPU util. (\%) & 50.2                      & 50.4                   & 50.7      & 6.9             & 50.4                \\
GPU util. (\%) & 44.7                      & 49.2                   & 49.5          & 0         & 49.1                \\
RAM (GB)             & 6.06                      & 6.05                   & 6.05      & 1.79             & 6.04                \\
GPU Mem (GB)    & 4.02                      & 4.06                   & 4.07         & 0          & 4.03                \\
Time (sec.)          & 421                       & 421                    & 426        & 10.18            & 428                 \\ \bottomrule
\end{tabular}}
\label{tab:resource_objectives}
\end{table}

\begin{figure}[]
    \centering
    \scalebox{0.27}{\includegraphics{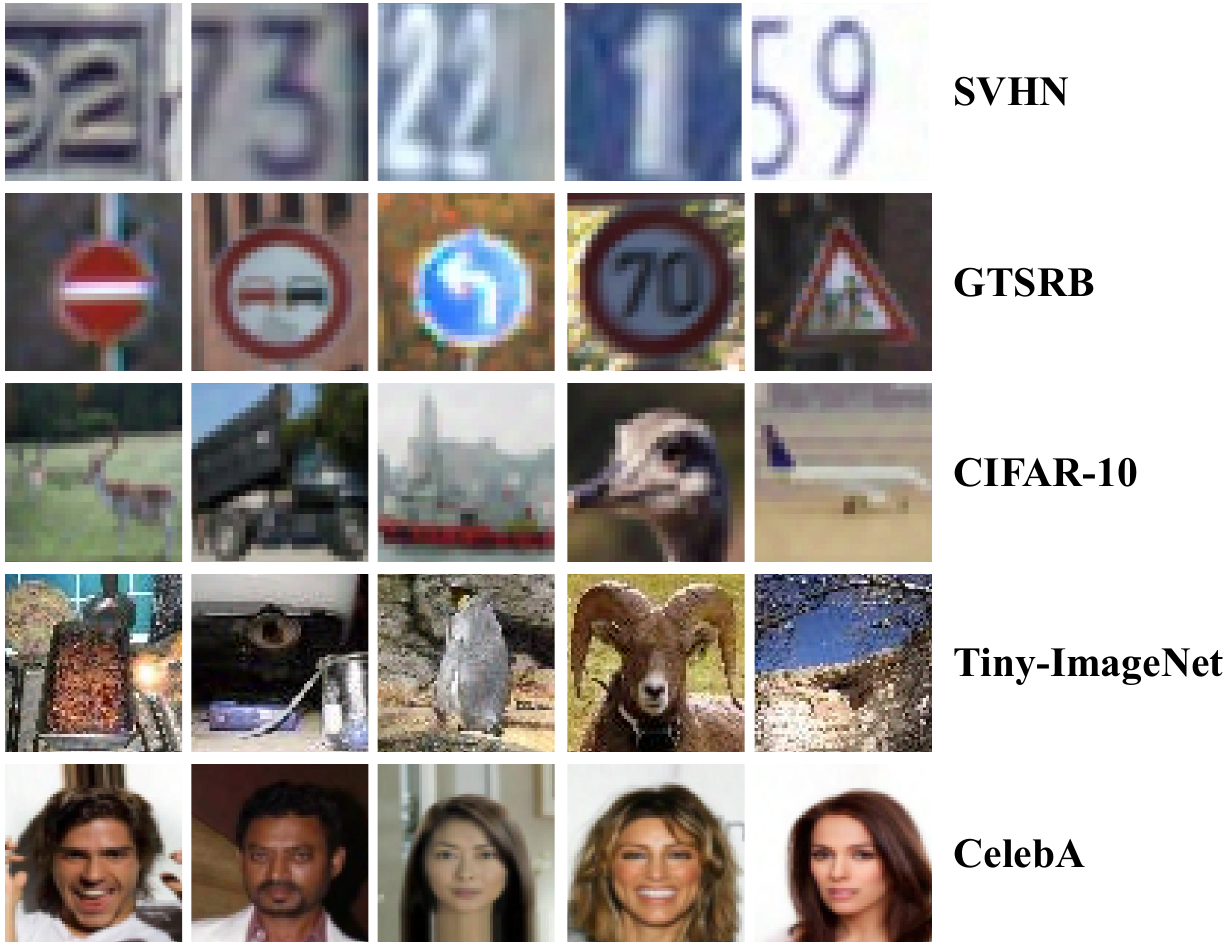}}
    \caption{Poisoned images produced by LADDER.}
    \label{fig:visual_inspect_images}
\end{figure}

\subsection{Scability Analysis}
\label{sec:complexity_analysis}
We investigate the LADDER's scalability in terms of time and resource usage, including CPU/GPU utilization (\%), RAM/GPU memory (GB) across 5 datasets, 5 models and various objectives.
\Cref{tab:resource_dataset} showcases the time and resource usage across datasets on ResNet18, in which small datasets such as SVHN and CIFAR-10 require around 50\% of CPU and GPU utilization while large datasets such as Tiny-ImageNet requires less CPU but more GPU usage.
Also, the time cost across datasets is positively correlated with the number of samples (see \Cref{tab:task_dataset_model}). 
We present the time and resource usage of LADDER across models in \Cref{tab:resource_models}.
The results show that CPU utilization increases while GPU utilization decreases as the number of model parameters grows (as shown from left to right in \Cref{tab:resource_models}, where the number of model parameters increases).
Note GoogLeNet contains Inception modules with a large number of convolution filters, which slows down the training speed and requires more GPU memory.
The time and resource usage across objectives is in \Cref{tab:resource_objectives}, where we run LADDER with different objectives (\underline{eff}ectiveness, \underline{rob}ustness and \underline{ste}althiness) on CIFAR-10 and ResNet18.
Our results show that objectives influence resource consumption.  
Specifically, evaluating effectiveness involves additional training on surrogate models and this consumes a significant proportion of resource usage (see Eff results in \Cref{tab:resource_objectives}). 
In contrast, robustness and stealthiness only yield constant complexity (see Ste+Rob results in \Cref{tab:resource_objectives}, \Cref{eq:moo_o2,eq:moo_o3}).

\subsection{Ablation Study of Trigger Design} 
\label{sec:ablation}
The trigger design of LADDER captures  {\underline{ste}}althiness, {\underline{rob}ustness}, and {\underline{eff}ectiveness} in the spectral domain. 
In \Cref{tab:ablation_trigger_components}, we showcase the results of  leveraging a subset of attack objectives and implementing a spatial variant. 
For example, Rob+Eff achieves superior robustness (93.84\%), but it falls short in providing practical stealthiness ($l_2$ = 3.5095).
Also, spatial trigger obtains 35.04\% of robustness although taking 2.2$\times$ more perturbation magnitude than the original LADDER.
This study confirms that LADDER can provide the most practical trigger considering all the objectives in the spectral domain.

\begin{table}[]
\centering
    \caption{Effectiveness (ASR), stealthiness and robustness of variants compared to the \underline{ori}ginal version of LADDER on CIFAR-10.}
\label{tab:ablation_trigger_components}
\scalebox{0.92}{
\begin{tabular}{ccccccc} \toprule
{\diagbox{Metrics}{Trigger}} & Spatial & Ste+Eff & Rob+Eff & Ste+Rob & Eff & Ori \\ \midrule
  Effectiveness (\%)& 99.99  & 99.99 & 99.85  & 94.83  & 99.88  & 99.99\\
  Stealthiness ($l_2$) & 0.6916 & 0.4007 & 3.5095 & 0.2020 & 2.9437 & 0.3183 \\
  Robustness (\%) & 35.04  & 24.94  & 93.84 & 64.62 & 11.42 & 82.52\\ \bottomrule
\end{tabular}}
\end{table}

\subsection{Evaluating LADDER against SOTA Backdoor Defenses} 
\label{sec:asd_defense}
Although successfully evading several classic backdoor defenses (see \Cref{Robustness Against Defenses}), LADDER is not perfectly robust against some (white-box) backdoor defenses especially defenses requiring model and training manipulation.
Recall that, in the strict black-box setting, attackers are not allowed to access the victim model (such as parameters, structures and gradient information) nor manipulate the training process. 

We test LADDER against six new SOTA backdoor defenses, ASD \cite{ASD}, CBD \cite{CBD}, DBD \cite{DBD}, FIP \cite{fip}, MM-BD \cite{MM-BM} and MOTH \cite{moth}. 
ASD adaptively splits clean from the poisoned dataset during training so as to defend backdoors. 
CBD leverages statistical effect among variables on the image to mitigate attacks. 
DBD proposes a three-stage mechanism, which involves learning on label-removed data, credible sample filtering and fine-tuning the trained model.
FIP proposes a backdoor purification technique that leverages the Fisher Information Matrix of a DNN to effectively eliminate backdoor imprints.
MM-BD introduces a maximum-margin-based backdoor
detection method by capturing the atypicality of the landscape of the classifier’s outputs (pre-softmax) induced by backdoor
attacks.
MOTH introduces an add-on training step that enhances class separation, making backdoor attacks more difficult by requiring larger and more detectable patterns compared to the original trigger perturbation.

\begin{table}[h]
\centering
\caption{Attack performance measured by ACC (\%) and ASR (\%) for 7 backdoor attacks against CIFAR-10 dataset and ResNet18 implemented in ASD.}
\begin{tabular}{@{}ccccc@{}} \toprule
Defense & \multicolumn{2}{c}{No Defense} & \multicolumn{2}{c}{ASD \cite{ASD}} \\ \cmidrule(l){2-3} \cmidrule(l){4-5} 
Attack  & ACC            & ASR           & ACC        & ASR     \\ \midrule
BadNets \cite{badnets} & 94.9           & 100.0           & 93.4       & 1.2       \\
Blend \cite{Chen2017TargetedBA}  & 94.1           & 98.3          & 93.7       & 1.6       \\
WaNet \cite{wanet}  & 93.6           & 99.9          & 93.1       & 1.7        \\
SIG \cite{sig} & 93.4           & 100.0          & 87.8      & 0.7          \\
DUBA \cite{DUBA} 				   & 93.2   & 99.1  &92.6   &4.0 \\
Narcissus-D \cite{Narcissus} 	   & 92.2   & 99.9  & 93.8 & 0.0  \\
Ours  & 93.8           & 99.9          & 92.4       & 25.0      \\ \bottomrule
\end{tabular}
\label{tab:asd_defense}
\end{table}

\begin{table}[h]
\centering
\caption{Attack performance measured by ACC (\%) and ASR (\%) for 7 backdoor attacks against CIFAR-10 dataset and WideResNet (WRN-16-1) implemented in CBD.}
\begin{tabular}{@{}ccccc@{}} \toprule
Defense & \multicolumn{2}{c}{No Defense} & \multicolumn{2}{c}{CBD \cite{CBD}} \\ \cmidrule(l){2-3} \cmidrule(l){4-5} 
Attack  & ACC        & ASR           & ACC        & ASR    \\ \midrule
BadNets \cite{badnets} & 85.4           & 100.0           & 87.5       & 1.1       \\
Blend \cite{Chen2017TargetedBA}  & 84.6           & 100.0          & 87.5       & 2.0     \\
WaNet \cite{wanet}  & 86.8           & 98.6          & 86.6       & 4.2     \\
SIG \cite{sig} & 84.1 & 99.3 & 87.3 & 0.3\\
% ReFool \cite{refool} & 1           & 2          & 3       & 4        \\
DUBA \cite{DUBA}  & 85.3 & 99.3 & 87.6 & 5.0 \\
Narcissus-D \cite{Narcissus} 	   & 87.1   & 100.0  & 87.9 & 4.1 \\
Ours  &  85.3          & 97.5         & 86.6       & 5.2      \\ \bottomrule
\end{tabular}
\label{tab:CBD_defense}
\end{table}

\begin{table}[h]
\centering
\caption{Attack performance measured by ACC (\%) and ASR (\%) for 7 backdoor attacks against CIFAR-10 dataset and ResNet18 implemented in DBD.}
\begin{tabular}{@{}ccccc@{}} \toprule
Defense & \multicolumn{2}{c}{No Defense} & \multicolumn{2}{c}{DBD \cite{DBD}} \\ \cmidrule(l){2-3} \cmidrule(l){4-5} 
Attack  & ACC        & ASR           & ACC        & ASR    \\ \midrule
BadNets \cite{badnets} & 95.0           & 100.0           & 92.4       & 1.0       \\
Blend \cite{Chen2017TargetedBA}  & 94.1           & 98.3          & 92.2       & 1.7     \\
WaNet \cite{wanet}  & 94.3           & 98.6          & 91.2       & 0.4     \\
SIG \cite{sig} & 94.2 & 100.0 & 91.6 & 0.3 \\
DUBA \cite{DUBA} & 94.4 & 99.4 & 90.8 & 0.2 \\
Narcissus-D \cite{Narcissus} 	   & 93.9   & 100.0  & 91.1 & 0.0 \\
Ours  &  94.1        &  100.0        & 90.9       & 0.0      \\ \bottomrule
\end{tabular}
\label{tab:DBD_defense}
\end{table}

\begin{table}[h]
\centering
\caption{Attack performance measured by ACC (\%) and ASR (\%) for 9 backdoor attacks against CIFAR-10 dataset and PreActResNet18 implemented in FIP.}
\begin{tabular}{@{}ccccc@{}} \toprule
Defense & \multicolumn{2}{c}{No Defense} & \multicolumn{2}{c}{FIP \cite{fip}} \\ \cmidrule(l){2-3} \cmidrule(l){4-5} 
Attack  & ACC        & ASR           & ACC        & ASR    \\ \midrule
BadNets \cite{badnets} & 93.0           & 100.0           & 89.3       & 1.9       \\
Blend \cite{Chen2017TargetedBA}  & 94.1           & 100.0          & 92.2       & 0.4     \\
WaNet \cite{wanet}  & 92.3           & 98.6          & 89.7       & 2.4     \\
SIG \cite{sig} & 88.6 & 100.0 & 86.7 & 0.9 \\
DUBA \cite{DUBA} & 94.6 & 99.5 & 86.7 & 8.8 \\
Narcissus-D \cite{Narcissus} 	   & 94.2   & 99.7  & 87.2 & 9.9 \\
HCB \cite{watchout} & 94.8           & 98.3          & 89.4       & 15.3        \\
BELT \cite{BELT} & 94.6   & 99.6          & 85.5       & 10.8        \\
Ours  &  93.6        &  99.0        & 90.4       & 10.4      \\  \bottomrule
\end{tabular}
\label{tab:FIP_defense}
\end{table}

\begin{table}[h]
\centering
\caption{Attack performance measured by ACC (\%) and ASR (\%) for 9 backdoor attacks against CIFAR-10 dataset and ResNet18 implemented in MM-BD.}
\begin{tabular}{@{}ccccc@{}} \toprule
Defense & \multicolumn{2}{c}{No Defense} & \multicolumn{2}{c}{MM-BD \cite{MM-BM}} \\ \cmidrule(l){2-3} \cmidrule(l){4-5} 
Attack  & ACC        & ASR           & ACC        & ASR    \\ \midrule
BadNets \cite{badnets} & 91.6           & 99.4           & 87.2       & 3.1       \\
Blend \cite{Chen2017TargetedBA}  & 91.4           & 96.4          & 88.7       & 10.8     \\
WaNet \cite{wanet}  & 91.0           & 96.5          & 86.7       & 8.7     \\
SIG \cite{sig} & 90.7 & 100.0 & 89.2 & 82.2 \\
DUBA \cite{DUBA} & 91.6 & 100.0 & 89.4 & 98.0 \\
Narcissus-D \cite{Narcissus} 	   & 90.7   & 100.0  & 88.6 & 98.3 \\
HCB \cite{watchout} & 91.6           & 97.9          & 88.4       & 12.4        \\
BELT \cite{BELT} & 91.9   & 100.0          & 89.1       & 10.2        \\
Ours  &  91.0        &  99.4        & 87.8       & 94.9      \\  \bottomrule
\end{tabular}
\label{tab:MMBD_defense}
\end{table}

\begin{table}[h]
\centering
\caption{Attack performance measured by ACC (\%) and ASR (\%) for 9 backdoor attacks against CIFAR-10 dataset and ResNet20 implemented in MOTH.}
\begin{tabular}{@{}ccccc@{}} \toprule
Defense & \multicolumn{2}{c}{No Defense} & \multicolumn{2}{c}{MOTH \cite{moth}} \\ \cmidrule(l){2-3} \cmidrule(l){4-5} 
Attack  & ACC        & ASR           & ACC        & ASR    \\ \midrule
BadNets \cite{badnets} & 94.1           & 100.0           & 91.5    & 0.7       \\
Blend \cite{Chen2017TargetedBA}  & 94.0           & 91.9          & 91.2       & 82.5     \\
WaNet \cite{wanet}  & 93.1           & 93.7          & 90.1       & 14.0     \\
SIG \cite{sig} & 93.8 & 82.7 & 90.8 & 65.8 \\
DUBA \cite{DUBA} & 93.6 & 100.0 & 90.4 & 0.4 \\
Narcissus-D \cite{Narcissus} 	   & 93.4   & 100.0  & 90.1 & 9.8 \\
HCB \cite{watchout} & 93.7          & 99.1          & 91.2       & 0.0        \\
BELT \cite{BELT} & 93.6   & 100.0          & 90.8       & 100.0        \\
Ours  &  93.3    &  98.3        & 90.1       & 0.8      \\ \bottomrule
\end{tabular}
\label{tab:MOTH_defense}
\end{table}

We evaluate attack effectiveness of LADDER and other attacks against ASD, CBD  DBD, FIP, MM-BD and MOTH, using their default parameter settings. 
The results are in \Cref{tab:asd_defense}, \Cref{tab:CBD_defense} and \Cref{tab:DBD_defense},  respectively.
In ASD, CBD, DBD and FIP, all the attacks achieve the ASRs above 97\% without defenses; otherwise they fall drastically (ranging from $0\%\sim25.0\%$ in ASD, $0.3\%\sim5.2\%$ in CBD, $0\%\sim1.7\%$ in DBD, $0.4\%\sim15.3\%$ in FIP). 
Under defenses, LADDER provides better ASRs of 25.0\% under ASD and 5.22\% under CBD, indicating a slightly advantage on evading defenses than others.  
This is because LADDER triggers require smaller perturbations (only 0.3183 $l_2$-norm on CIFAR-10). 
As a result, the poisoned samples are more likely to be split into the clean data pool by ASD and seldom being detected by CBD with their statistical effect, thus delivering relatively higher ASR of LADDER than others. 
In DBD, the attack effectiveness of LADDER is eliminated, while the ACC is with a roughly 12\% drop down.
DBD eliminates the attack of LADDER because of its fine-pruning process.
Since the trigger produced by LADDER is "weaker",  the trigger injected into the model is gradually pruned and eventually erased after a large number of fine-pruning iterations. 
However, LADDER achieves the ASRs of only 10.4\% and 0.8\% in FIP and MOTH respectively. 
This is because LADDER neither alters the model parameters nor manipulates the backdoor training process, whereas MM-BD and MOTH have full access to both models and datasets.
Consequently, the anomaly features of LADDER trigger can be effectively detected and mitigated by the two defenses.

\subsection{Adaptive Defense}
\label{sec:possible_data_defense}
Several image-level anomaly detectors that have been proposed in the spatial domain can be used to eliminate the threat of LADDER on DNNs in black-box environment.  

We propose a frequency domain anomaly detector that locates poisoned images by exploring the statistical information of the spectrum. 
Specifically, given the averaged spectra $M$ of a natural image $x$, 
the averaged magnitudes $\mathcal{A}$ of the frequency bands $f$ in $M$ have a relationship $\mathcal{A} \propto f^{s}$ on the double-logarithmic coordinates with a constant slope $s$=2 \cite{Burton:87, amplitudeSpectra, rethink}.
To obtain $M$ of a given RGB image $x$, we first convert $x$ to the spectrum $X$ using DCT in \Cref{eq:DCT}. 
Then, we compute the power (of magnitudes) in each channel of $X$, i.e., $X_c^{pow}$=$X_c\odot X_c$, where $c\in\{R,G,B\}$ and $\odot$ is the Hadamard product.
For each $X_c^{pow}$, we divide the frequency bands into groups $f=\{f_0, f_1, \cdots, f_{max-1}\}$ where $k\in[0,max)$ and $max$ is the dimension of the spectrum in $X$, so that the frequency bands in each group $f_k$ have the same distance to the upper left corner of the spectrum.
We calculate the averaged magnitude of each group of frequency bands $f_k$ to obtain the averaged spectra $M_{c}$ for each channel $c\in\{R,G,B\}$.
Finally, we obtain the logarithm of the averaged magnitude of the frequency bands from $M_{R}$, $M_{G}$ and $M_B$, i.e, $\log(M^{avg})$= $\log(\frac{(M_{R}+M_{G}+M_B)}{3})$, and fit the slope $s$ with $\log(f)$ and  $\log(M^{avg})$.

We show, in \Cref{tab:average_slope}, $s$ (averaged over 1000 randomly selected samples from CIFAR-10) obtained with clean and poisoned data by the black-box backdoor attacks. 
We see that $s$ is the smallest on clean samples compare to poisoned data. 
The slope $s$ is a feasible indicator to distinguish poisoned data.

\begin{table}[h]
\caption{The averaged $s$ and standard deviation on 1000 randomly chosen images from CIFAR-10 under  attacks.}
\scalebox{0.86}{
\begin{tabular}{cccccccc} \toprule
Attacks & Clean & BadNets & SIG & Blend & FTrojan & FIBA & Ours \\ \midrule
\multirow{2}{*}{Slope} & -1.8922 & -1.6882 & -1.5922 & -1.7602 & -1.8236 & -1.7826 & -1.8238 \\
 & (0.3810) & (0.3803 & (0.3645) & (0.3509) & (0.3591) & (0.3456) & (0.3803) \\ \bottomrule
\end{tabular}}
\label{tab:average_slope}
\end{table}

\subsection{Trade-offs among Attack Objectives}
\label{sec:objconflict_rob_stealth}

We illustrate the conflict between attack effectiveness and stealthiness in \Cref{fig:objective_conflicting}.
To further investigate the trade-off between stealthiness and robustness, we generate random noise of size $3\times3$ with an initial $l_2$-norm of 0.25. 
We create two additional variants by scaling the $l_2$-norm of the original noise by 2$\times$ and 4$\times$.
These noises are used as triggers and injected into the low-, mid-, and high-frequency regions. 
We evaluate attack effectiveness and robustness of each noise under different levels of stealthiness and injection regions on CIFAR-10 using ResNet18. 
Attack robustness is measured by averaging ASRs after the preprocessings (see \Cref{preprocess_defense}).

In \Cref{tab:objconflict_rob_ste_eff},
increasing the $l_2$-norm (i.e., reducing stealthiness) enhances attack robustness in both low- and high-frequency regions, though it has a minimal effect in the mid-frequency region.
For instance, raising the $l_2$-norm from 0.25 to 1.0 improves attack robustness by 15.37\% in the low-frequency region while yielding only a slight increase of 1.44\% in the mid-frequency region. 
A closer examination of the Eff-to-Rob ratio reveals that, in the low-frequency region, increasing the $l_2$-norm has a minimal impact on robustness, with a max. difference of 3.17\%.
Moreover, in the mid-frequency region, the ratio closely aligns with Robs across different $l_2$-norm values.
In the high-frequency region, the ratio rises by 11.79\%.
The results indicate a distinct trade-off between stealthiness and attack robustness in both low- and high-frequency regions.

\Cref{tab:objconflict_rob_ste_eff} also indicates that under the same stealthiness level, inserting trigger patterns in different spectral regions has a modest impact on effectiveness.
For example, with an $l_2$-norm of 0.25, moving the trigger from low-frequency to high-frequency region increases attack effectiveness by 13.88\%. 
But this adjustment significantly harms robustness, resulting in a 50.89\% decrease.
We conclude that designing triggers in the low-frequency region has a minimal impact on attack effectiveness while significantly enhancing trigger robustness.

\begin{table}[]
\caption{The attack effectiveness (Eff) (\%), robustness (Rob) (\%) and Eff-to-Rob ratio (\%) on CIFAR-10 and ResNet18, evaluated by injecting  noises into \underline{L}ow-, \underline{M}id- and \underline{H}igh-frequency regions, across different levels of stealthiness.}
\centering
\begin{tabular}{@{}ccccc@{}}
\toprule
\multirow{2}{*}{$l_2$-norm} & \multirow{2}{*}{Metric} & \multicolumn{3}{c}{Region of Injection} \\ \cmidrule(l){3-5} 
 &  & L & M & H \\ \midrule
\multirow{3}{*}{0.25} & Eff & 86.12 & 99.96 & 100.0 \\
 & Rob & 81.51 & 30.65 & 30.62 \\
 & Ratio & 94.64 & 30.66 & 30.62 \\ \midrule
\multirow{3}{*}{0.5} & Eff & 95.89 & 100.0 & 100.0 \\
 & Rob & 91.70 & 31.50 & 33.31 \\
 & Ratio & 95.63 & 31.50 & 33.31 \\ \midrule
\multirow{3}{*}{1.0} & Eff & 99.04 & 100.0 & 100.0 \\
 & Rob & 96.88 & 32.10 & 42.41 \\
 & Ratio & 97.81 & 32.10 & 42.41 \\ \bottomrule
\end{tabular}
\label{tab:objconflict_rob_ste_eff}
\end{table}

\subsection{Overview of Model parameters}
\label{sec:model_parameters}
\Cref{table:model_total_parameters} briefly summarizes the number of parameters. 
We adopt various models acting as the surrogate and the victim in different scales.

\begin{table}[h]
\centering
\caption{Total parameters (million) per neural network.} 
\begin{tabular}{cc} 
\toprule
Model                   &  Number of parameters \\ 
\midrule
CNN                & 3.52               \\
GoogLeNet           & 6.80               \\
ResNet18 & 11.69               \\
PreAct-ResNet18          & 11.69               \\
ViT & 86.0 \\
VGG11              &    132.87            \\
VGG16              &   138.37             \\
VGG19              &  143.68              \\
\bottomrule
\end{tabular}
\label{table:model_total_parameters}
\end{table}

\subsection{Time Cost for Models and Optimization Hyperparameters}
\label{sec:complexity_analysis_append}

We investigate the time cost against population size and the number of trigger perturbations, two critical parameters for optimization.
\begin{figure}[h]
     \centering
     \begin{subfigure}[b]{0.2\textwidth}
         \centering
         \includegraphics[width=\textwidth]{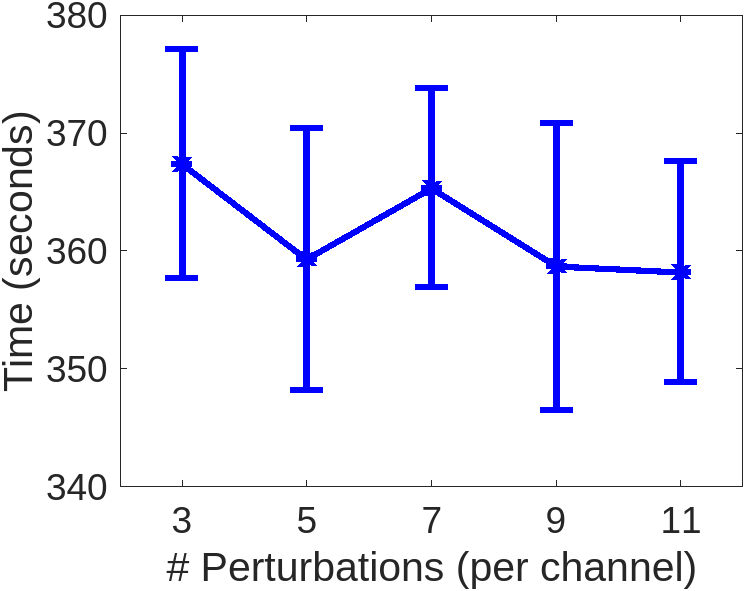}
         \caption{Time cost against the number of perturbations.}
         \label{fig:time_pixelsPerChannel}
     \end{subfigure} 
     \begin{subfigure}[b]{0.2\textwidth}
         \centering
         \includegraphics[width=\textwidth]{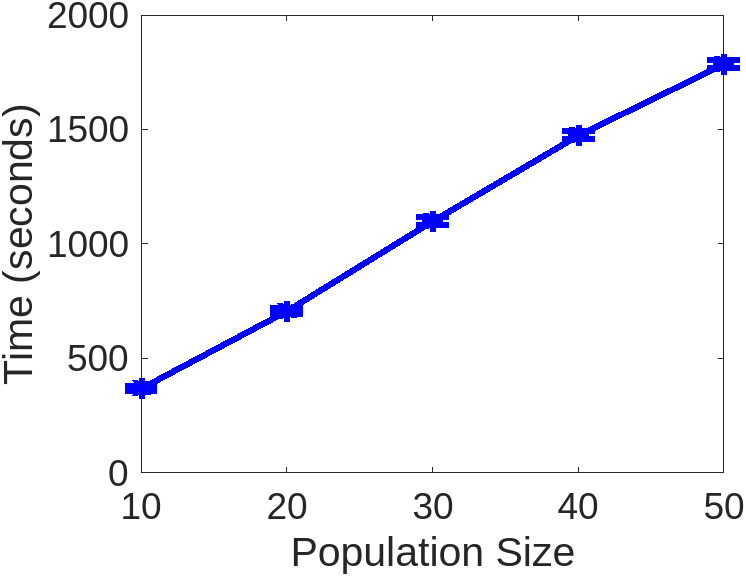}
         \caption{Time cost against the population size.}
         \label{fig:time_popsize}
     \end{subfigure}
    \caption{The time cost (averaged over 5 runs) of LADDER versus (a) the number of manipulated frequency bands $n$ and (b) population size $p$ on ResNet18 and CIFAR-10 using the default parameter settings and hardware specifications.}
    \label{fig:time_cost_pop_pixels}
\end{figure}
In \Cref{fig:time_pixelsPerChannel}, under the default experimental settings when $n=3$, i.e., 3 perturbations are manipulated, the average time cost is 370 seconds with a standard deviation of about 10 seconds.
Increasing $n$ to 11, however, does not have a significant effect on the cost. 
This is because the variation operations affect a fixed number of perturbations per optimization iteration, no matter how many perturbations are included in the trigger.
In this sense, the time to execute the variation on each trigger is almost consistent.

\Cref{fig:time_popsize} shows LADDER's time cost against the population size $p$ between 10 and 50.
Starting from a time cost of 370 seconds for $n$=10, the cost increases by about 300 seconds for every 10 more triggers in the population, reaching 700, 1050, 1400, and 1700 seconds when $p$ is set ot 20, 30, 40, and 50, respectively.
Thus, the cost of LADDER is linearly proportional to the population size. 
The evolutionary algorithm treats each individual (i.e., a trigger) in the population sequentially.
Thus, the time to process triggers is linearly proportional to the number of triggers in the population.

\begin{figure}[]
    \centering
    \scalebox{0.3}{
\includegraphics{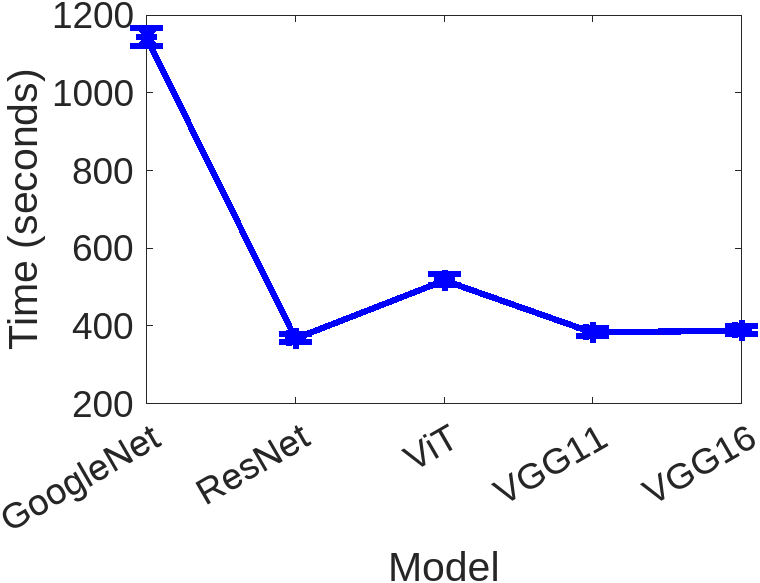}}
    \caption{The time cost (averaged on 5 runs) against GoogleNet, ResNet18, ViT, VGG11 and VGG16 on CIFAR-10 under default parameter settings.}
    \label{fig:time_models}
\end{figure}

The most expensive step of LADDER is to evaluate the attack effectiveness on a semi-converged surrogate model.
We run LADDER with 5 surrogate models and include the results in \Cref{fig:time_models}. 
The costs of LADDER range from 360 to 1100 seconds against GoogleNet, ResNet18, ViT, VGG11 and VGG16,  which have parameter counts ranging from 6.8 to 138.37 million (see \Cref{table:model_total_parameters}).
GooglNet is the most time-consuming surrogate model, requiring 1100 seconds, despite the fact that it has the minimum number of parameters (6.8 million).
Recall that VGGs have fewer parameters for the conv layers but more for the fully-connected layers compared to others. 
According to \cite{convExpensive}, %the computation on 
the conv layers request more time costs on computation than the fully connected layers. 
Accordingly, VGGs (with less conv structures) outperforms GoogleNet in the results.

\subsection{Impact of Frequency Bands $n$ and Perturbations $\epsilon$ }
\label{sec:abla_freq_band_perturb_pixels}

\begin{figure}[]
    \centering
    \begin{subfigure}[b]{0.23\textwidth}
    \centering
    \includegraphics[width=\textwidth]{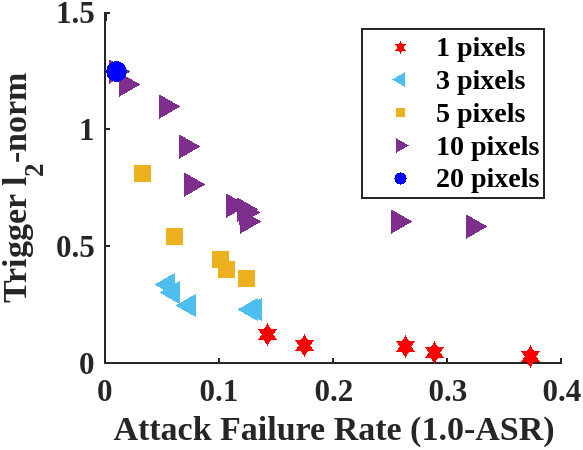}
         \caption{Sensitivity of the number of perturbations.}
         \label{fig:abla_num_pixels}
    \end{subfigure} 
    \begin{subfigure}[b]{0.21\textwidth}
    \centering
    \includegraphics[width=\textwidth]{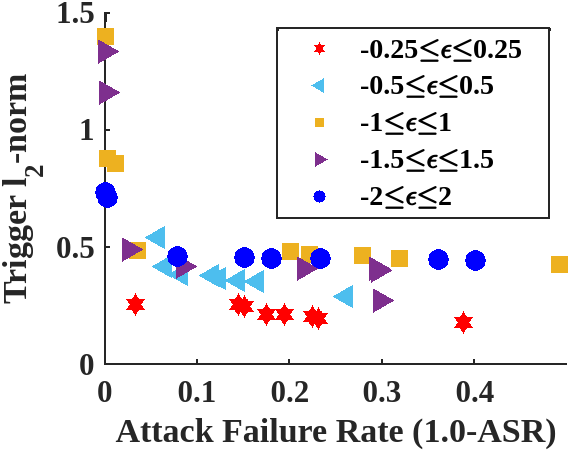}
         \caption{Sensitivity of the range of perturbation.}
    \label{fig:abla_epsilon}
    \end{subfigure} 
\caption{Sensitivity of (a) the number of perturbations and (b) the range of perturbations towards trigger optimization.}
\label{fig:ablation_epsilon_pixels}
\end{figure}

LADDER iteratively optimizes a set of triggers on a given dataset until the triggers achieve satisfactory dual-domain stealthiness and attack effectiveness.
A trigger consists of a set of frequency bands to perturb and the magnitudes of those perturbations.
During the optimization process,
$\epsilon$ restrains the maximum perturbation of each frequency band while $n$ controls the number of manipulated frequency bands.  
To investigate the impact of $\epsilon$ and $n$ towards the attack effectiveness and stealthiness, we showcase their sensitivity investigation towards the trigger searching process in \Cref{fig:ablation_epsilon_pixels}.

In \Cref{fig:abla_num_pixels}, we search triggers for CIFAR-10 dataset with $n$ of 1, 3, 5, 10 and 20 pixels in the spectrum under the default perturbation range $|\epsilon|\leq$0.5, 50 optimization iterations for trigger search and 200 epochs for target model training, and show the objective values of the respective triggers obtained by LADDER.
We can see that with the increase of the number of perturbations, the convergence of triggers (i.e. the objective values of triggers toward ideal) gets worse. 
This is because a large $n$ results in an extensive search space, and consequently leads to a slower convergence under the same optimization budgets.
When $n\leq$5, the attack failure rate of triggers shifts from 0 to 0.4. 
By increasing $n$, a stronger backdoor pattern is captured, which brings better attack effectiveness yet results in impractically bad stealthiness (high $l_2$-norm). 
In contrast, when $n$=1, the triggers' pattern is too tiny to be captured by CNN models. 
Consequently, such triggers fail to achieve practical attack effectiveness. 
According to the experiments, $n$=3 is a practical option to guarantee a proper ASR while remaining stealthy without significantly extending the search space.

Under the default $n$=3, we investigate the sensitivity of $\epsilon$ by setting it to $|\epsilon|\leq$ 0.25, 0.5, 1, 1.5 and 2 under CIFAR-10 dataset, and draw the objective values of the obtained triggers in \Cref{fig:abla_epsilon}.
A larger $\epsilon$ results in a slower convergence speed yet the obtained triggers cover a wider range of objective values than a smaller $\epsilon$.
This occurs because a large value of $\epsilon$ enables LADDER to search triggers with more possible patterns.
However, a trigger should ideally maintain a practical stealthiness while minimally sacrificing attack effectiveness.
From this perspective, an expanded range of $\epsilon$ easily leads to the triggers that either are not stealthy (e.g. $l_2$-norm$>$0.5) or yield an unsatisfied ASR (e.g. AFR$\geq$10\%).
In this work, $|\epsilon|\leq$0.5 is adopted as the default setting as the triggers obtained under this are generally practical w.r.t. stealthiness and attack effectiveness.

\subsection{Searching Triggers with Lagrange Multipliers+SGD in the Frequency Domain}
\label{sec:lagrange+SGD_freq}

\begin{figure}[]
    \centering
    \scalebox{0.34}{\includegraphics{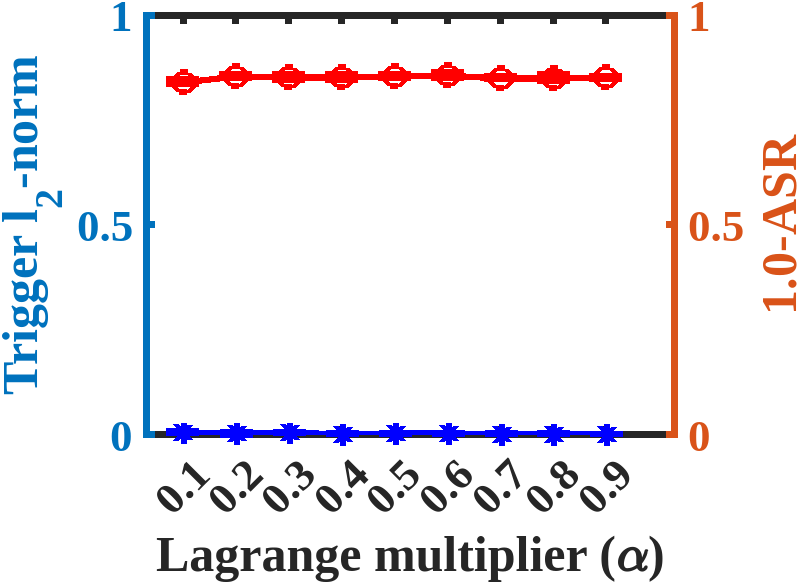}}
    \caption{The impact of Lagrange coefficient $\alpha$ in backdoor attack formulated with Lagrange multipliers and solved by SGD concerning trigger perceptibility (measured with $l_2$-norm) and attack failure rate.}
    \label{fig:Lagrange+SGD_Frequency}
\end{figure}

To confirm the existence of conflicts among objectives in the frequency domain, we conduct experiments by optimizing a frequency trigger $t$ under the same experimental setup as in the main body, except that a trigger function designed specifically for the frequency domain is used to produce $D_{bd}$. 
We show the results in \Cref{fig:Lagrange+SGD_Frequency} which averaged by 10 repetitions on CIFAR-10 and PreAct-ResNet18, where the error bar indicates the standard deviation.

Trigger stealthiness (in blue) and attack effectiveness (in red) are mutually exclusive for all the values of Lagrange coefficient $\alpha$, confirming the conflict between the objectives. 
In detail, the triggers optimized with $\alpha$ between 0.1 and 0.9 produce an extremely excellent stealthy ($l_2$-norm below 0.01) trigger. 
But their ASR values are all below 20\% (1.0-\emph{ASR} is in between 0.83 and 0.87) since the triggers' pattern is too "weak" to be captured by the CNN model.
Thus, they are impractical in real-world attack scenarios.

\subsection{The Consistency of \texorpdfstring{$\mathcal{L}_{2}$}{Lg} Norm-based Stealthiness Measurement between Spatial and Spectral Domains}
\label{sec:LpDifConsistency}
This work achieves trigger stealthiness in \textit{dual domains}.
In our trigger design, the stealthiness in dual domains can both be restricted by the $l_p$-norm of disparity between the clean and poisoned images, and the restriction in one domain naturally guarantees the stealthiness in the other when $p$=2.
We theoretically prove this conclusion below. 

\begin{lemma}
A frequency signal \emph{X}=$\left\{X^{0},X^{1}, \cdots, X^{N-1}\right\}$ and the corresponding spatial signal $\emph{x}=\left\{x^{0},x^{1}, \cdots, x^{N-1}\right\}$, where $x=\mathcal{D}^{-1}(X)$ is obtained with type-II IDCT, has the same inner product, i.e., 
$\left\langle\ X,X \right\rangle \equiv \left\langle\ \mathcal{D}^{-1}(X),\mathcal{D}^{-1}(X)\right\rangle$ 
\label{lemma:innerproduct}
\end{lemma}
\proof By expanding the standard type-II DCT expression, we have:
\begin{equation}
\begin{aligned}
    & [X_{0}, X_{1}, \cdots, X_{N-1}] = \\
    & [x_{0}, x_{1}, \cdots, x_{N-1}]  
    \begin{bmatrix}
     C_{0,0} & \cdots   & C_{0,N-1} \\ 
     & \vdots  & \ddots   & \vdots    \\
     C_{N-1,0} & \cdots & C_{N-1,N-1} 
    \end{bmatrix} \\
\end{aligned}
    \label{eq:typeIIDCT_matrixexp}
\end{equation}
The above \Cref{eq:typeIIDCT_matrixexp} can be concisely denoted as:
\begin{equation}
    \mathcal{D}^{-1}(\emph{X}) = C^{T}XC
\end{equation}
where \emph{C} is orthogonal matrix, so that \emph{C}$^{T}$\emph{C}=\emph{I}. Based on \Cref{eq:typeIIDCT_matrixexp}, we can rewrite the inner product $\left\langle \mathcal{D}^{-1}(X),\mathcal{D}^{-1}(X) \right\rangle$ as: 
\begin{equation}
\begin{aligned}
\label{eq:IDCT_innerproduct_trace}
    \left\langle \mathcal{D}^{-1}(X),\mathcal{D}^{-1}(X)\right\rangle & = tr((C^{T}XC)^{T}(C^{T}XC)) \\
    & = tr(C^{T}X^{T}CC^{T}XC) \\
    & = tr(C^{T}X^{T}XC)
\end{aligned}
\end{equation}
where \emph{tr}($\cdot$) denotes the trace of matrix and ($\cdot$)$^{T}$ is the matrix transpose. 
Since $C$ is orthogonal matrix, we have 
\begin{equation}
\label{eq:Ctranspose=Cinverse}
    C^{T}=C^{-1}
\end{equation}
by plugging \Cref{eq:Ctranspose=Cinverse} into \Cref{eq:IDCT_innerproduct_trace} and considering the cyclic property of trace, we have
\begin{equation}
\begin{aligned}
    \left\langle \mathcal{D}^{-1}(X),\mathcal{D}^{-1}(X)\right\rangle & = tr(X^{T}X)
     = \left\langle X,X \right\rangle
\end{aligned}
\end{equation}
the lemma is therefore proved.

\begin{theorem}
Given a clean image $\emph{x}_{c}$ and its  poisoned sample $\emph{x}_{bd}$ in the spatial domain, where $\emph{x}_{bd}$ is crafted by adding a frequency trigger pattern $\delta$ in the spectrum of $\emph{x}_{c}$, the $l_{2}$-norm of disparity between $\emph{x}_{c}$ and $\emph{x}_{bd}$ in the spatial domain and spectral domain are consistent, i.e., 
\begin{equation}
    \left\|DCT(x_{c})-DCT(x_{bd})\right\|_{2} \equiv \left\| x_{c}- x_{bd} \right\|_{2}.
    \label{eq_theorem_l2normconsistency}
\end{equation}
\label{theorem_l2normconsistency}
\end{theorem}
\proof Based on the definition of our data poisoning process where triggers are designed and patched to the images in the frequency domain, we have:
\begin{equation}
    \delta \triangleq \mathcal{D}(x_{bd})-\mathcal{D}(x_{c}) 
\end{equation}
That's to say:
\begin{equation}
    \left\|\mathcal{D}(x_{bd})-\mathcal{D}(x_{c})\right\|_{2}=\left\| \delta \right\|_{2}
    \label{theorem_l2normconsistencyLeft}
\end{equation}
based on the method we generate poisoned image $x_{bd}$ (in the spatial domain) from the corresponding clean image $x_{c}$ (in the spatial domain) and perturbation $\delta$ (in the frequency domain), we have:
\begin{equation}
    \begin{aligned}
        x_{bd} 
        & =  \mathcal{D}^{-1}(\mathcal{D}(x_{c})+\delta) \\
        & = \mathcal{D}^{-1}(\mathcal{D}(x_{c}))+\mathcal{D}^{-1}(\delta) 
         = x_{c} + \mathcal{D}^{-1}(\delta)
    \end{aligned}
\end{equation}
in other words,  
\begin{equation}
    \begin{aligned}
        \left\| x_{bd}-x_{c} \right\|_{2} 
        & = \left\| \mathcal{D}^{-1}(\delta) \right\|_{2}
    \end{aligned}
    \label{theorem_l2normconsistencyRight}    
\end{equation}
Taking \Cref{eq_theorem_l2normconsistency}, \Cref{theorem_l2normconsistencyLeft} and \Cref{theorem_l2normconsistencyRight} into consideration, the task of proving \Cref{theorem_l2normconsistency} is equivalent to the proof of the following relationship:
\begin{equation}
    \left\|\delta\right\|_{2} = \left\| \mathcal{D}^{-1}(\delta) \right\|_{2}    \label{eq:spatial_delta_L2Norm_eq_delta_L2Norm}    
\end{equation}
the $L_{2}$-norm of adversarial perturbation $\delta$ in spatial domain is equivalent to its inner product, with \cref{lemma:innerproduct}, we have:
\begin{equation}
    \begin{aligned}
        \left\|\mathcal{D}^{-1}(\delta)\right\|_{2} & = \left\langle \mathcal{D}^{-1}(\delta),\mathcal{D}^{-1}(\delta) \right\rangle 
         = \left\langle \delta,\delta \right\rangle 
         = \left\| \delta \right\|_{2}
    \end{aligned}
\label{eq:spatial_delta_L2Norm_eq_delta_L2Norm_proof}
\end{equation}
The fact of \Cref{eq:spatial_delta_L2Norm_eq_delta_L2Norm_proof} proves that \Cref{eq:spatial_delta_L2Norm_eq_delta_L2Norm} is satisfied. That's to say, \Cref{eq_theorem_l2normconsistency} is a true statement. Hereby \Cref{theorem_l2normconsistency} is proved.

\subsection{Discussions for LADDER}
\label{sec:appendix_limitations}

\subsubsection{Extending LADDER to the White-box setting}
\label{sec:appendix_discussion_extendEMOBAF_whitebox} 
The performance of LADDER to evade SOTA white-box backdoor defenses can be enhanced if it can access models and control the training process, i.e., extension to white-box scenario.
Following the idea by \cite{latentbackdoorattack} which performs an imperceptible attack on the feature representation (extracted from the victim model), 
we introduce a new attack objective by limiting the distance of latent features between clean and poisoned data in our multi-objective backdoor attack problem.
In detail, we capture the latent representation $L_{cln}$ with clean data and $L_{bd}$ with poisoned data from the last conv layer.
Then, we construct an objective $O_{latent}$=$dist(L_{cln}-L_{bd})$ to reduce the anomaly of the latent representation, where $dist(\cdot)$ is a distance measurement such as cosine distance and $L_2$-norm.
Finally, we replace the objective function by including $O_{latent}$, i.e, 
\begin{equation}
(\delta^*,\nu^*) = \underset{\delta,\nu}{\operatorname{argmin}}  \ O(\delta,\nu) = (O_{1}, O_{2}, O_{3}, O_{latent}),     
\end{equation} 

\subsubsection{Extending Current Backdoor Attacks to Dual Domains}
\label{sec:appendix_discussion_extendfreqBD_dualdomain}
Most attacks, e.g., \cite{inputAware,badnets,sig,refool,wanet, ft,fiba}, 
design triggers merely for single domain.
They cannot be trivially extended to produce dual-domain stealthy triggers. 
This is so because: 
\\ 
\noindent
\textit{a. Fixed trigger pattern from spatial and frequency backdoor attacks cannot be optimized.} 
Several works \cite{badnets,sig} formulate backdoor tasks by placing a fixed pattern on images in the spatial domain, and directly optimize the loss function w.r.t. the accuracy on the clean and backdoor tasks.
Similar ideas \cite{ft,fiba} are used in the frequency domain. 
Wang et al. \cite{ft} put fixed trigger pattern in two frequency bands of a given image, while Feng et al. \cite{fiba} insert triggers by directly overlaying the spectrum of a predefined trigger image to a clean image. 
Since the triggers of the above approaches are fixed, he objective value related to stealthiness becomes a constant and cannot be further optimized.  
\\
\noindent
\textit{b. Achieving stealthiness in the spectral domain could harm the original spatial stealthiness.}
Some papers \cite{refool, wanet} adopt semantically meaningful images as a trigger pattern and further overlap them on clean data to bypass defenses.
Refool \cite{refool} leverages natural phenomena to plant lifelike reflections as a pattern on clean data with mathematical modeling of physical models.
This approach is designed in particular to imitate the reflection in the natural environment rather than seeking minimum trigger perturbations. 
Recall the $l_2$-norm of disparity between clean and poisoned images achieved by Refool is the largest among other attacks.  
In this case, limiting the spectrum disparity between clean and poisoned data can weaken the quality of reflection-based triggers in the spatial domain, hereby harming spatial stealthiness.

%% file: 0_NDSS2024_main.bbl
% Generated by IEEEtranS.bst, version: 1.12 (2007/01/11)
\begin{thebibliography}{10}
\providecommand{\url}[1]{#1}
\csname url@samestyle\endcsname
\providecommand{\newblock}{\relax}
\providecommand{\bibinfo}[2]{#2}
\providecommand{\BIBentrySTDinterwordspacing}{\spaceskip=0pt\relax}
\providecommand{\BIBentryALTinterwordstretchfactor}{4}
\providecommand{\BIBentryALTinterwordspacing}{\spaceskip=\fontdimen2\font plus
\BIBentryALTinterwordstretchfactor\fontdimen3\font minus
  \fontdimen4\font\relax}
\providecommand{\BIBforeignlanguage}[2]{{%
\expandafter\ifx\csname l@#1\endcsname\relax
\typeout{** WARNING: IEEEtranS.bst: No hyphenation pattern has been}%
\typeout{** loaded for the language `#1'. Using the pattern for}%
\typeout{** the default language instead.}%
\else
\language=\csname l@#1\endcsname
\fi
#2}}
\providecommand{\BIBdecl}{\relax}
\BIBdecl

\bibitem{abad2024sneaky}
G.~Abad, O.~Ersoy, S.~Picek, and A.~Urbieta, ``Sneaky spikes: Uncovering
  stealthy backdoor attacks in spiking neural networks with neuromorphic
  data,'' in \emph{Network and Distributed System Security Symposium}, 2024.

\bibitem{DCT}
N.~Ahmed, T.~Natarajan, and K.~Rao, ``Discrete cosine transform,'' \emph{IEEE
  Transactions on Computers}, vol. C-23, no.~1, pp. 90--93, 1974.

\bibitem{sgd}
S.-i. Amari, ``Backpropagation and stochastic gradient descent method,''
  \emph{Neurocomputing}, vol.~5, no. 4-5, pp. 185--196, 1993.

\bibitem{sig}
M.~Barni, K.~Kallas, and B.~Tondi, ``A new backdoor attack in cnns by training
  set corruption without label poisoning,'' in \emph{IEEE International
  Conference on Image Processing}, 2019, pp. 101--105.

\bibitem{autodriving}
M.~Bojarski, D.~Del~Testa, D.~Dworakowski, B.~Firner, B.~Flepp, P.~Goyal, L.~D.
  Jackel, M.~Monfort, U.~Muller, J.~Zhang, X.~Zhang, and J.~Zhao, ``End to end
  learning for self-driving cars,'' \emph{arXiv preprint arXiv:1604.07316},
  2016.

\bibitem{Burton:87}
G.~J. Burton and I.~R. Moorhead, ``Color and spatial structure in natural
  scenes,'' \emph{Applied Optics}, vol.~26, no.~1, pp. 157--170, 1987.

\bibitem{activation_clustering}
B.~Chen, W.~Carvalho, N.~Baracaldo, H.~Ludwig, B.~Edwards, T.~Lee, I.~Molloy,
  and B.~Srivastava, ``Detecting backdoor attacks on deep neural networks by
  activation clustering,'' \emph{arXiv preprint arXiv:1811.03728}, 2018.

\bibitem{deepinspect}
H.~Chen, C.~Fu, J.~Zhao, and F.~Koushanfar, ``Deepinspect: A black-box trojan
  detection and mitigation framework for deep neural networks,'' in
  \emph{Proceedings of the International Joint Conference on Artificial
  Intelligence}, 2019, pp. 4658--4664.

\bibitem{Chen2017TargetedBA}
X.~Chen, C.~Liu, B.~Li, K.~Lu, and D.~Song, ``Targeted backdoor attacks on deep
  learning systems using data poisoning,'' \emph{arXiv preprint
  arXiv:1712.05526}, 2017.

\bibitem{dfst}
S.~Cheng, Y.~Liu, S.~Ma, and X.~Zhang, ``Deep feature space trojan attack of
  neural networks by controlled detoxification,'' in \emph{Proceedings of the
  AAAI Conference on Artificial Intelligence}, vol.~35, no.~2, 2021, pp.
  1148--1156.

\bibitem{watermark}
I.~Cox, J.~Kilian, F.~Leighton, and T.~Shamoon, ``Secure spread spectrum
  watermarking for multimedia,'' \emph{IEEE Transactions on Image Processing},
  vol.~6, no.~12, pp. 1673--1687, 1997.

\bibitem{nsgaii}
K.~Deb, A.~Pratap, S.~Agarwal, and T.~Meyarivan, ``A fast and elitist
  multiobjective genetic algorithm: Nsga-ii,'' \emph{IEEE Transactions on
  Evolutionary Computation}, vol.~6, no.~2, pp. 182--197, 2002.

\bibitem{Deb1995SimulatedBC}
K.~Deb and R.~B. Agrawal, ``Simulated binary crossover for continuous search
  space,'' \emph{Complex System}, vol.~9, 1995.

\bibitem{Deb1996ACG}
K.~Deb and M.~Goyal, ``A combined genetic adaptive search (geneas) for
  engineering design,'' \emph{Computer Science and Informatics}, vol.~26, pp.
  30--45, 1996.

\bibitem{wb}
K.~Doan, Y.~Lao, and P.~Li, ``Backdoor attack with imperceptible input and
  latent modification,'' \emph{Advances in Neural Information Processing
  Systems}, vol.~34, pp. 18\,944--18\,957, 2021.

\bibitem{lira}
K.~Doan, Y.~Lao, W.~Zhao, and P.~Li, ``Lira: Learnable, imperceptible and
  robust backdoor attacks,'' in \emph{Proceedings of the IEEE/CVF International
  Conference on Computer Vision}, 2021, pp. 11\,966--11\,976.

\bibitem{marksman}
K.~D. Doan, Y.~Lao, and P.~Li, ``Marksman backdoor: Backdoor attacks with
  arbitrary target class,'' \emph{Advances in Neural Information Processing
  Systems}, vol.~35, pp. 38\,260--38\,273, 2022.

\bibitem{ViT}
A.~Dosovitskiy, L.~Beyer, A.~Kolesnikov, D.~Weissenborn, X.~Zhai,
  T.~Unterthiner, M.~Dehghani, M.~Minderer, G.~Heigold, S.~Gelly, J.~Uszkoreit,
  and N.~Houlsby, ``An image is worth 16x16 words: Transformers for image
  recognition at scale,'' in \emph{International Conference on Learning
  Representations}, 2021.

\bibitem{esteva2017dermatologist}
A.~Esteva, B.~Kuprel, R.~A. Novoa, J.~Ko, S.~M. Swetter, H.~M. Blau, and
  S.~Thrun, ``Dermatologist-level classification of skin cancer with deep
  neural networks,'' \emph{Nature}, vol. 542, no. 7639, pp. 115--118, 2017.

\bibitem{fiba}
Y.~Feng, B.~Ma, J.~Zhang, S.~Zhao, Y.~Xia, and D.~Tao, ``Fiba:
  Frequency-injection based backdoor attack in medical image analysis,'' in
  \emph{Proceedings of the IEEE/CVF Conference on Computer Vision and Pattern
  Recognition}, 2022, pp. 20\,876--20\,885.

\bibitem{ASD}
K.~Gao, Y.~Bai, J.~Gu, Y.~Yang, and S.-T. Xia, ``Backdoor defense via
  adaptively splitting poisoned dataset,'' in \emph{Proceedings of the IEEE/CVF
  Conference on Computer Vision and Pattern Recognition}, 2023, pp. 4005--4014.

\bibitem{strip}
Y.~Gao, C.~Xu, D.~Wang, S.~Chen, D.~C. Ranasinghe, and S.~Nepal, ``Strip: A
  defence against trojan attacks on deep neural networks,'' in
  \emph{Proceedings of the Annual Computer Security Applications Conference},
  2019, pp. 113--125.

\bibitem{DUBA}
Y.~Gao, H.~Chen, P.~Sun, J.~Li, A.~Zhang, Z.~Wang, and W.~Liu, ``A dual
  stealthy backdoor: From both spatial and frequency perspectives,'' in
  \emph{Proceedings of the AAAI Conference on Artificial Intelligence},
  vol.~38, no.~3, 2024, pp. 1851--1859.

\bibitem{badnets}
T.~Gu, B.~Dolan-Gavitt, and S.~Garg, ``Badnets: Identifying vulnerabilities in
  the machine learning model supply chain,'' \emph{arXiv preprint
  arXiv:1708.06733}, 2017.

\bibitem{LFAP}
C.~Guo, J.~S. Frank, and K.~Q. Weinberger, ``Low frequency adversarial
  perturbation,'' in \emph{Uncertainty in Artificial Intelligence}, 2020, pp.
  1127--1137.

\bibitem{cyo}
H.~A. A.~K. Hammoud and B.~Ghanem, ``Check your other door! creating backdoor
  attacks in the frequency domain,'' \emph{arXiv preprint arXiv:2109.05507},
  2021.

\bibitem{resnet}
K.~He, X.~Zhang, S.~Ren, and J.~Sun, ``Deep residual learning for image
  recognition,'' in \emph{Proceedings of the IEEE/CVF Conference on Computer
  Vision and Pattern Recognition}, 2016, pp. 770--778.

\bibitem{stealthy_freq}
R.~Hou, T.~Huang, H.~Yan, L.~Ke, and W.~Tang, ``A stealthy and robust backdoor
  attack via frequency domain transform,'' \emph{World Wide Web}, pp. 1--17,
  2023.

\bibitem{gtsrb}
S.~Houben, J.~Stallkamp, J.~Salmen, M.~Schlipsing, and C.~Igel, ``Detection of
  traffic signs in real-world images: The german traffic sign detection
  benchmark,'' in \emph{Proceedings of the International Joint Conference on
  Neural Networks}, 2013, pp. 1--8.

\bibitem{DBD}
K.~Huang, Y.~Li, B.~Wu, Z.~Qin, and K.~Ren, ``Backdoor defense via decoupling
  the training process,'' in \emph{International Conference on Learning
  Representations}, 2022.

\bibitem{jahne2005digital}
B.~J{\"a}hne, \emph{Digital Image Processing}.\hskip 1em plus 0.5em minus
  0.4em\relax Springer Science \& Business Media, 2005.

\bibitem{color}
W.~Jiang, H.~Li, G.~Xu, and T.~Zhang, ``Color backdoor: A robust poisoning
  attack in color space,'' in \emph{Proceedings of the IEEE/CVF Conference on
  Computer Vision and Pattern Recognition}, 2023, pp. 8133--8142.

\bibitem{fip}
N.~Karim, A.~A. Arafat, A.~S. Rakin, Z.~Guo, and N.~Rahnavard, ``Fisher
  information guided purification against backdoor attacks,'' in
  \emph{Proceedings of the ACM SIGSAC Conference on Computer and Communications
  Security}, 2024.

\bibitem{dualspace_concept2}
C.~Knaus and M.~Zwicker, ``Dual-domain image denoising,'' in \emph{IEEE
  International Conference on Image Processing}, 2013, pp. 440--444.

\bibitem{ulp}
S.~Kolouri, A.~Saha, H.~Pirsiavash, and H.~Hoffmann, ``Universal litmus
  patterns: Revealing backdoor attacks in cnns,'' in \emph{Proceedings of the
  IEEE/CVF Conference on Computer Vision and Pattern Recognition}, 2020, pp.
  301--310.

\bibitem{cifar10}
A.~Krizhevsky and G.~Hinton, ``Learning multiple layers of features from tiny
  images,'' 2009.

\bibitem{lan2024flowmur}
J.~Lan, J.~Wang, B.~Yan, Z.~Yan, and E.~Bertino, ``Flowmur: A stealthy and
  practical audio backdoor attack with limited knowledge,'' in \emph{IEEE
  Symposium on Security and Privacy}, 2024, pp. 1646--1664.

\bibitem{tiny}
Y.~Le and X.~Yang, ``Tiny imagenet visual recognition challenge,'' \emph{CS
  231N}, vol.~7, no.~7, p.~3, 2015.

\bibitem{litheoretical}
B.~Li and W.~Liu, ``A theoretical analysis of backdoor poisoning attacks in
  convolutional neural networks,'' in \emph{International Conference on Machine
  Learning}, 2024, pp. 8296--8316.

\bibitem{Li2019InvisibleBA}
S.~Li, M.~Xue, B.~Z.~H. Zhao, H.~Zhu, and X.~Zhang, ``Invisible backdoor
  attacks on deep neural networks via steganography and regularization,''
  \emph{IEEE Transactions on Dependable and Secure Computing}, vol.~18, pp.
  2088--2105, 2019.

\bibitem{nad}
Y.~Li, X.~Lyu, N.~Koren, L.~Lyu, B.~Li, and X.~Ma, ``Neural attention
  distillation: Erasing backdoor triggers from deep neural networks,'' in
  \emph{International Conference on Learning Representations}, 2021.

\bibitem{Rethinking}
Y.~Li, T.~Zhai, B.~Wu, Y.~Jiang, Z.~Li, and S.~Xia, ``Rethinking the trigger of
  backdoor attack,'' \emph{arXiv preprint arXiv:2004.04692}, 2020.

\bibitem{issba}
Y.~Li, Y.~Li, B.~Wu, L.~Li, R.~He, and S.~Lyu, ``Invisible backdoor attack with
  sample-specific triggers,'' in \emph{Proceedings of the IEEE/CVF
  International Conference on Computer Vision}, 2021, pp. 16\,463--16\,472.

\bibitem{fine_pruning}
K.~Liu, B.~Dolan-Gavitt, and S.~Garg, ``Fine-pruning: Defending against
  backdooring attacks on deep neural networks,'' in \emph{International
  Symposium on Research in Attacks, Intrusions, and Defenses}, 2018, pp.
  273--294.

\bibitem{refool}
Y.~Liu, X.~Ma, J.~Bailey, and F.~Lu, ``Reflection backdoor: A natural backdoor
  attack on deep neural networks,'' in \emph{European Conference on Computer
  Vision}, 2020, pp. 182--199.

\bibitem{celeba}
Z.~Liu, P.~Luo, X.~Wang, and X.~Tang, ``Deep learning face attributes in the
  wild,'' in \emph{Proceedings of the IEEE/CVF International Conference on
  Computer Vision}, 2015, pp. 3730--3738.

\bibitem{lv2023data}
P.~Lv, C.~Yue, R.~Liang, Y.~Yang, S.~Zhang, H.~Ma, and K.~Chen, ``A data-free
  backdoor injection approach in neural networks,'' in \emph{USENIX Security
  Symposium}, 2023, pp. 2671--2688.

\bibitem{watchout}
H.~Ma, S.~Wang, Y.~Gao, Z.~Zhang, H.~Qiu, M.~Xue, A.~Abuadbba, A.~Fu, S.~Nepal,
  and D.~Abbott, ``Watch out! simple horizontal class backdoor can trivially
  evade defense,'' in \emph{Proceedings of the ACM SIGSAC Conference on
  Computer and Communications Security}, 2024.

\bibitem{SVHN}
Y.~Netzer, T.~Wang, A.~Coates, A.~Bissacco, B.~Wu, and A.~Y. Ng, ``Reading
  digits in natural images with unsupervised feature learning,'' in
  \emph{Neural Information Processing Systems Workshop on Deep Learning and
  Unsupervised Feature Learning}, 2011.

\bibitem{inputAware}
T.~A. Nguyen and A.~Tran, ``Input-aware dynamic backdoor attack,'' in
  \emph{Advances in Neural Information Processing Systems}, vol.~33, 2020, pp.
  3454--3464.

\bibitem{wanet}
T.~A. Nguyen and A.~T. Tran, ``Wanet - imperceptible warping-based backdoor
  attack,'' in \emph{International Conference on Learning Representations},
  2021.

\bibitem{o2015introduction}
K.~O'shea and R.~Nash, ``An introduction to convolutional neural networks,''
  \emph{arXiv preprint arXiv:1511.08458}, 2015.

\bibitem{paszke2019pytorch}
A.~Paszke, S.~Gross, F.~Massa, A.~Lerer, J.~Bradbury, G.~Chanan, T.~Killeen,
  Z.~Lin, N.~Gimelshein, L.~Antiga, A.~Desmaison, A.~Köpf, E.~Yang, Z.~DeVito,
  M.~Raison, A.~Tejani, S.~Chilamkurthy, B.~Steiner, L.~Fang, J.~Bai, and
  C.~Soumith, ``Pytorch: An imperative style, high-performance deep learning
  library,'' \emph{Advances in Neural Information Processing Systems}, vol.~32,
  pp. 8026--8037, 2019.

\bibitem{generative_distribution_modeling}
X.~Qiao, Y.~Yang, and H.~Li, ``Defending neural backdoors via generative
  distribution modeling,'' \emph{Advances in Neural Information Processing
  Systems}, vol.~32, pp. 14\,027--14\,036, 2019.

\bibitem{deepsweep}
H.~Qiu, Y.~Zeng, S.~Guo, T.~Zhang, M.~Qiu, and B.~Thuraisingham, ``Deepsweep:
  An evaluation framework for mitigating dnn backdoor attacks using data
  augmentation,'' in \emph{Proceedings of the ACM Asia Conference on Computer
  and Communications Security}, 2021, pp. 363--377.

\bibitem{BELT}
H.~Qiu, J.~Sun, M.~Zhang, X.~Pan, and M.~Yang, ``Belt: Old-school backdoor
  attacks can evade the state-of-the-art defense with backdoor exclusivity
  lifting,'' in \emph{IEEE Symposium on Security and Privacy}, 2024, pp.
  2124--2141.

\bibitem{hiddentriggerbd}
A.~Saha, A.~Subramanya, and H.~Pirsiavash, ``Hidden trigger backdoor attacks,''
  in \emph{Proceedings of the AAAI Cconference on Artificial Intelligence},
  vol.~34, no.~07, 2020, pp. 11\,957--11\,965.

\bibitem{dynamic}
A.~Salem, R.~Wen, M.~Backes, S.~Ma, and Y.~Zhang, ``Dynamic backdoor attacks
  against machine learning models,'' in \emph{IEEE European Symposium on
  Security and Privacy}, 2022, pp. 703--718.

\bibitem{gradcam}
R.~R. Selvaraju, M.~Cogswell, A.~Das, R.~Vedantam, D.~Parikh, and D.~Batra,
  ``Grad-cam: Visual explanations from deep networks via gradient-based
  localization,'' in \emph{Proceedings of the IEEE/CVF International Conference
  on Computer Vision}, 2017, pp. 618--626.

\bibitem{sharma2019OnTheEffe}
Y.~Sharma, G.~W. Ding, and M.~A. Brubaker, ``On the effectiveness of low
  frequency perturbations,'' in \emph{Proceedings of the International Joint
  Conference on Artificial Intelligence}, 2019, pp. 3389--3396.

\bibitem{NEURIPS2023_b36554b9}
Y.~Shi, M.~Du, X.~Wu, Z.~Guan, J.~Sun, and N.~Liu, ``Black-box backdoor defense
  via zero-shot image purification,'' in \emph{Advances in Neural Information
  Processing Systems}, 2023, pp. 57\,336--57\,366.

\bibitem{vgg}
K.~Simonyan and A.~Zisserman, ``Very deep convolutional networks for
  large-scale image recognition,'' in \emph{International Conference on
  Learning Representations}, 2015.

\bibitem{googlenet}
C.~Szegedy, W.~Liu, Y.~Jia, P.~Sermanet, S.~Reed, D.~Anguelov, D.~Erhan,
  V.~Vanhoucke, and A.~Rabinovich, ``Going deeper with convolutions,'' in
  \emph{Proceedings of the IEEE/CVF Conference on Computer Vision and Pattern
  Recognition}, 2015, pp. 1--9.

\bibitem{bypassing}
T.~J.~L. Tan and R.~Shokri, ``Bypassing backdoor detection algorithms in deep
  learning,'' in \emph{IEEE European Symposium on Security and Privacy}, 2020,
  pp. 175--183.

\bibitem{moth}
G.~Tao, Y.~Liu, G.~Shen, Q.~Xu, S.~An, Z.~Zhang, and X.~Zhang, ``Model
  orthogonalization: Class distance hardening in neural networks for better
  security,'' in \emph{IEEE Symposium on Security and Privacy}, 2022, pp.
  1372--1389.

\bibitem{amplitudeSpectra}
D.~J. Tolhurst, Y.~Tadmor, and T.~Chao, ``Amplitude spectra of natural
  images,'' \emph{Ophthalmic and Physiological Optics}, pp. 229--232, 1992.

\bibitem{spectral}
B.~Tran, J.~Li, and A.~Madry, ``Spectral signatures in backdoor attacks,''
  \emph{Advances in Neural Information Processing Systems}, vol.~31, pp.
  8011--8021, 2018.

\bibitem{nc}
B.~Wang, Y.~Yao, S.~Shan, H.~Li, B.~Viswanath, H.~Zheng, and B.~Y. Zhao,
  ``Neural cleanse: Identifying and mitigating backdoor attacks in neural
  networks,'' in \emph{IEEE Symposium on Security and Privacy}, 2019, pp.
  707--723.

\bibitem{MM-BM}
H.~Wang, Z.~Xiang, D.~J. Miller, and G.~Kesidis, ``{MM-BD: Post-Training
  Detection of Backdoor Attacks with Arbitrary Backdoor Pattern Types Using a
  Maximum Margin Statistic},'' in \emph{IEEE Symposium on Security and
  Privacy}, 2024, pp. 1994--2012.

\bibitem{Wang2023RobustBA}
R.~Wang, H.~Chen, Z.~Zhu, L.~Liu, and B.~Wu, ``Versatile backdoor attack with
  visible, semantic, sample-specific, and compatible triggers,'' \emph{arXiv
  preprint arXiv:2306.00816}, 2023.

\bibitem{ft}
T.~Wang, Y.~Yao, F.~Xu, S.~An, H.~Tong, and T.~Wang, ``An invisible black-box
  backdoor attack through frequency domain,'' in \emph{European Conference on
  Computer Vision}, 2022, pp. 396--413.

\bibitem{dualspace_concept1}
Z.~Wang, D.~Liu, S.~Chang, Q.~Ling, Y.~Yang, and T.~S. Huang, ``D3: Deep
  dual-domain based fast restoration of jpeg-compressed images,'' in
  \emph{Proceedings of the IEEE/CVF Conference on Computer Vision and Pattern
  Recognition}, 2016, pp. 2764--2772.

\bibitem{latentbackdoorattack}
Y.~Yao, H.~Li, H.~Zheng, and B.~Y. Zhao, ``Latent backdoor attacks on deep
  neural networks,'' in \emph{Proceedings of the ACM SIGSAC Conference on
  Computer and Communications Security}, 2019, pp. 2041--2055.

\bibitem{Narcissus}
Y.~Zeng, M.~Pan, H.~A. Just, L.~Lyu, M.~Qiu, and R.~Jia, ``Narcissus: A
  practical clean-label backdoor attack with limited information,'' in
  \emph{Proceedings of the ACM SIGSAC Conference on Computer and Communications
  Security}, 2023, pp. 771--785.

\bibitem{rethink}
Y.~Zeng, W.~Park, Z.~M. Mao, and R.~Jia, ``Rethinking the backdoor attacks'
  triggers: A frequency perspective,'' in \emph{Proceedings of the IEEE/CVF
  International Conference on Computer Vision}, 2021, pp. 16\,473--16\,481.

\bibitem{zhang2024badmerging}
J.~Zhang, J.~Chi, Z.~Li, K.~Cai, Y.~Zhang, and Y.~Tian, ``Badmerging: Backdoor
  attacks against model merging,'' \emph{arXiv preprint arXiv:2408.07362},
  2024.

\bibitem{lpips}
R.~Zhang, P.~Isola, A.~A. Efros, E.~Shechtman, and O.~Wang, ``The unreasonable
  effectiveness of deep features as a perceptual metric,'' in \emph{Proceedings
  of the IEEE/CVF Conference on Computer Vision and Pattern Recognition}, 2018,
  pp. 586--595.

\bibitem{convExpensive}
X.~Zhang, X.~Zhou, M.~Lin, and J.~Sun, ``Shufflenet: An extremely efficient
  convolutional neural network for mobile devices,'' in \emph{Proceedings of
  the IEEE/CVF Conference on Computer Vision and Pattern Recognition}, 2018,
  pp. 6848--6856.

\bibitem{CBD}
Z.~Zhang, Q.~Liu, Z.~Wang, Z.~Lu, and Q.~Hu, ``Backdoor defense via
  deconfounded representation learning,'' in \emph{Proceedings of the IEEE/CVF
  Conference on Computer Vision and Pattern Recognition}, 2023, pp.
  12\,228--12\,238.

\bibitem{defeat}
Z.~Zhao, X.~Chen, Y.~Xuan, Y.~Dong, D.~Wang, and K.~Liang, ``Defeat: Deep
  hidden feature backdoor attacks by imperceptible perturbation and latent
  representation constraints,'' in \emph{Proceedings of the IEEE/CVF Conference
  on Computer Vision and Pattern Recognition}, 2022, pp. 15\,213--15\,222.

\bibitem{IBA}
N.~Zhong, Z.~Qian, and X.~Zhang, ``Imperceptible backdoor attack: From input
  space to feature representation,'' in \emph{Proceedings of the International
  Joint Conference on Artificial Intelligence}, 2022, pp. 1736--1742.

\bibitem{NEURIPS2023_03df5246}
M.~Zhu, S.~Wei, H.~Zha, and B.~Wu, ``Neural polarizer: A lightweight and
  effective backdoor defense via purifying poisoned features,'' in
  \emph{Advances in Neural Information Processing Systems}, vol.~36, 2023, pp.
  1132--1153.

\end{thebibliography}
